\documentclass[a4paper,11pt]{article}
\pdfoutput=1
\usepackage{jheppub}
\usepackage{multirow}
\usepackage{graphicx}
\usepackage{amssymb}
\usepackage{amsmath}
\usepackage{amsfonts}
\usepackage[utf8]{inputenc}
\usepackage{bm}
\usepackage{xcolor}
\usepackage{soul}
\newcommand{\ii}{\mathrm{i}}
\newcommand{\rme}{\mathrm{e}}

\newcommand{\vev}[1]{\langle #1 \rangle}

\DeclareMathOperator{\Tr}{Tr}
\DeclareMathOperator{\tr}{tr}
\newcommand{\trp}{\Tr_{\cR}^\prime}

\newcommand{\adj}{\mathrm{adj}}
\newcommand{\bone}{\mathbf{1}}
\makeatletter
\newcommand*{\letterdef@}{}
\newcommand*{\letterdef}[3]{%
	\def\letterdef@##1{\expandafter\newcommand\csname #1\endcsname{#2{##1}}}%
	\@tfor\@tempa :=#3\do{\expandafter\letterdef@\expandafter{\@tempa}}}
\makeatother
\letterdef{c#1} {\mathcal}{ABCDEFGHIJKLMNOPQRSTUVWXYZ} % \cX = \mathcal{X}
\letterdef{rm#1}{\mathrm} {dDmM} % \rmX = \mathrm{X} for X in {dDmM}

\newdimen\tableauside\tableauside=1.0ex
\newdimen\tableaurule\tableaurule=0.4pt
\newdimen\tableaustep
\def\phantomhrule#1{\hbox{\vbox to0pt{\hrule height\tableaurule
			width#1\vss}}}
\def\phantomvrule#1{\vbox{\hbox to0pt{\vrule width\tableaurule
			height#1\hss}}}
\def\sqr{\vbox{%
		\phantomhrule\tableaustep
		\hbox{\phantomvrule\tableaustep\kern\tableaustep\phantomvrule\tableaustep}%
		\hbox{\vbox{\phantomhrule\tableauside}\kern-\tableaurule}}}
\def\squares#1{\hbox{\count0=#1\noindent\loop\sqr
		\advance\count0 by-1 \ifnum\count0>0\repeat}}
\def\tableau#1{\vcenter{\offinterlineskip
		\tableaustep=\tableauside\advance\tableaustep by-\tableaurule
		\kern\normallineskip\hbox
		{\kern\normallineskip\vbox
			{\gettableau#1 0 }%
			\kern\normallineskip\kern\tableaurule}%
		\kern\normallineskip\kern\tableaurule}}
\def\gettableau#1 {\ifnum#1=0\let\next=\null\else
	\squares{#1}\let\next=\gettableau\fi\next}
\newcommand{\Yfund}{\tableau{1}}
\newcommand{\Ysymm}{\tableau{2}}
\newcommand{\Yasymm}{\tableau{1 1}}

\title{\boldmath BPS Wilson loops in generic conformal
	$\cN=2$ SU($N$) SYM theories}

\author[a,b]{M.~Bill\`o,}
\affiliation[a]{Universit\`a di Torino,
	Dipartimento di Fisica\\
	Via P. Giuria 1, I-10125 Torino, Italy\\}
\affiliation[b]{I.\,N.\,F.\,N. - sezione di Torino, \\
	Via P. Giuria 1, I-10125 Torino, Italy\\}
\emailAdd{billo@to.infn.it}

\author[a,b]{F. Galvagno,}
\emailAdd{galvagno@to.infn.it}

\author[c,b]{and A.~Lerda\,}
\affiliation[d]{Universit\`a del Piemonte Orientale,\\
	Dipartimento di Scienze e Innovazione Tecnologica,\\
	Viale T. Michel 11, I-15121 Alessandria, Italy\\}
\emailAdd{lerda@to.infn.it}

\abstract{We consider the 1/2 BPS circular Wilson loop in a generic $\cN=2$ SU$(N)$ SYM theory 
	with conformal matter content. We study its vacuum expectation value, both at finite $N$ 
	and in the large-$N$ limit, using the interacting matrix model provided by localization 
	results. We single out some families of theories for which the Wilson loop vacuum expectation values 
	approaches the $\cN=4$ result in the large-$N$ limit, in agreement with the fact that they 
	possess a simple holographic dual. At finite $N$ and in the generic case, we explicitly 
	compare the matrix model result with the field-theory perturbative expansion up to 
	order $g^8$ for the terms proportional
	to the Riemann value $\zeta(5)$, finding perfect agreement. 
	Organizing the Feynman diagrams as suggested by the 
	structure of the matrix model turns out to be very convenient for this computation.}


\begin{document}
	\maketitle
	\flushbottom

\section{Introduction}
\label{sec:intro}
An ambitious goal in theoretical physics is to obtain exact results that are valid for all values of parameters and couplings.
This goal, however, is still out of reach for realistic quantum field
theories describing the elementary particles in our world.
It is therefore natural to study models obeying stronger symmetry constraints, such as 
supersymmetric and/or conformal theories. Moreover, some progress can be achieved 
by considering special regimes, like for instance the large-$N$ limit in SU$(N)$ gauge theories, 
or by restricting to some specific sectors of observables. 
The hope is that the methods developed and the results obtained in this way 
could improve our understanding of more realistic situations.      

A paradigmatic case, which sits at the crossroad of many approaches, is represented by the 
BPS Wilson loops of the $\cN=4$ Super-Yang-Mills (SYM) theory in four dimensions. This theory
has the maximum possible amount of supersymmetry allowed for non-gravitational models; 
it is exactly conformal also at the quantum level, 
and many sub-sectors of its observables are integrable. 
Moreover, it admits a holographic dual description \cite{Maldacena:1997re} as Type IIB superstring 
theory on $\mathrm{AdS}_5\times S^5$.
In this theory, it is possible to construct BPS Wilson loops which preserve part 
of the supersymmetry. In particular, a 1/2 BPS straight Wilson loop vanishes identically, but 
a circular one is non-trivial.
Its vacuum expectation value was computed in the planar limit in 
\cite{Erickson:2000af} by resumming the rainbow diagrams that contribute to 
it. The result has a holographic interpretation as the area of the surface bordered by the 
loop in the $\mathrm{AdS}_5\times S^5$ background \cite{Berenstein:1998ij}.
This computation was extended to finite $N$ in \cite{Drukker:2000rr} where 
it was observed that the perturbative expansion is 
captured by a Gaussian matrix model. Many extensions and generalizations have 
been studied in the $\cN=4$ context with either field-theoretic or holographic methods 
or through relations to integrability \cite{Semenoff:2001xp,Pestun:2002mr,Zarembo:2002an,Drukker:2006ga,Semenoff:2006am,Giombi:2006de,Drukker:2007yx,Drukker:2007dw,Drukker:2007qr,Gomis:2008qa,Bassetto:2008yf,Giombi:2009ds,Bassetto:2009rt,Bassetto:2009ms,Giombi:2012ep,Bonini:2014vta}.
Wilson loops that preserve a subgroup of the superconformal symmetry of the $\cN=4$ theory 
are also instances \cite{Kapustin:2005py} of a defect conformal field theory
\cite{McAvity:1993ue,McAvity:1995zd,Billo:2016cpy}
and have been investigated also from this point of view \cite{Cooke:2017qgm,Kim:2017sju,Giombi:2018hsx}.

The matrix model description of the 1/2 BPS circular Wilson loop has been derived in 
\cite{Pestun:2007rz} from the localization approach.
Actually, the localization methods are valid not 
only for the $\cN=4$ SYM theory, but for any $\cN=2$ SYM theory, in which case
the resulting matrix model is not Gaussian any longer
but contains interaction terms.
This has been very useful in the study of the AdS/CFT duality in 
the $\mathcal{N}=2$ setting \cite{Rey:2010ry,Passerini:2011fe,Russo:2013sba,Fiol:2015mrp}, 
since the interacting matrix model allows one to study the large-$N$ limit in an efficient way, 
also in the strong coupling regime. 

In this context, the localization is realized on a spherical space manifold $S^4$, 
but when the theory is conformal it also reproduces the results in flat space.
In fact, it has been shown to provide information about correlators of chiral operators 
\cite{Baggio:2014ioa,Baggio:2014sna,Gerchkovitz:2014gta,Baggio:2015vxa,Baggio:2016skg,Gerchkovitz:2016gxx,Rodriguez-Gomez:2016ijh,Rodriguez-Gomez:2016cem,Billo:2017glv} 
and about one-point functions of chiral operators in presence of a Wilson loop 
\cite{Billo:2018oog}. In non-conformal cases, one expects a conformal anomaly in relating 
the localization results obtained on $S^4$ to flat space quantities; there are however 
strong indications \cite{Billo:2019job} that this anomaly, at least for correlators of 
chiral operators, is rather mild and that the interacting matrix model still contains a lot of
information about perturbation theory in flat space. 
Localization also provides exact results for important observables related to the 
Wilson loop, such as the Bremsstrahlung function and the cusp 
anomalous dimension \cite{Correa:2012at,Correa:2012hh,Lewkowycz:2013laa,Fiol:2015spa,Mitev:2015oty,Bonini:2015fng,Bianchi:2018zpb,Gomez:2018usu}.

For $\cN=2$ superconformal theories, 
the first check of the agreement between matrix model predictions from
localization in $S^4$ and explicit calculations using Feynman diagrams in $\mathbb{R}^4$
has been presented in \cite{Andree:2010na}. Here
the authors considered $\mathcal{N}=2$ SQCD with gauge group
SU($N$) and $2N$ flavors, and explicitly showed that the terms 
proportional to $g^6\,\zeta(3)$ in the vacuum expectation value of a circular BPS 
Wilson loop predicted by the Pestun matrix model exactly matched the $g^6\,\zeta(3)$-terms 
arising from Feynman diagrams in flat space at three loops. 
In particular they performed their check by
considering the difference between the
Wilson loop computed in $\cN = 4$ SYM and in $\cN = 2$
SU($N$) SQCD, finding in this way an enormous reduction in the number of 
Feynman diagrams to be evaluated. Focusing on the ``difference theory", namely
computing only the diagrammatic difference with respect to
$\cN=4$ SYM, is highly convenient and indeed this method has been extensively used in many subsequent 
developments in this context (see for example \cite{Pomoni:2011jj,Pomoni:2013poa,Fiol:2015spa,Mitev:2015oty,Billo:2017glv,Billo:2018oog,Billo:2019job,Gomez:2018usu}).

In this paper we present an extension of the work of \cite{Andree:2010na}
in two respects. Firstly, we consider the vacuum expectation value of the fundamental 1/2 BPS 
circular loop in conformal $\cN=2$ SU($N$) theories that are more general than SQCD, 
namely in theories with matter transforming 
in a generic SU($N$) representation subject only to 
the requirement that the $\beta$-function vanishes. Secondly, 
we perform our calculations at one loop-order higher than in \cite{Andree:2010na}, {\it{i.e.}} we compute the terms proportional to $g^8\,\zeta(5)$ at four loops. Our motivations are several.

First of all, by considering theories with a generic matter content 
we can gain a better understanding of how the matrix model diagrams are packaged color-wise. Indeed, we show that the interaction terms in the matrix model can be expressed
as the trace of the 
logarithm of the fluctuation operator around the fixed points selected in the localization 
computation. The color structure of such operator is that of multiple 
insertions of adjoint generators in a loop where the hypermultiplets run - the matter 
ones contributing with a positive sign and the adjoint ones, which would be present in 
the $\cN=4$ theory, with a negative sign. This fact indicates that the matrix model itself 
naturally organizes its outcomes in terms of the ``difference theory", thus suggesting
to organize in the same fashion also the Feynman diagrams arising in the corresponding
field-theoretic computations. Furthermore, the matrix model also suggests 
that the lowest-order contributions to the 
vacuum expectation value of the circular Wilson loop proportional to 
a given Riemann $\zeta$-value, namely the terms of the type $g^{2n+2}\,\zeta(2n-1)$, are 
entirely due to the $n$-th loop correction to a single propagator inserted in the 
Wilson loop in all possible ways. This is indeed what we find up to $n=3$, thus extending
the result at $n=2$ of \cite{Andree:2010na}.

By working at one loop-order higher than in \cite{Andree:2010na} we can
put the agreement between the matrix model predictions and the field-theory results on a
more solid ground. Indeed, at order $g^6$ all the numerous diagrams computed in \cite{Andree:2010na} using the component formalism, actually collapse to just two superdiagrams if one uses the
$\mathcal{N}=1$ superfield formalism in the Fermi-Feynman gauge. 
One of these two superdiagrams trivially vanishes since it is proportional to the $\beta$-function coefficient, and thus the check with the matrix model predictions reduces to the comparison of a single coefficient. On the contrary, at order $g^8$ 
even in the $\mathcal{N}=1$ superfield formalism one finds many different 
non-vanishing contributions corresponding to superdiagrams with different topologies, different
combinatorial coefficients and different color structures. Therefore, obtaining an agreement 
with the matrix model results in this case is much more challenging and not at all obvious
since many independent factors have to conspire in the right way. Moreover, differently from what
happens at three loops, at order $g^8$ the color factors in the matrix model expressions 
have a different trace structure as compared to the Feynman diagrams at four loops, 
and the agreement between the two can be obtained only by using group-theoretic identities. Dealing with a matter content in a generic representation allows us to have full control on the color
and combinatorial factors, thus avoiding accidental simplifications.

A further motivation for our work is that being able to treat conformal $\cN=2$ theories with 
a generic matter content allows us to select special cases that exhibit a 
particular behavior in the large-$N$ limit. 
For instance, we consider theories in which the matter content consists of $N_F$ hypermultiplets in the fundamental, $N_S$ in the rank-two symmetric and $N_A$ in the rank-two anti-symmetric representations of SU($N$)\,%
\footnote{We thank J. Russo for suggesting to us to study the case
with $N_F=0$ and $N_S=N_A=1$, from which we started our investigation.}.
By requiring the vanishing of the $\beta$-function coefficient one obtains five classes 
of theories that exist for arbitrary $N$ \cite{Koh:1983ir}, one of which is the $\cN=2$ SQCD. For two other classes we show that the difference of the Wilson loop vacuum expectation value with respect to 
the $\cN=4$ case is sub-leading in the large-$N$ limit 
and thus vanishes in the planar approximation. 
In fact, these two classes of theories were shown to have a holographic dual \cite{Ennes:2000fu} of the type $\mathrm{AdS}_5\times S^5/\mathbb{Z}$ for an appropriate discrete group 
$\mathbb{Z}$, which is a simple modification of the $\mathrm{AdS}_5\times S^5$ geometry
corresponding to the $\cN=4$ SYM theory. Since the circular Wilson loop only sees the Anti-de Sitter
factor, one should expect no deviations from the $\cN=4$ case, and this is indeed what our
results indicate.

We hope that our analysis might be useful also to study the vacuum expectation value 
of a Wilson loop in a generic representation and its behavior in the limit where the dimension of such a representation is large, along the lines recently discussed for example in 
\cite{Bourget:2018obm,Beccaria:2018xxl,Beccaria:2018owt}.

This paper is organized as follows. In section~\ref{sec:mm} we review the matrix model obtained in 
\cite{Pestun:2007rz} via localization, and formulate it for an $\cN=2$ theory with gauge group
SU($N$) and a generic matter content. In section~\ref{secn:propWilson} we first compute the quantum 
correction to the ``propagator'' of  the interacting matrix model up to three loops, and then use it to 
obtain the leading terms of the vacuum expectation value of the 1/2 BPS circular Wilson loop in the 
fundamental representation. We also derive the exact expressions in $g$ and $N$ for the corrections 
proportional to $\zeta(3)$ and $\zeta(5)$ in this vacuum expectation value, and exploit them to 
study the large-$N$ limit. In section~\ref{secn:fieldtheory} we perform a perturbative field-theory
computation in the $\cN=2$ superconformal theories at order $g^8$ using the $\cN=1$ superfield
formalism. By computing (super) Feynman diagrams in the ``difference theory", we show the 
perfect agreement with the matrix model results. Finally in section~\ref{secn:concl} we briefly present
our conclusions.

A lot of technical material is contained in the appendices.
In particular, appendix~\ref{app:group} contains our group theory notations and conventions for
SU$(N)$, while appendix~\ref{app:grassmann} describes our notations and conventions regarding 
the spinor algebra and Grassmann variables. Appendix~\ref{app:grass-super} describes a 
method to carry out the Grassmann integrations appearing in $\cN=1$ superdiagrams 
with chiral/anti-chiral multiplet and vector multiplet lines. 
We have found this method, which follows a different route from the use of the $D$-algebra 
proposed long ago in \cite{Grisaru:1979wc}, quite efficient in dealing with the type of diagrams 
involved in our computation. 
Finally, in appendix~\ref{app:diagrams} we give the details on the various
three-loop diagrams contributing at order $g^6\,\zeta(5)$ to the adjoint scalar propagator.

\section{The matrix model for $\cN=2$ SYM theories}
\label{sec:mm}
Localization techniques have been exploited to compute exactly certain observables in 
$\cN=2$ SYM theories, such as the partition function on a 4-sphere $S^4$ or the vacuum
expectation value of BPS Wilson loops \cite{Pestun:2007rz}. 
Here we consider $\cN=2$ SYM theories with gauge group SU($N$) and matter hypermultiplets transforming in a generic representation $\cR$. 

\subsection{The $S^4$ partition function}
\label{subsec:S4part}
The partition function on a 4-sphere $S^4$ with unit radius\,\footnote{The dependence on 
the radius $R$ can be trivially recovered by replacing $a$ with $R\,a$.}, 
computed via localization, can be expressed as follows:
\begin{equation}
	\cZ_{S^4}=\int \prod_{u=1}^N da_u \,\,\Delta(a) \,
	\big| Z(\ii a,g)\big|^2\,\delta\Big(\sum_{u=1}^N a_u\Big)
	\label{ZS4}
\end{equation}
where $a$ is a Hermitean $N\times N$ matrix with (real) 
eigenvalues $a_u$ ($u=1,\ldots,N$), $\Delta$ is the
Vandermonde determinant
\begin{equation}
	\Delta(a)=\prod_{u<v=1}^N (a_u-a_v)^2~,
	\label{vandermonde}
\end{equation}
and $Z(\ii a,g)$ is the partition function for a gauge theory with coupling $g$ defined 
on $\mathbb{R}^4$ with $a$ parametrizing the Coulomb branch. 
Note that in non-conformal theories the gauge coupling $g$ has to be interpreted as the
renormalized coupling at a scale inversely proportional to radius of the 4-sphere.

Before considering $Z(\ii a,g)$ in more detail, let us remark that the integration
over the eigenvalues $a_u$ in (\ref{ZS4}) can be rewritten simply as the integral over all 
components of the Hermitean traceless matrix $a$, namely
\begin{equation}
	\label{intda}
		\cZ_{S^4} = \int da\,\big| Z(\ii a,g)\big|^2~. 
\end{equation}
The matrix $a$ can be decomposed over a basis of generators $t_a$ of $\mathfrak{su}(N)$:
\begin{equation}
	\label{aont}
	a = a^b \,t_b~,~~~b = 1,\ldots, N^2-1~;
\end{equation}
we will normalize these generators so that the index of the fundamental representation equals $1/2$:
\begin{equation}
	\label{normtatb}
	\tr \,t_a t_b = \frac 12\, \delta_{ab}~.
\end{equation}
In Appendix \ref{app:group} we collect our group theory conventions and other useful 
formulas. The integration measure is then simply proportional to ${\prod_b da^b}$.

The $\mathbb{R}^4$ partition function $Z(\ii a,g)$ can be written as
\begin{equation}
	Z= Z_{\mathrm{tree}}\,Z_{\mathrm{1-loop}}\,
	Z_{\mathrm{inst}}~.
\end{equation}
In perturbation theory, we can neglect the instanton contributions and put 
$Z_{\mathrm{inst}}=1$. The tree-level term is given by
\begin{equation}
\big|Z_{\mathrm{tree}}\big|^2 = \rme^{-\frac{8\pi^2}{g^2}
	\,{\mathrm{tr}\,a^2}}~,
\end{equation} 
providing a free matrix model with a Gaussian term. 
The 1-loop part contains interaction terms, which we write as follows: 
\begin{equation}
	\label{Z1ltoS}
	\big|Z_{\mathrm{1-loop}}\big|^2 \equiv \rme^{-\widehat{S}(a)}~.
\end{equation}
The matrix model corresponding to the $\cN=4$ SYM theory has $\widehat{S}(a)=0$ and is 
purely Gaussian.
For $\cN=2$ SYM theories, instead, there are interaction terms.
In general, let us denote by $\mathbf{a}$ the $N$-dimensional vector of 
components $a_u$, and by $W(\cR)$ the set of the weights $\mathbf{w}$ of the 
representation $\cR$ and by
$W(\mathrm{adj})$ is the set of weights of the adjoint representation. Then, 
\begin{equation}
	\big|Z_{\mathrm{1-loop}}\big|^2 =
	\frac{\prod_{\mathbf{w}\in W(\mathrm{adj}) } H(\ii \mathbf{w}\cdot\mathbf{a})}
	{\prod_{\mathbf{w}\in W(\mathcal{R})} H(\ii \mathbf{w}\cdot\mathbf{a})})
	~,
	\label{Z1loop}
\end{equation} 
where
\begin{equation}
H(x)=G(1+x)\,G(1-x)
\end{equation}
and $G$ is the Barnes $G$-function. 

\subsection{The interaction action}
\label{subsec:intSa}
Let us now consider the interaction action $\widehat{S}(a)$. {From} (\ref{Z1ltoS}) it follows that 
\begin{align}
	\label{StologH}
	\widehat{S}(a) & =\!  \sum_{\mathbf{w}\in W(\mathcal{R})}\!\! \log H(\ii \mathbf{w}\cdot
	\mathbf{a})
	~~- \sum_{\mathbf{w}\in W(\mathrm{adj})}\!\! \log H(\ii \mathbf{w}\cdot\mathbf{a}) 
	\notag\\[1mm]
	&= \Tr_{\cR} \log H(\ii a) - \Tr_{\mathrm{adj}} \log H(\ii a)  \,=\, \Tr_{\cR}^\prime \log 
	H(\ii a)~, 
\end{align}
where in the last step  we introduced the notation   
\begin{align}
	\label{deftrp}
	\Tr_{\cR}^\prime \bullet \,= \,\Tr_{\cR} \bullet - \Tr_{\mathrm{adj}} \bullet~.
\end{align}
This indeed vanishes for the $\cN=4$ SYM theory, where the representation $\cR$
of the hypermultiplets is the adjoint. For $\cN=2$ models, this combination of 
traces is non-vanishing and precisely accounts for the matter content of the ``difference 
theory'' which is often used in field theory computations \cite{Andree:2010na}, where one 
removes from the $\cN=4$ result the diagrams with the adjoint hypermultiplets 
running in internal lines and replaces them with the corresponding diagrams involving
the matter hypermultiplets in the representation $\cR$.

Using the properties of the Barnes $G$-function, one can prove that
\begin{equation}
	\label{logHexp}
		\log H(x)=-(1+\gamma_{\mathrm{E}})\,x^2-\sum_{n=1}^\infty
		\frac{\zeta(2n+1)}{n+1}\,
		x^{2n+2}
\end{equation}
where $\zeta(n)$ are the Riemann $\zeta$-values. Then, we can rewrite (\ref{StologH}) 
as follows 
\begin{equation}
	\label{Saexp}
		\widehat{S}(a) = (1+\gamma_{\mathrm{E}})\trp a^2 + \sum_{n=2}^\infty(-1)^{n}
		\frac{\zeta(2n-1)}{n} \trp a^{2n}~.
\end{equation} 
With the rescaling 
\begin{equation}
	\label{rescaling}
		a\to \sqrt{\frac{g^2}{8\pi^2}}~a~,
\end{equation}
we bring the partition function on $S^4$ to the form 
\begin{equation}
	\label{ZS4resc}
		\cZ_{S^4} =  \Big(\frac{g^2}{8\pi^2}\Big)^{\frac{N^2-1}{2}}
		\int da~\rme^{-\mathrm{tr}\, a^2 - S(a)}~,
\end{equation}
where
\begin{align}
	\label{Sint}
		S(a)&=\trp \log H\Big(\ii\,\sqrt{\frac{g^2}{8\pi^2}}\,a\Big)\nonumber\\
		&=\frac{g^2}{8\pi^2}\,(1+\gamma_{\mathrm{E}})\,\trp a^{2}
		-\left(\frac{g^2}{8\pi^2}\right)^2\frac{\zeta(3)}{2}\,\trp a^{4}
		+\left(\frac{g^2}{8\pi^2}\right)^3\frac{\zeta(5)}{3}\,\trp a^{6}
		+\ldots
\end{align}
The overall $g$-dependent pre-factor in (\ref{ZS4resc}) is irrelevant in computing 
matrix model correlators, and thus can be discarded. Using the expansion (\ref{aont}),
the traces appearing in $S(a)$ can be expressed as 
\begin{equation}
	\label{tra2ntoC}
		\trp a^{2k} = C^\prime_{(b_1\ldots b_k)} \, a^{b_1}\ldots a^{b_k}~,	
\end{equation}
where
\begin{equation}
	\label{defC}
		C^\prime_{b_1\ldots b_n} = \trp T_{b_1}\ldots T_{b_n}~.
\end{equation}
These tensors are cyclic by definition.
In particular, we have
\begin{equation}
	\label{C2is} 
		C^\prime_{b_1 b_2} = \left(i_{\cR} - i_{\mathrm{adj}}\right) \delta_{b_1 b_2} = 
		\left(i_{\cR} - N\right) \delta_{b_1 b_2} = - \frac{\beta_0}{2}\, \delta_{b_1 b_2}
\end{equation}
where $i_{\cR}$ is the index of the representation $\cR$ and $\beta_0$ the one-loop coefficient of the $\beta$-function of the corresponding $\cN=2$ gauge theory. 
In superconformal models, one has $\beta_0=0$. This implies that $\trp a^{2}=0$ 
so that the interaction action $S(a)$ starts at order $g^4$, {\it{i.e.}} at two loops.

\subsection{Expectation values in the interacting matrix model}
\label{subsec:mmvev}
Other observables of the $\cN=2$ gauge theory, beside its partition function 
on $S^4$, can be evaluated via localization and mapped to suitable 
expectation values in this matrix model. For any observable represented by a 
function $f(a)$ in the matrix model, its vacuum expectation value is  
\begin{align}
	\label{vevmat}
		\big\langle f(a) \big\rangle\, 
		= \,\frac{\displaystyle{ \int \!da ~\rme^{-\tr a^2-S(a)}\,f(a)}}
		{ \displaystyle{\int \!da~\rme^{-\tr a^2-S(a)}} }\,	= \,\frac{\big\langle\,
		\rme^{- S(a)}\,f(a)\,\big\rangle_0\phantom{\Big|}}
		{\big\langle\,\rme^{- S(a)}\,\big\rangle_0
		\phantom{\Big|}}~,
\end{align} 
where the subscript 0 in the right-hand-side 
indicates that the vacuum expectation value is evaluated 
in the free Gaussian model describing the $\mathcal{N}=4$ theory. 
These free vacuum expectation values can be computed in a straightforward way 
via Wick's theorem in terms of the propagator\,%
\footnote{We  normalize the flat measure as 
$da = \prod_{b} \left(da^b/\sqrt{2\pi}\right)$,
so that  $\int da\, \rme^{-\tr a^2}=1$.
In this way the contraction (\ref{wickabc}) immediately follows.}
\begin{equation}
	\label{wickabc}
		\big\langle a^b \,a^c\big\rangle_0 \,=\, \delta^{bc}~.
\end{equation}
As discussed in \cite{Billo:2017glv,Billo:2018oog,Billo:2019job}, using the basic 
contraction (\ref{wickabc}) and the so-called fusion/fission relations for traces in the 
fundamental representation of SU($N$), it is possible to recursively evaluate the quantities
\begin{equation}
	\label{tn}
		t_{k_1,k_2,\cdots} = \big\langle \tr a^{k_1}\,\tr a^{k_2} \cdots\big\rangle_0
\end{equation}
and obtain explicit expressions for generic values of $k_1,k_2,\ldots$.
 
To compute perturbatively the vacuum expectation value $\big\langle f(a)\big\rangle$ 
in the interacting theory, one starts from the right-hand-side of (\ref{vevmat}) 
and expands the action $S(a)$ as in (\ref{Sint}). Proceeding in this way, for conformal theories where the $g^2$-term vanishes, one gets
\begin{align}
	\label{espvev}
		\big\langle f(a)\big\rangle
		& =   \big\langle f(a)\big\rangle_0 
		+ \left(\frac{g^2}{8\pi^2}\right)^2 \frac{\zeta(3)}{2}\,
		\big\langle f(a)\, \trp a^4\big\rangle_{0,\mathrm{c}}
		- \left(\frac{g^2}{8\pi^2}\right)^3 \frac{\zeta(5)}{3}\, \big\langle
		f(a)\, \trp a^6\big\rangle_{0,\mathrm{c}}\nonumber\\[2mm]
		&~~ + \ldots~,
\end{align}
where the notation $\langle ~\rangle_{0,\mathrm{c}}$ stands for
the connected part of a free correlator, namely
\begin{align}
	\label{defccor}
		\big\langle f(a)\, g(a)\big\rangle_{0,\mathrm{c}}  \,=\,  
		\big\langle f(a) \, g(a)\big\rangle_0 \,-\, \big\langle f(a)\big\rangle_0 
		\,\,\big\langle g(a)\big\rangle_0~.
\end{align}
We may regard (\ref{espvev}) as an expansion in ``trascendentality'', in the sense that each term 
in the sum has a given power of Riemann $\zeta$-values since it comes from the expansion of the
exponential of the interaction action (\ref{Sint}). 
For example the second term is the only one proportional to $\zeta(3)$, the third term is the only one 
proportional to $\zeta(5)$, while the ellipses stand for terms proportional to $\zeta(7)$, 
$\zeta(3)^2$ and so on.

Often $f(a)$ is  a ``gauge-invariant'' quantity, expressed in terms of traces of powers 
of $a$ in some representations. Also the quantities $\trp a^{2k}$ are traces of this type.
As shown in Appendix \ref{app:group}, relying on the Frobenius theorem it is possible 
to express such traces in terms of traces in the fundamental representation. 
At this point, the vacuum expectation value (\ref{espvev}) is reduced to a 
combinations of the quantities $t_{k_1,k_2,\ldots}$ introduced in (\ref{tn}). 
This is the computational strategy we adopt in the following sections. 

\subsection{A class of conformal $\cN=2$ theories}
\label{subsec:CN2class}
Let us consider a class of theories with $N_F$ matter hypermultiplets transforming in the fundamental 
representation, $N_S$ in the symmetric and $N_A$ in the anti-symmetric representation of order 2. 
This corresponds to taking 
\begin{equation}
	\label{RNFNASNA}
		\cR = N_F\, \Yfund \oplus N_S\, \Ysymm \oplus N_A\, \Yasymm~.
\end{equation}
The traces $\trp a^{2k}$ appearing in the interaction action $S(a)$ 
can be re-expressed in terms of traces in the fundamental representation, as discussed in 
appendix~\ref{app:group}. 

For example, for $k=1$ one has
\begin{equation}
	\label{trpa2}
		\trp a^2 = 2 \left(i_{\cR} - i_{\mathrm{adj}}\right) \tr a^2 
		= - \beta_0 \tr a^2~, 
\end{equation}
with
\begin{equation}
	\label{b0is}
		\beta_0=2N-N_F-N_S(N+2)-N_A(N-2)~.
\end{equation}
Superconformal theories must have $\beta_0=0$. It is easy to see that imposing 
this condition leads to five families of $\mathcal{N}=2$ superconformal field theories 
with gauge group SU($N$), and matter in the fundamental, symmetric or anti-symmetric 
representations. They were identified long ago in \cite{Koh:1983ir} and recently reconsidered in
\cite{Fiol:2015mrp,Bourget:2018fhe}. They are displayed in table \ref{tab:scft}.
\renewcommand{\arraystretch}{1}
\begin{table}[ht]
	\begin{center}
		{\small
		\begin{tabular}{c|c|c|c}
			\hline
			\hline
			\,\,theory \phantom{\bigg|}& $N_F$ & $N_S$ & $N_A$ \\
			\hline
			$\mathbf{A}\phantom{\Big|}$& $~~2N~~$ & $~~0~~$ & $~~0~~$  \\
			$\mathbf{B}\phantom{\Big|}$& $~~N-2~~$ & $~~1~~$ & $~~0~~$  \\
			$\mathbf{C}\phantom{\Big|}$& $~~N+2~~$ & $~~0~~$ & $~~1~~$  \\
			$\mathbf{D}\phantom{\Big|}$& $~~4~~$ & $~~0~~$ & $~~2~~$  \\
			$\mathbf{E}\phantom{\Big|}$& $~~0~~$ & $~~1~~$ & $~~1~~$  \\
			\hline
			\hline
			
		\end{tabular}
		}
	\end{center}
	\caption{The five families of $\cN=2$ superconformal theories with SU($N$) gauge 
	group and matter in fundamental, symmetric and anti-symmetric representations.}
	\label{tab:scft}
\end{table}

Theory $\mathbf{A}$ is the $\cN=2$ conformal SQCD which is often considered as
the prototypical example of a $\cN=2$ superconformal theory.
On the other hand, theories $\mathbf{D}$ and $\mathbf{E}$ are quite interesting:
for these superconformal models a holographic dual of the form 
$\mathrm{AdS}_5\times S^5/\mathbb{Z}$ with an appropriate discrete group $\mathbb{Z}$ 
has been identified \cite{Ennes:2000fu}. We will discuss some properties 
of these theories in the following.

For higher traces with $k>1$, one finds (see again Appendix \ref{app:group} for details)
\begin{align} 
	\label{S2n}
		\trp a^{2k} = &
		\frac{1}{2}\sum_{\ell=2}^{2k-2}
		\binom{2k}{\ell}\left(N_S+N_A-2\,(-1)^\ell\right)\,
		\tr a^\ell\,\tr a^{2k-\ell}\nonumber\\
		& + \left(\big(2^{k-1}-2\big)\, (N_S- N_A)-\beta_0\right)\, \tr a^{2k}~.
\end{align} 
Inserting this into the expansion (\ref{Sint}) we can express the interaction action 
in terms of traces in the fundamental representation. For the superconformal theories of table \ref{tab:scft} we find the results displayed in table \ref{tab:S4S6conf}.
\begin{table}[ht]
	\begin{center}
		{\small
		\begin{tabular}{c|c|c}
			\hline
			\hline
			\,\,theory \phantom{\bigg|}& $\trp a^{4}$ & 
			$\trp a^{6}$ 
			\\
			\hline
			$\mathbf{A}\phantom{\Big|}$& $~~6\,\big(\mathrm{tr} \,a^2\big)^2~~\phantom{\bigg|}$ & $~~10\,\Big[2\big(\mathrm{tr}\,a^3\big)^2-3\,\mathrm{tr}\,a^4\,\mathrm{tr}\,a^2
			\Big]~~$   \\
			$\mathbf{B}\phantom{\Big|}$& $~~3\,\Big[\big(\mathrm{tr}\,a^2\big)^2-2\,\mathrm{tr}\,a^4
			\Big]~~\phantom{\bigg|}$ & $~~15\,\Big[2\big(\mathrm{tr}\,a^3\big)^2
			-\mathrm{tr}\,a^4\,\mathrm{tr}\,a^2+2\,\mathrm{tr}\,a^6
			\Big]~~$  \\
			$\mathbf{C}\phantom{\Big|}$& $~~3\,\Big[\big(\mathrm{tr}\,a^2\big)^2+2\,\mathrm{tr}\,a^4
			\Big]~~\phantom{\bigg|}$ & $~~15\,\Big[2\big(\mathrm{tr}\,a^3\big)^2
			-\mathrm{tr}\,a^4\,\mathrm{tr}\,a^2-2\,\mathrm{tr}\,a^6
			\Big]~~$   \\
			$\mathbf{D}\phantom{\Big|}$& $~~12\,\mathrm{tr}\,a^4~~\phantom{\bigg|}$ & $~~20\,\Big[2\big(\mathrm{tr}\,a^3\big)^2-3\,\mathrm{tr}\,a^6
			\Big]~~$     \\
			$\mathbf{E}\phantom{\Big|}$& $~~0~~\phantom{\bigg|}$ & $~~40\,\big(\mathrm{tr}\,a^3\big)^2~~$   \\
			\hline
			\hline
	\end{tabular}
	}
\end{center}
\caption{The quartic and sextic interaction terms in the action $S(a)$
 for the five families of conformal 
theories defined in table \ref{tab:scft}.}
\label{tab:S4S6conf}
\end{table}

Notice that for theory $\mathbf{E}$ the quartic term vanishes and thus in this case the
effects of the interactions appear for the first time at order $g^6$, {\it{i.e.}} at three loops, 
and are proportional to $\zeta(5)$. This feature, which has been recently pointed out  
also in \cite{Bourget:2018fhe}, is a simple consequence of the properties of the quartic
trace in a representation $\cR$ formed by one symmetric and one anti-symmetric 
representation. Altogether, the matter hypermultiplets fill a generic $N\times N$ matrix; this is to
be compared with the $\cN=4$ case in the hypermultiplets are in the 
adjoint representation, which is equivalent to $N\times \overline{N}$ minus one singlet. The strong similarity of the two representations explains why theory
$\mathbf{E}$ is the $\cN=2$ model which is more closely related to the $\cN=4$ SYM theory.
For theory $\mathbf{D}$, instead, the quartic term is a single fundamental 
trace and thus is simpler than in the other theories. In the following we will see that these
features of theories $\mathbf{D}$ and $\mathbf{E}$ have a bearing on their large-$N$ behavior.

\section{Propagator and Wilson loops in superconformal matrix models}
\label{secn:propWilson}
We now discuss in detail two specific applications of the formula (\ref{espvev}): first 
the ``propagator'' $\vev{a^b\, a^c}$ and later the 1/2 BPS circular Wilson loops $\cW(a)$
in the fundamental representation.

\subsection{The propagator}
\label{subsec:propagator}
If in (\ref{espvev}) we take $f(a)=a^b\,a^c$, we get 
\begin{align}
	\label{corrbc}
		\big\langle a^b\, a^c\big\rangle & = \big\langle a^b\, a^c\big\rangle_0 
		+ \left(\frac{g^2}{8\pi^2}\right)^2 \frac{\zeta(3)}{2}\, C^\prime_{(d_1 d_2 d_3 d_4)}
		\vev{a^b\, a^c\, a^{d_1}\, a^{d_2}\, a^{d_3}\, a^{d_4}}_{0,\mathrm{c}}
		\nonumber\\
		&~~ 
		- \left(\frac{g^2}{8\pi^2}\right)^3 \frac{\zeta(5)}{3}\, C^\prime_{(d_1 d_2 d_3 d_4 d_5 d_6)}
		\vev{a^b\, a^c\, a^{d_1}\, a^{d_2}\, a^{d_3}\, a^{d_4}\, a^{d_5}\, a^{d_6}}_{0,\mathrm{c}} +\ldots~,
\end{align}
where inside each connected correlator we cannot contract $a^b$ with $a^c$. Doing all legitimate contractions we obtain
\begin{align}
	\label{corrbcres}
		\vev{a^b\, a^c} & = \delta^{bc} 
		+ \left(\frac{g^2}{8\pi^2}\right)^2\zeta(3)\times 6\, C^\prime_{(bcdd)}
		- \left(\frac{g^2}{8\pi^2}\right)^3 \zeta(5)\times 30\, C^\prime_{(bcddee)}
		+ \ldots~. 
\end{align}
The above contracted tensors are proportional to $\delta^{bc}$, and thus if define
\begin{align}
	\label{c4c6delta}
		6 \,C^\prime_{(bcdd)} & = \cC^\prime_4\, \delta^{bc}~,~~~
		30 \,C^\prime_{(bcddee)}  = \cC^\prime_6\, \delta^{bc}~,	
\end{align}
we can rewrite (\ref{corrbcres}) as
\begin{align}
	\label{corrbcres1}
		\big\langle a^b\, a^c\big\rangle& = \delta^{bc} \,\big(1 + \Pi\big)
\end{align}
with
\begin{align}
	\label{Piis}		 
		\Pi & = \left(\frac{g^2}{8\pi^2}\right)^2\zeta(3)\,\cC^\prime_4
		- \left(\frac{g^2}{8\pi^2}\right)^3 \zeta(5)\,\cC^\prime_6
		+ \ldots~. 
\end{align}
Using the expressions of the tensors $C^\prime$ for the five families of superconformal SU(N)
theories that can be obtained from the formul\ae~ in Appendix \ref{app:group} 
with the help of Form Tracer \cite{Cyrol:2016zqb}, one finds
\begin{equation}
	\begin{aligned}
		\cC^\prime_4 &= \frac{6\,C^\prime_{(ccdd)}}{N^2-1} = 3
		\Big[(N_S+N_A-2)\,\frac{N^2+1}{2}+(N_S-N_A)\,\frac{2N^2-3}{N}\Big]~,\\[2mm]
		\cC^\prime_6&=\frac{30\,C^\prime_{(ccddee)}}{N^2-1}=15\Big[
		(N_S+N_A-2)\,\frac{2N^4+5N^2-17}{4N}\\[1mm]
		&~~~\qquad\qquad\qquad\qquad+
		(N_S-N_A)\,\frac{5(N^4-3N^2+3)}{2N^2}+\frac{2(N^2-4)}{N}\Big]~.
		\end{aligned}
	\label{C4C6prime}
\end{equation}
These coefficients are tabulated in table
\ref{tab:C4C6conf}. 
\begin{table}[ht]
	\begin{center}
		{\small
		\begin{tabular}{c|c|c}
			\hline
			\hline
			\,\,theory \phantom{\bigg|}& $\cC^\prime_4 $ & 
			$\cC^\prime_6 $ 
			\\
			\hline
			$\mathbf{A}\phantom{\Big|}$& $~~-3(N^2+1)~~\phantom{\bigg|}$
			& $~~-\frac{15(N^2+1)(2N^2-1)}{2 N}~~\phantom{\bigg|}$   \\
			$\mathbf{B}\phantom{\Big|}$& $~~-\frac{3(N+1)(N-2)(N-3)}{2N}~~\phantom{\bigg|}$
			& $~~-\frac{15 (N-2) (2N^4-6N^3-15N^2+15)}{4 N^2}~~\phantom{\bigg|}$ \\
			$\mathbf{C}\phantom{\Big|}$& $~~-\frac{3(N-1)(N+2)(N+3)}{2N}~~\phantom{\bigg|}$
			& $~~-\frac{15 (N+2) (2N^4+6N^3-15N^2+15)}{4 N^2}	~~\phantom{\bigg|}$ \\
			$\mathbf{D}\phantom{\Big|}$&$~~-\frac{6(2N^2-3)}{N}~~\phantom{\bigg|}$
			& $~~-\frac{15(5N^4-2N^3-15N^2+8N+15)}{N^2}~~\phantom{\bigg|}$ \\
			$\mathbf{E}\phantom{\Big|}$&$~~0~~\phantom{\bigg|}$
			& $~~\frac{30 (N^2-4)}{N}~~\phantom{\bigg|}$\\
			\hline
			\hline
		\end{tabular}
		}
	\end{center}
	\caption{The coefficients $\cC^\prime_4 $ and $\cC^\prime_6$ 
	for the five families of conformal theories defined in table \ref{tab:scft}.}
	\label{tab:C4C6conf}
\end{table}

For the comparison with perturbative field theory calculations presented 
in section \ref{secn:fieldtheory}, it is useful to make explicit the symmetrization of the $C^\prime$-tensors appearing in (\ref{corrbcres}).  For the 4-index tensor, we have
\begin{align}
	\label{symmC4is}
		6\, C^\prime_{(bcdd)} & = 2\, \big(
		C^\prime_{bcdd} + C^\prime_{bdcd} + C^\prime_{bddc}\big) ~.
\end{align}
Indeed, due to the cyclic property and the fact that two indices are identified, a subgroup 
$\mathbb{Z}_4\times \mathbb{Z}_2$ of permutations leaves $C^\prime_{bcdd}$ invariant and one has to average only over the $4!/8=3$ permutations in the coset with respect to this stability subgroup. In a similar way, for the 6-index tensor we have
\begin{align}
	\label{symmC6is}
		30\, C^\prime_{(bcddee)} & = 2\, \big(C^\prime_{bcddee} + C^\prime_{bcdede} + C^   
		\prime_{bcdeed} + C^\prime_{bdcdee} + C^\prime_{bdcede} \notag\\[1mm]
		&~~~~~+ C^\prime_{bdceed} +\, C^\prime_{bddcee} + C^\prime_{bdecde} 
		+ C^\prime_{bdeced} + C^\prime_{bddece}\notag\\[1mm]
		&~~~~~+ C^\prime_{bdedce} + C^\prime_{bdeecd}+ C^\prime_{bddeec} 
		+ C^\prime_{bdedec} + C^\prime_{bdeedc} \big)~.		
\end{align}     
In this case, the stability subgroup is $\mathbb{Z}_6\times \mathbb{Z}_2\times 
\mathbb{Z}_2\times \mathbb{Z}_2$ and the coset has $6!/48 = 15$ elements. 

We would like to remark that even if we have considered theories with SU$(N)$ gauge group
and matter in the fundamental, symmetric and anti-symmetric representations, the color tensors
$C^\prime_{b_1\ldots b_n}$ in (\ref{defC}) and the corresponding coefficients $\cC^\prime_n$
can be defined also for other representations of SU($N$) (or U$(N)$) using the Frobenius theorem,
as indicated in appendix~\ref{secn:frob}, and also for other gauge groups. Thus, the structure
of the propagator corrections in (\ref{corrbcres1}) is very general.

\subsection{Wilson loops}
\label{subsec:wilson}
As a second example, we consider the 1/2 BPS circular Wilson loop in the fundamental representation.
If this operator is inserted on the equator of $S_4$, in the matrix model we can represent it 
by the operator \cite{Pestun:2007rz}
\begin{equation}
	\label{WLa}
		\mathcal{W}(a)=\frac{1}{N}\,\mathrm{tr}\exp\Big(\frac{g}{\sqrt{2}}\,a\Big)
		=\frac{1}{N}\,\sum_{k=0}^\infty\frac{1}{k!}\,\frac{g^k}{2^{\frac{k}{2}}}\,\mathrm{tr}\,a^k~.
\end{equation}
Its vacuum expectation value is computed starting from (\ref{espvev}), following the strategy outlined in section \ref{subsec:mmvev}. We write
\begin{align}
	\label{Wexpchis}
		\Delta\cW\,\equiv\, \big\langle\cW(a)\big\rangle - \big\langle\cW(a)\big\rangle_0 = \cX_3 + \cX_5 + \ldots~,
\end{align}
where
\begin{align}
\cX_3&=\left(\frac{g^2}{8\pi^2}\right)^2 \frac{\zeta(3)}{2}\,
		\big\langle \cW(a)\, \trp a^4\big\rangle_{0,\mathrm{c}}~,\label{chi3is}\\
\cX_5&=- \left(\frac{g^2}{8\pi^2}\right)^3 \frac{\zeta(5)}{3}\, \big\langle
		\cW(a)\, \trp a^6\big\rangle_{0,\mathrm{c}}~,\label{chi5is}
\end{align}
and so on. {From} these expressions it is easy to realize that for each Riemann $\zeta$-value
(or product thereof) the term with the lowest power of $g$ in $\Delta\cW$ arises from 
the quadratic term in the expansion of the Wilson loop operator. Indeed, we have
\begin{align}
\cX_3&=\left(\frac{g^2}{8\pi^2}\right)^2\frac{\zeta(3)}{2}\,\frac{g^2}{4N}\,
		\big\langle \mathrm{tr}\,a^2\, \trp a^4\big\rangle_{0,\mathrm{c}}+O(g^8)\notag\\[1mm]
		&=\frac{g^6\,\zeta(3)}{512\pi^4	}\,\frac{N^2-1}{N}\,\cC_4^\prime+O(g^8)
		\label{chi3expg}
\end{align} 
where $\cC_4^\prime$ is the coefficient of the two-loop correction of the ``propagator'' of the
matrix model defined in (\ref{C4C6prime}). This result is valid for any superconformal theory, and in particular for the five families introduced in 
section~\ref{subsec:CN2class}. Clearly, for theory $\mathbf{E}$ the correction is zero; actually the whole $\cX_3$ vanishes in this case. 
In a similar way we find
\begin{align}
\cX_5&=-\left(\frac{g^2}{8\pi^2}\right)^3\frac{\zeta(5)}{3}\,\frac{g^2}{4N}\,
		\big\langle \mathrm{tr}\,a^2\, \trp a^6\big\rangle_{0,\mathrm{c}}+O(g^{10})\notag\\[1mm]
		&=-\frac{g^8\,\zeta(5)}{4096\pi^6}\,\frac{N^2-1}{N}\,\cC_6^\prime+O(g^{10})
		\label{chi5expg}
\end{align}
where $\cC_6^\prime$ is the three-loop correction of the matrix model ``propagator''.
Combining (\ref{chi3expg}) and (\ref{chi5expg}) we see that at the lowest orders in $g$ 
the difference of the vacuum expectation value of the Wilson loop with respect to the 
$\cN=4$ expression is given by
\begin{equation}
\Delta\cW= \frac{N^2-1}{8N}\,g^2\,\Pi+\ldots
\label{DeltaWPi}
\end{equation}
where $\Pi$ is the quantum correction to the ``propagator" given in (\ref{Piis}).
In the following sections we will prove that these results are in perfect agreement with perturbative
field theory calculations using ordinary (super) Feynman diagrams.

Actually, as explained in \cite{Billo:2018oog}, within the matrix model it is possible to evaluate 
$\cX_3$, $\cX_5$ and so on, without making any expansion in $g$. To obtain these 
{\emph{exact}} results, one has to write the traces $\trp a^{2k}$ in terms of the traces
in the fundamental representation by means of (\ref{S2n}). In this way everything is reduced 
to combinations of the quantities $t_{k_1,k_2,\ldots}$ defined in (\ref{tn}), 
which in turn can be evaluated in an algorithmic way using the fusion/fission identities
\cite{Billo:2017glv}. In the end, this procedure allows one to express the result 
in terms of the exact vacuum expectation value of the Wilson loop in the $\mathcal{N}=4$ theory
given by
\begin{align}
	\label{defWN4}
		W(g)\, \equiv\, \big\langle \cW(a)\big\rangle_0 = 
		\sum_{k=0}^\infty
		\frac{1}{k!}\,\frac{g^k}{2^{\frac{k}{2}}}\,t_k~.
\end{align}	
This expression can be resummed to obtain \cite{Erickson:2000af,Drukker:2000rr}:	
\begin{align}
	\label{WN4exact}
		W(g) =
		\frac{1}{N}\,L_{N-1}^1\Big(\!-\frac{g^2}{4}\Big)\,
		\exp\Big[\frac{g^2}{8}\Big(1-\frac{1}{N}\Big)\Big]~,
\end{align}
where $L_{n}^m(x)$ is the generalized Laguerre polynomial of degree $n$.
Applying this procedure to the five families of superconformal theories introduced in section~\ref{subsec:CN2class}, we find
\begin{align}
	\label{X3exact}
		\mathcal{X}_3& =	\left(\frac{g^2}{8\pi^2}\right)^2\frac{3\,\zeta(3)}{16N^2}
		\bigg[2\big(N_S+N_A-2\big)N^2
		\Big(\big(2N^2+1\big)\,g\,\partial_g W(g)+g^2\,\partial_g^2 W(g)\Big)
		\nonumber\\[1mm]
	& ~~
	\qquad\quad\qquad\quad
		+\big(N_S-N_A\big)\Big(\big(N^2-1\big)\,g^2\,W(g)+
		\big(g^2+8N^3 - 12N\big)\,g\,\partial_g W(g)
\nonumber \\[1mm]
		& \qquad\qquad\quad\qquad\quad\qquad\quad
		-4N g^2\,\partial_g^2 W(g)+16N^2\,g\,\partial_g^3 W(g)\Big)\bigg]
		~.
\end{align}
Expanding in $g$, it is easy to check the validity of (\ref{chi3expg}).
The case of theory $\mathbf{A}$, namely $N_S=N_A=0$, 
was already described in \cite{Billo:2018oog}.  For theory $\mathbf{E}$, as we have already remarked, $\cX_3=0$ since $\trp a^4=0$. Therefore, in this case the first correction with respect to 
the $\cN=4$ result for the Wilson loop is $\cX_5$, which turns
out to be
\begin{align}
	\label{X5exact}
		\mathcal{X}_5\,\big|_{\mathbf{E}}& =	
		-\left(\frac{g^2}{8\pi^2}\right)^3\frac{5\,\zeta(5)}{12N^2}
		\bigg[\big(N^4+5N^2-6\big)\,g^2\,W(g)\nonumber\\
		&\qquad\qquad~~\qquad
		+\big(2g^2N^2+6g^2-8N^3-48N\big)\,g\,\partial_g W(g)\nonumber\\[1mm]
		&\qquad\qquad~~\qquad+\big(g^2-8N^3-48N\big)\,g^2\,\partial_g^2W(g)\nonumber
		\\
		&\qquad\qquad~~\qquad
-8N\big(g^2-10N\big)\,g\,\partial_g^3W(g)+16N^2\,g^2\,\partial_g^4W(g)\bigg]~.
\end{align}
Similar formul\ae~ can be easily obtained for the other families of superconformal theories. We have derived them but we do not report their explicit expressions since for theories $\mathbf{A}$,
$\mathbf{B}$, $\mathbf{C}$, and $\mathbf{D}$ the leading term in the difference with respect 
to the $\cN=4$ result is given by $\cX_3$.

We stress once more that this procedure allows one to obtain in an algorithmic way the {\emph{exact}} expression in $g$ and $N$ of any term of the vacuum expectation value 
of the circular Wilson loop with a fixed structure of Riemann $\zeta$-values. This fact will now
be used to study the behavior of the matrix model in the large-$N$ limit.

\subsection{The large-$N$ limit}
\label{subsec:largeN}
The large-$N$ limit is defined by taking $N\to\infty$ and keeping the 't Hooft coupling
\begin{equation}
\lambda=g^2\,N
\label{lambda}
\end{equation}
fixed. In this limit the perturbative correction $\Pi$ to the ``propagator'' given in (\ref{Piis}) becomes
\begin{align}
\Pi&=\big(N_S+N_A-2\big)\Big(\frac{3\zeta(3)\,\lambda^2}{128\pi^4}-\frac{15\zeta(5)\,\lambda^3}{1024\pi^6}+O(\lambda^4)\Big)\notag\\[1mm]
&~~~+\big(N_S-N_A\big)\Big(\frac{3\zeta(3)\,\lambda^2}{32\pi^4}-\frac{75\zeta(5)\,\lambda^3}{1024\pi^6}+O(\lambda^4)\Big)\,\frac{1}{N}\label{PilargeN}\\[1mm]
&~~~+\bigg[
\big(N_S+N_A-2\big)\Big(\frac{3\zeta(3)\,\lambda^2}{128\pi^4}-\frac{75\zeta(5)\,\lambda^3}{2048\pi^6}\Big)-\frac{15\zeta(5)\,\lambda^3}{256\pi^6}
+O(\lambda^4)\bigg]\,\frac{1}{N^2}+ O\Big(\frac{1}{N^3}\Big)\notag~.
\end{align}
{From} this expression we easily see that in the planar limit $\Pi$ is non-zero for 
theories $\mathbf{A}$, $\mathbf{B}$ and $\mathbf{C}$, whereas it 
vanishes for theories $\mathbf{D}$ and $\mathbf{E}$:
\begin{equation}
\lim_{N\to\infty} \Pi\,\big|_{\mathbf{D}~\mathrm{or}~\mathbf{E}} = 0~.
\label{PiplanarDE}
\end{equation}
In particular for theory $\mathbf{D}$ the correction to the ``propagator'' goes
like $1/N$, whereas for theory $\mathbf{E}$ it goes like $1/N^2$:
\begin{align}
\Pi\,\big|_{\mathbf{D}}&
=-\Big(\frac{3\zeta(3)\,\lambda^2}{16\pi^4}-\frac{75\zeta(5)\,\lambda^3}{512\pi^6}+O(\lambda^4)\Big)\,\frac{1}{N}+O\Big(\frac{1}{N^2}\Big)~,\label{PiD}\\[1mm]
\Pi\,\big|_{\mathbf{E}}&=-\Big(\frac{15\zeta(5)\,\lambda^3}{256\pi^6}+O(\lambda^4)\Big)\,\frac{1}{N^2}+O\Big(\frac{1}{N^3}\Big)~.\label{PiE}
\end{align}
Therefore, in the planar limit, the ``propagator'' of the matrix model for these two families is
identical to that of the free matrix model describing the $\cN=4$ SYM theory.

Let us now consider the vacuum expectation value of the circular Wilson loop. Taking the 
large-$N$ limit in the $\cN=4$ expression (\ref{WN4exact}) one obtains 
\cite{Erickson:2000af}
\begin{equation}
\lim_{N\to\infty} W\big(\sqrt{\lambda/N}\big)\,
=\,\frac{2}{\sqrt{\lambda}}\,I_1\big(\sqrt{\lambda}\big)
\label{Wlambda}
\end{equation}
where $I_\ell$ is the modified Bessel function of the first kind.

Using this result in the $\zeta(3)$-correction (\ref{X3exact}), we get
\begin{equation}
\cX_3 = \big(N_S+N_A-2\big)\,\frac{3\zeta(3)\,\lambda^2}{128\pi^4}\,
I_2\big(\sqrt{\lambda}\big)+ O\big(\frac{1}{N}\big)~.
\label{X3largeN}
\end{equation}
This is a generalization of the formula obtained in \cite{Billo:2018oog} for the SQCD 
theory. With the same procedure we have also derived the planar limit of the $\zeta(5)$-
correction, finding
\begin{equation}
\cX_5 = -\big(N_S+N_A-2\big)\,\frac{5\zeta(5)\,\lambda^3}{1024\pi^6}\,\Big(
3I_2\big(\sqrt{\lambda}\big)+I_4\big(\sqrt{\lambda}\big)\Big)+ O\big(\frac{1}{N}\big)~.
\label{X5largeN}
\end{equation}
These results indicate that for theories $\mathbf{A}$, $\mathbf{B}$ and $\mathbf{C}$ the 
vacuum expectation value of the circular Wilson loop in the planar limit is different from 
the one of the $\cN=4$ SYM theory. On the other hand, for theories $\mathbf{D}$ 
and $\mathbf{E}$ this difference vanishes, namely
\begin{equation}
\lim_{N\to\infty} \Delta\cW \,\big|_{\mathbf{D}~\mathrm{or}~\mathbf{E}} = 0
\label{DeltaDE}
\end{equation}
in analogy with the ``propagator'' result (\ref{PiplanarDE}).
Working out the details at the next-to-leading order for theory $\mathbf{D}$, we find
\begin{align}
\Delta\cW \,\big|_{\mathbf{D}} &= -\bigg[
\frac{3\zeta(3)\,\lambda^2}{32\pi^4}\,\Big(
2I_2\big(\sqrt{\lambda}\big)+I_4\big(\sqrt{\lambda}\big)\Big)\label{DeltaWD}\\[1mm]
&\quad\qquad-\frac{15\zeta(5)\,\lambda^3}{256\pi^6}\,\Big(
5I_2\big(\sqrt{\lambda}\big)+4I_4\big(\sqrt{\lambda}\big)+
I_6\big(\sqrt{\lambda}\big)\Big)+\ldots
\bigg]\,\frac{1}{N}+O\Big(\frac{1}{N^2}\Big)\notag
\end{align}
where the ellipses stand for terms with higher Riemann $\zeta$-values (or product 
thereof). Similarly, at the next-to-next-to-leading order for theory $\mathbf{E}$, we find
\begin{align}
\Delta\cW \,\big|_{\mathbf{E}} &= -\bigg(
\frac{15\zeta(5)\,\lambda^{7/2}}{1024\pi^6}\,
I_1\big(\sqrt{\lambda}\big)+\ldots\bigg)\,\frac{1}{N^2}+O\Big(\frac{1}{N^3}\Big)~.
\label{DeltaWE}
\end{align}

Our findings have been obtained with a weak-coupling analysis at small $\lambda$. They 
are, however, in agreement with the strong-coupling results at large $\lambda$ presented 
in \cite{Fiol:2015mrp}, in the sense that also at strong coupling the vacuum expectation
value of the circular Wilson loop in the planar limit is different from that of the 
$\cN=4$ SYM theory for theories $\mathbf{A}$, $\mathbf{B}$ and $\mathbf{C}$, while
it is the same for theories $\mathbf{D}$ and $\mathbf{E}$. This observation suggests that
also the interpolating function between weak and strong coupling shares the same features
for the various theories.
The fact that for theories $\mathbf{D}$ and $\mathbf{E}$ the vacuum expectation value
of the circular Wilson loop is identical to that of the $\cN=4$ SYM theory in the planar 
limit is also in agreement with the fact that the holographic dual of 
theories $\mathbf{D}$ and $\mathbf{E}$ is of the form $\mathrm{AdS}_5\times S^5/\mathbb{Z}$ 
with an appropriate discrete group $\mathbb{Z}$ \cite{Ennes:2000fu}. Indeed, 
for the 1/2 BPS circular Wilson loop, the relevant part of the geometry is the 
Anti-de Sitter factor $\mathrm{AdS}_5$,
which is the same one that appears in the famous $\mathrm{AdS}_5\times S^5$ holographic dual of 
the $\cN=4$ SYM theory \cite{Maldacena:1997re}. It would be interesting 
to identify other observables that have this property in the planar limit and check 
the holographic correspondence, and also to find which observables 
of the theories $\mathbf{D}$ and $\mathbf{E}$ instead 
feel the difference with the $\cN=4$ SYM theory in the planar limit. Investigating
which sectors of our $\cN=2$ theories are planar equivalent to those  of the
$\cN=4$ SYM theory would be useful to better clarify the relations among the
various models and also to understand to which extent the arguments discussed
for example in \cite{Armoni:2004uu} in the so-called orientifold models 
can be applied to our case. We leave however this issue for future work.

We conclude by observing that the coefficients $\big(N_S+N_A-2\big)$ and
$\big(N_S-N_A\big)$ appearing in the planar limit results (see, for example, 
(\ref{PilargeN}), (\ref{X3largeN}) and
(\ref{X5largeN})) have an interesting meaning in terms of the central charges of 
the $\cN=2$ superconformal gauge theories corresponding to the matrix model. 
Indeed, taking into account the matter content corresponding to the representation
(\ref{RNFNASNA}) and using the formul\ae ~for the $c$ and $a$ central charges derived in
\cite{Anselmi:1997ys}, we find
\begin{equation}
\begin{aligned}
c&=-\frac{1}{24}\Big(\big(N_S+N_A-8\big)N^2+3\big(N_S-N_A\big)N+4\Big)~,\\
a&=-\frac{1}{48}\Big(\big(N_S+N_A-14\big)N^2+3\big(N_S-N_A\big)N+10\Big)~,
\end{aligned}
\label{ca}
\end{equation}
implying that 
\begin{align}
\frac{48(a-c)}{N^2}=\big(N_S+N_A-2\big)+\frac{3\big(N_S-N_A)}{N}-\frac{2}{N^2}
\label{c-a}
\end{align}
Using this, we can rewrite our results for the Wilson loop in the following way
\begin{equation}
\Delta\cW=\frac{a-c}{N^2}\bigg[
\frac{9\zeta(3)\,\lambda^2}{8\pi^4}\,I_2\big(\sqrt{\lambda}\big)
-\frac{15\zeta(5)\,\lambda^3}{64\pi^6}\,\Big(
3I_2\big(\sqrt{\lambda}\big)+I_4\big(\sqrt{\lambda}\big)
\Big)+\ldots\bigg]\,+\,O\Big(\frac{1}{N}\Big)~.
\label{DeltaWc-a}
\end{equation}
It would be nice to have an interpretation of this formula, and in particular of its prefactor, based on general principles.

\section{Field theory checks}
\label{secn:fieldtheory}

In this section we consider the field-theoretic counterpart of the computations we performed in section \ref{secn:propWilson} using the matrix model.

\subsection{Action and Feynman rules}
\label{subsec:action-fr}
We compute Feynman superdiagrams, working in $\cN=1$ superspace formalism and considering 
the diagrammatic difference of the $\cN=2$ SYM theory with respect to the 
$\cN=4$ theory. We now briefly review these techniques; this serves also to explain our conventions.

Our $\cN=2 $ theory contains both gauge fields, organized in an $\cN=2$ vector multiplet, and matter fields, organized in hypermultiplets. In terms of $\cN=1$ superfields the $\cN=2$ vector multiplet contains a vector superfield $V$ and a chiral superfield $\Phi$, both in the adjoint
representation of SU$(N)$. The adjoint complex scalar $\varphi$ of the $\cN=2$ gauge multiplet 
is the lowest component of the chiral superfield $\Phi$, while the gauge field $A_\mu$ is the 
$\big(\bar{\theta}\sigma^\mu\theta\big)$-component of $V$ (we refer to 
appendix~\ref{app:grassmann} for our conventions on spinors, Pauli matrices and
Grassmann variables).

\begin{figure}[ht]
\begin{center}
    \includegraphics[scale=0.7]{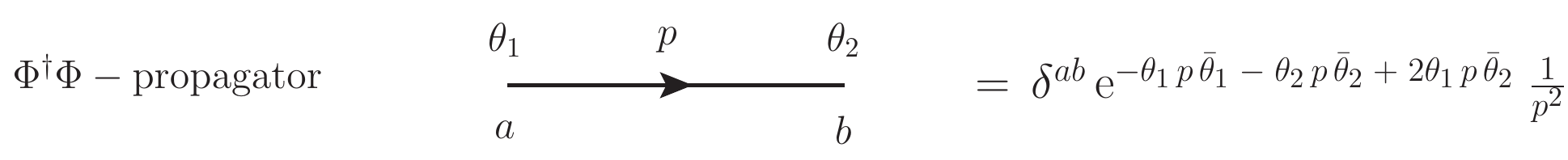}\\ \vspace{0.2cm}
	\hspace{-2.3cm}\includegraphics[scale=0.7]{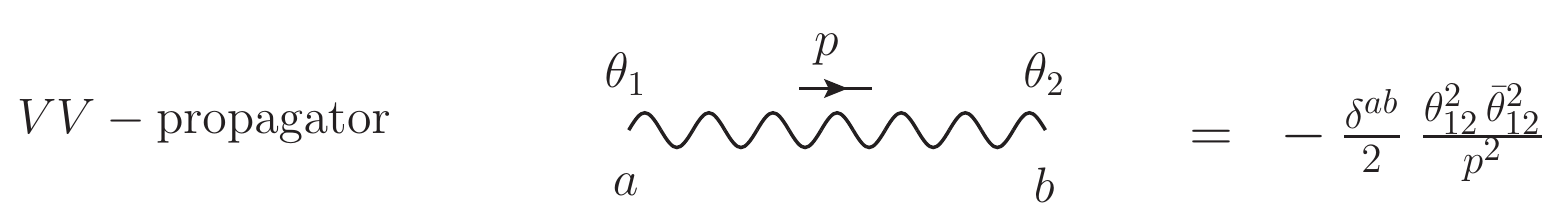} \\
	\includegraphics[scale=0.6]{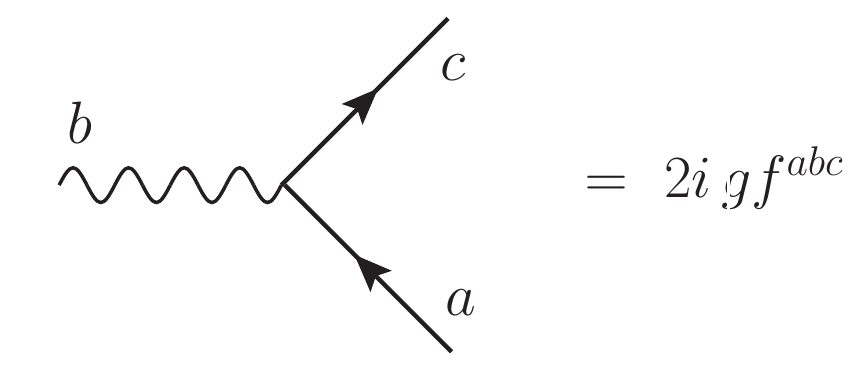}\hspace{1.2cm}
	\includegraphics[scale=0.6]{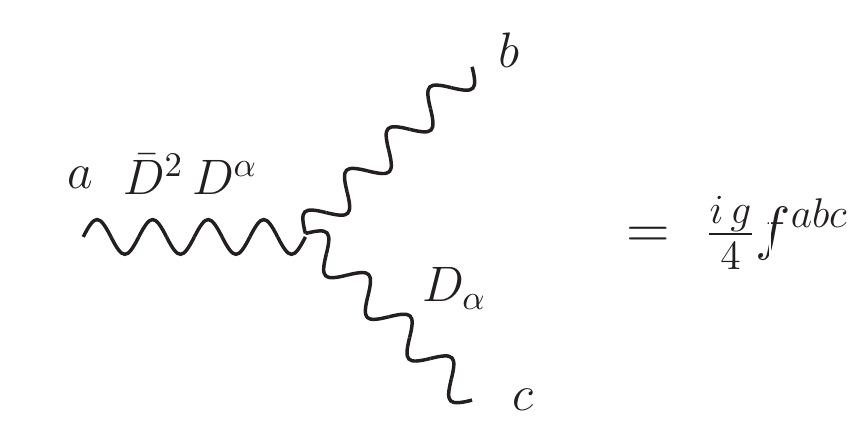} \\
	\hspace{1.2cm}\includegraphics[scale=0.6]{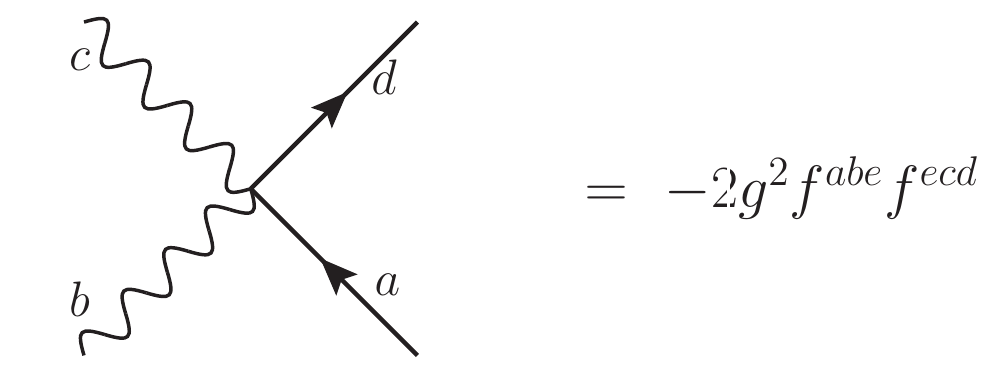}
	\caption{Feynman rules for the gauge part  of the $\cN=2$ theory that are relevant
	for our calculations.}
	\label{fig:Feyngauge}
	\end{center}
\end{figure}

In the Fermi-Feynman gauge the part of the action which only involves these adjoint fields is
\begin{align}
	\label{Sgauge}
		S_{\mathrm{gauge}}&=\!\int\!d^4x\,d^2\theta\,d^2\bar{\theta}\,\Big(
		- V^a\square V^a+\Phi^{\dagger a}\Phi^a
		+\frac{\ii}{4}g f^{abc}\,\big[\bar{D}^2(D^{\alpha} V^a)\big]\, V^b\, (D_{\alpha} V^c) 
		\nonumber\\
		& \qquad\qquad 
		-\frac{1}{8}\,g^2 f^{abe}f^{ecd}\, V^a (D^{\alpha} V^b) (\bar{D}^2 V^c)(D_{\alpha} V^d) \nonumber\\
		& \qquad\qquad+ 2\,\ii g f^{abc}\,\Phi^{\dagger a}V^b\Phi^c
		- 2g^2 f^{abe}f^{ecd}\,\Phi^{\dagger a}V^b V^c \Phi^d
		+\cdots\Big)
\end{align}
where the dots stand for higher order vertices of the schematic form 
$g^k\: \Phi^{\dagger}V^k \Phi$ with $k\geq 3$. 
Here $f^{abc}$ are the structure constants of SU$(N)$ (see
appendix~\ref{app:group} for our group-theory conventions). 
The Feynman rules following from this action are displayed in figure \ref{fig:Feyngauge}.

\begin{figure}[ht]
\begin{center}
	\includegraphics[scale=0.7]{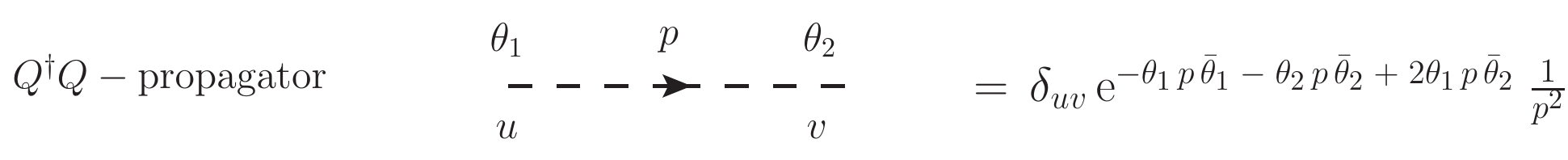}\\ \vspace{0.2cm}
	\includegraphics[scale=0.7]{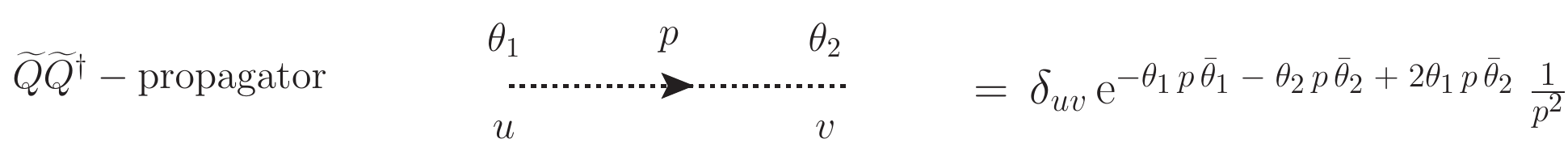} \\
	\includegraphics[scale=0.6]{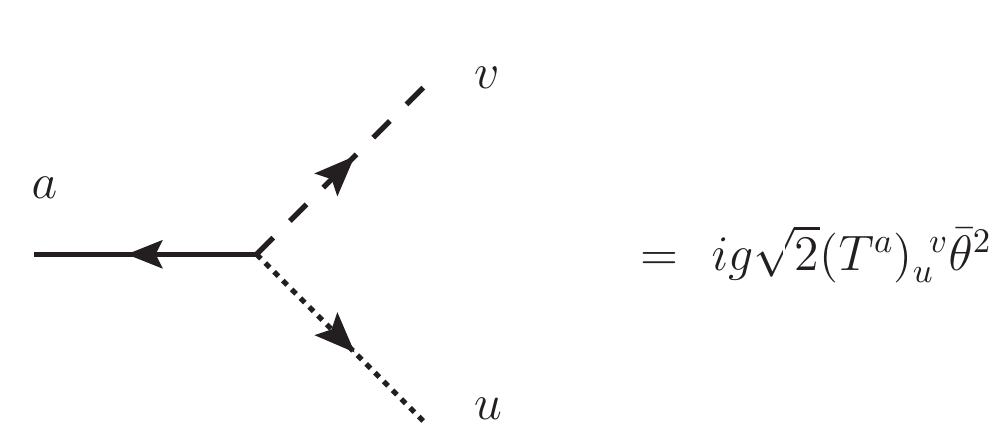}\hspace{1.5cm}
	\includegraphics[scale=0.6]{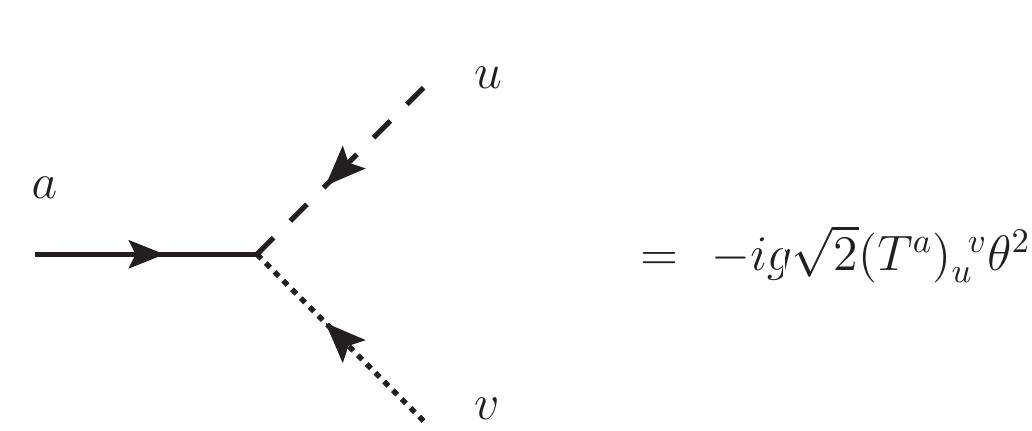} \\
	\includegraphics[scale=0.6]{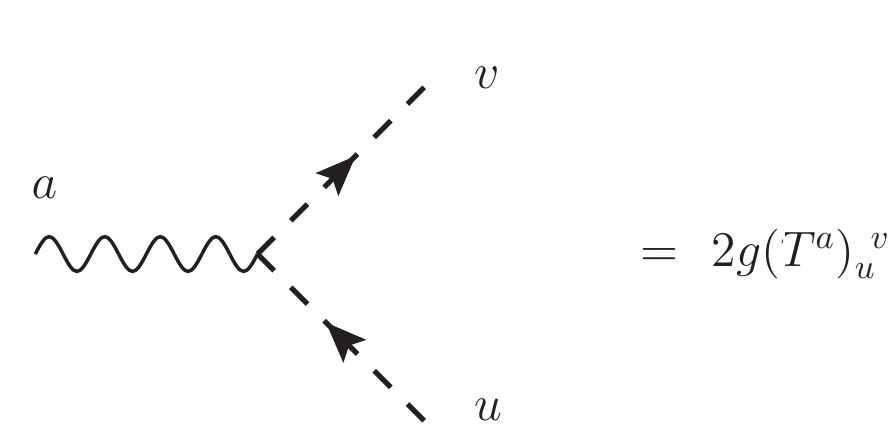}\hspace{2cm}
	\includegraphics[scale=0.6]{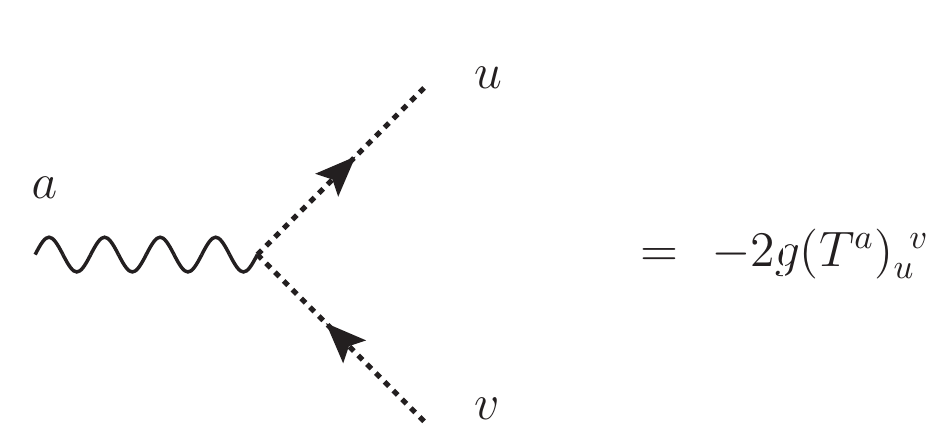} \\ \vspace{0.2cm}
	\includegraphics[scale=0.6]{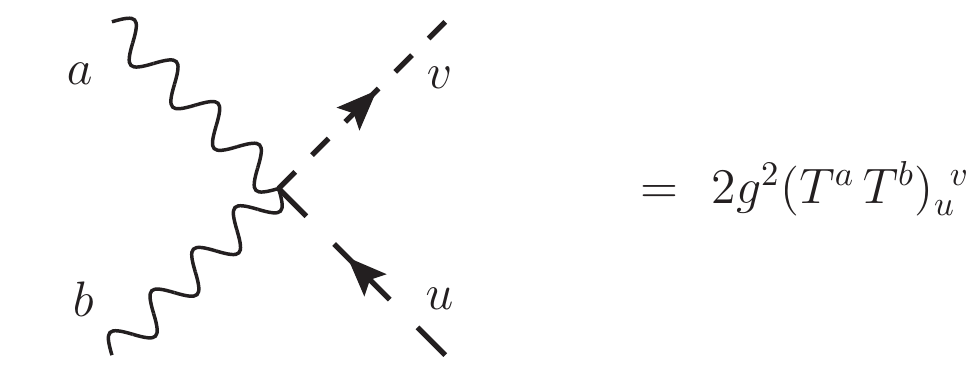}\hspace{1.5cm}
	\includegraphics[scale=0.6]{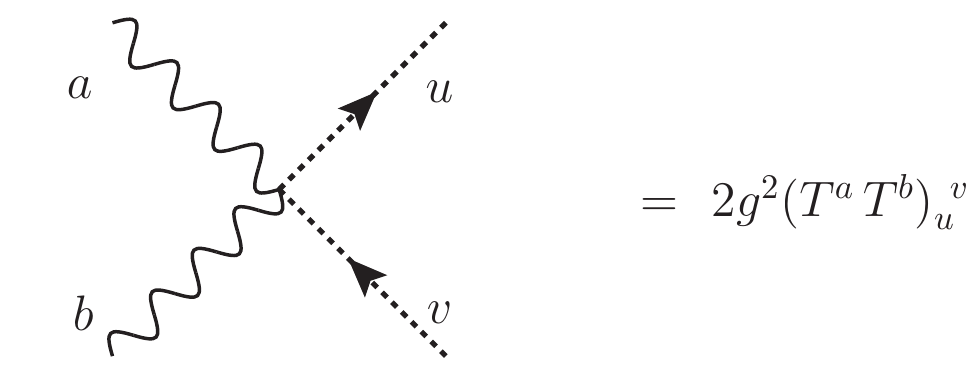}
	\caption{Feynman rules involving the matter superfields that are relevant
	for our calculations.}
	\label{fig:Feynmatter}
\end{center}
\end{figure}

An $\cN=2 $ hypermultiplet in a representation $\cR$ contains two $\cN=1$ chiral multiplets, $Q$ 
transforming in the representation $\cR$ and $\widetilde Q$ transforming in the conjugate 
representation $\bar{\cR}$; we denote by $Q_u$, $u=1,\ldots d_\cR$ and $\widetilde{Q}^u$ their 
components\,%
\footnote{This is a compact notation which encompasses also the cases in which $\cR$ is reducible, 
and in particular the cases in which it contains several copies of a given irreducible representation. 
For instance, if $\cR$ is the direct sum of $N_F$ fundamental representations, we use here an 
index $u=1,\ldots N_F N$, instead of a double index, $m,i$ with $m=1,\ldots n$ for the color and 
$i=1,\ldots N_F$ for the flavor.}. The action for these matter fields, again in the Fermi-Feynman
gauge, is
\begin{align}
	\label{Smatter}
		S_{\mathrm{matter}}& =
		\!\int\!d^4x\,d^2\theta\,d^2\bar{\theta}\,\Big(Q^{\dagger\,u} Q_{u}
		+ 2g\,Q^{\dagger\,u} V^a (T^a)_{u}^{\,v}\,Q_{v}
		+  2g^2\,Q^{\dagger\,u}
		V^a\,V^b(T^a \,T^b)_{u}^{\,v} \, Q_{v}\notag\\
		&\qquad\qquad+ \widetilde{Q}^{u}\,\widetilde{Q}^\dagger_{u} 
		- 2g\,\widetilde{Q}^{u} \,V^a (T^a)_{u}^{\,v}\,\widetilde{Q}^\dagger_{v}
		+2g^2\,\widetilde{Q}^{u} V^a \,V^b (T^a\,T^b)_{u}^{\,v}
		\,\widetilde{Q}^\dagger_{v}+\cdots\notag\\
		&\qquad\qquad
		+ \ii\sqrt{2}g\,\widetilde{Q}^{u}\Phi^a (T^a)_{u}^{\, v} Q_{v}\,\bar{\theta}^2
		- \ii\sqrt{2}g\,Q^{\dagger\,u}\Phi^{\dagger\,a}(T^a)_{u}^{\,v} 
		\widetilde{Q}^\dagger_{v }\,\theta^2
		\Big)
\end{align}
where by $T^a$ we denote the SU$(N)$ generators in the representation $\cR$.
The Feynman rules that are derived from this action are illustrated in figure \ref{fig:Feynmatter}.

Therefore, the total action for the $\cN=2$ theory is simply
\begin{equation}
	\label{Stot} 
		 S = S_{\mathrm{gauge}} +S_{\mathrm{matter}}~.
\end{equation}
The $\cN=4$ SYM theory can be seen as a particular $\cN=2$ theory containing a vector multiplet and an hypermultiplet, both in the adjoint representation of the gauge group. 
So it corresponds simply to the case in which $\cR$ is the adjoint representation. 
In terms of $\cN=1$ superfields, beside $V$ and $\Phi$, it contains also two adjoint
chiral multiplets that we call $H$ and $\widetilde H$ (note that the adjoint representation 
is self-conjugate). Their components are denoted as $H_a$, $\tilde{H}_a$, with $a= 1,\ldots N^2-1$,
and their action $S_H$ has the same structure as $S_{\mathrm{matter}}$ 
with $Q_u$ and $\tilde Q^u$ replaced by $H_a$ and $\tilde H_a$ and the generator components 
$(T_a)_u^{\, v}$ by the structure constants $\ii f_{abc}$.
Thus we can write
\begin{equation}
	\label{S4}
		S_{\cN=4}= S_{\mathrm{gauge}} + S_H~.
\end{equation}
Doing the same substitutions on the Feynman rules of figure \ref{fig:Feynmatter} yields the Feynman rules for the $H$ and $\widetilde{H}$ superfields.

{From} (\ref{Stot}) and (\ref{S4}) it is easy to realize that the total action of our $\cN=2$ theory can be written as
\begin{equation}
S = S_{\cN=4} - S_H + S_{\mathrm{matter}}~.
\end{equation}
Actually, given any observable $\cA$ of the $\cN=2$ theory, which also exists in the
$\cN=4$ theory, we can write
\begin{equation}
\label{difference}
\Delta\cA= \mathcal{A} - \mathcal{A}_{\cN=4} = \mathcal{A}_{\mathrm{matter}} - \mathcal{A}_H~.
\end{equation}
Thus, if we just compute the difference with respect to the $\cN=4$ result, we have to consider only 
diagrams where the hypermultiplet fields, either of the $Q$, $\widetilde Q$ type or of
the $H$, $\widetilde{H}$ type, propagate in the internal lines, and then consider the difference
between the $(Q,\widetilde{Q})$ and the $(H,\widetilde{H})$ diagrams.
This procedure, which was originally used in \cite{Andree:2010na}, reduces in a significant way
the number of diagrams to be computed. Moreover, as we remarked in section~\ref{subsec:intSa},
it is suggested by the structure of the matrix model.
 
We will apply this method to explicitly evaluate by means of Feynman superdiagrams two 
quantities: the propagator of the scalar $\varphi$ and the vacuum expectation value
of the 1/2 BPS circular Wilson loop. From now on we assume that our theory is conformal at the quantum level, namely that the $\beta$-function coefficient $\beta_0$ vanishes. This amounts 
to ask that the index of the representation $\cR$ be equal to $N$, see (\ref{C2is}).  

\subsection{The scalar propagator}
\label{subsec:scalprop}
The tree level propagator for the adjoint scalar field $\varphi$ of
the vector multiplet can be extracted from the propagator of the superfield $\Phi$ given
in the first line of figure~\ref{fig:Feynmatter} by imposing $\theta_1 = \theta_2 =0$:
\begin{equation}
	\label{prop-tree}
		\Delta^{bc}_{(0)}(q)= \frac{\delta^{bc}}{q^2}~.
\end{equation}
Since we consider conformal  $\cN=2$ theories, the quantum corrected propagator will 
depend on the momentum only through the factor $1/q^2$, 
and by gauge symmetry it can only be proportional to $\delta^{bc}$. So we will have
\begin{equation}
	\label{propPi}
		\Delta^{bc}(q)= \frac{\delta^{bc}}{q^2}\,\big(1 + \Pi\big)
\end{equation}
where $\Pi$ is a $g$-dependent constant describing the effect of the perturbative 
corrections.  This 
constant should be captured by the matrix model and thus should be the same as the 
quantity $\Pi$ defined in (\ref{corrbcres1}). We will now check explicitly that this is indeed 
the case, up to the three-loop order corrections proportional to $\zeta(5)$. 

\subsubsection*{One loop} 
\begin{figure}[ht]
	\begin{center}
		\includegraphics[scale=0.5]{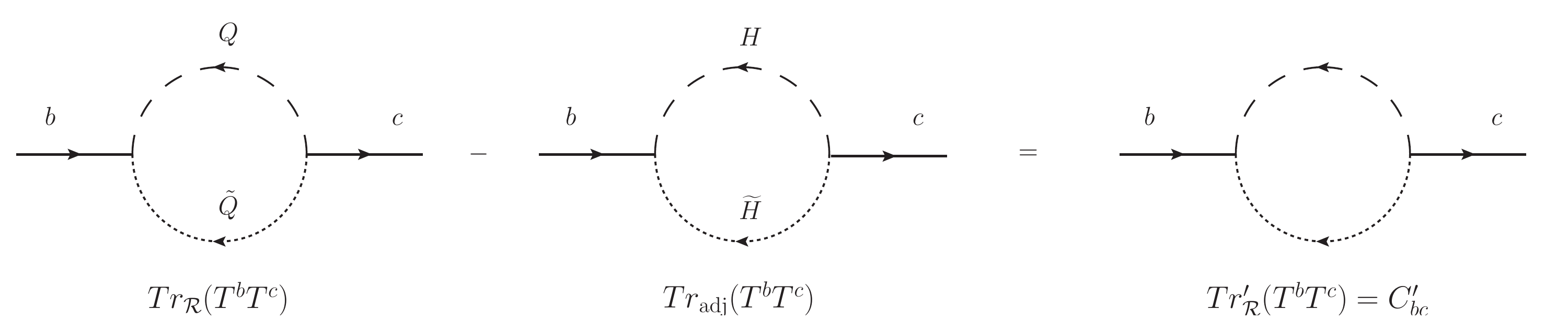}
	\end{center}
	\caption{One-loop correction to $\Phi$ propagator in the difference theory. }
	\label{fig:1loop}
\end{figure}

At order $g^2$ the first diagram we have to consider is
\begin{equation}
	\label{1loopQ}
		\parbox[c]{.4\textwidth}{\includegraphics[width 
		= .4\textwidth]{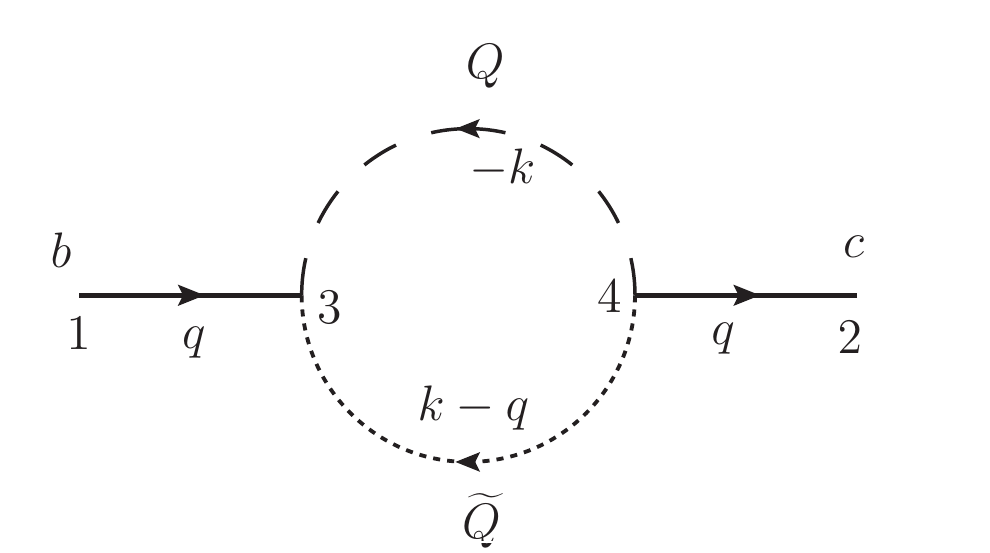}}
		\hspace{-0.6cm}= 2 g^2 \times \Tr_\cR(T^b T^c)\, \times
		\int \frac{d^dk}{(2\pi)^d} \frac{1}{(q^2)^2}\frac{1}{k^2(k-q)^2}\, \cZ(k,q)~.
\end{equation}
Here, and in all following diagrams, we adopt the notation explained in detail in 
appendix~\ref{app:diagrams} (see in particular (\ref{gen-diag}) and the following 
sentences): we write the diagram as the product of a normalization factor, $2 g^2$ 
in this case, which takes into account the combinatoric factor and the strength 
of the vertices, a color factor, and an integral over the internal momenta. 
The factor $\cZ(k,q)$ is the result of the integration over the Grassmann variables at each 
internal vertex\,%
\footnote{The Grassmann variables in the external points 1 and 2 are set to zero to pick up 
the lowest component $\varphi$ of the superfield, namely we have 
$\theta_1=\bar\theta_2=0$. 
Note that if we do not do this and consider the propagator of the full superfield $\Phi$ the 
color factor remains the same.} and, according to the rules in figures \ref{fig:Feyngauge} 
and \ref{fig:Feynmatter} reads
\begin{align}
	\label{Z1loopQ}
		\cZ(k,q) = \int d^4\theta_3 \,d^4\theta_4\, (\theta_3)^2 ({\bar\theta}_4)^2\,
		\exp{\big(\!-2 \,\theta_4 \,q\, \bar{\theta}_3\big)} = - q^2~.
\end{align}
The momentum integral in (\ref{1loopQ}) is divergent for $d\to 4$; however in the 
difference theory we have to subtract an identical diagram in which the adjoint superfields 
$H$ and $\widetilde H$ run in the loop instead of $Q$ and $\widetilde Q$. This diagram 
has the same expression except for the color factor which is now given by 
$\Tr_\mathrm{adj}(T^b T^c)$. The difference of the two diagrams is therefore proportional 
to 
\begin{align}
	\label{Trprbc}
		\Tr_\cR(T^b T^c) - \Tr_\mathrm{adj}(T^b T^c) = 	\Tr_\cR^\prime(T^b T^c) 
		= C^\prime_{bc}~. 
\end{align}
{From} now on, we will use the graphical notation introduced in figure~\ref{fig:1loop}, 
according to which a hypermultiplet loop stands for the difference between 
the $(Q,\widetilde Q)$ and the $(H,\widetilde H)$ diagrams, with a color factor 
that is directly given by a primed trace.

As already stated in (\ref{C2is}), the color factor (\ref{Trprbc}) for the one-loop correction, 
being  proportional to the $\beta_0$ coefficient, vanishes for conformal theories. 
Thus the constant $\Pi$ in (\ref{propPi}) starts at order $g^4$ and all diagrams including 
the one-loop correction to the $\Phi$ propagator as a sub-diagram vanish.  

\subsubsection*{Building blocks for higher order diagrams}
Let us now consider higher order diagrams in the difference theory. 
Similarly to what happens at one-loop as shown in figure~\ref{fig:1loop}, the 
contributions of the $(Q,\widetilde Q)$ and $(H,\widetilde H)$ hypermultiplets always 
have a color factor that contains a ``primed'' trace of generators, 
{\it{i.e.}} they contain the tensor $C^\prime_{b_1 \ldots b_n}$ defined in (\ref{defC}). 
We will use the symbol $C^\prime_{(n)}$ to denote such a tensor 
when we do not need to specify explicitly its $n$ indices. Notice that, according to the 
Feynman rules, each insertion of a generator on the hypermultiplet loop carries a factor 
of $g$, so that the color factor $C^\prime_{(n)}$ is always accompanied by a factor of 
$g^n$.

\begin{figure}[ht]
	\begin{center}
		\includegraphics[scale=0.7]{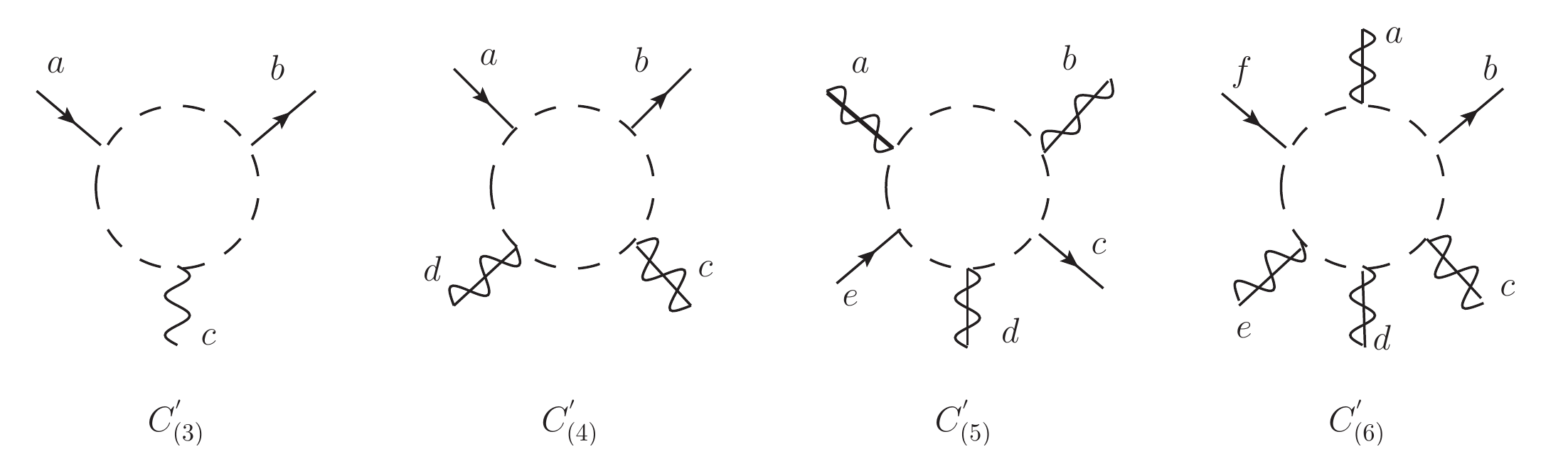} \\[6mm]
		\includegraphics[scale=0.7]{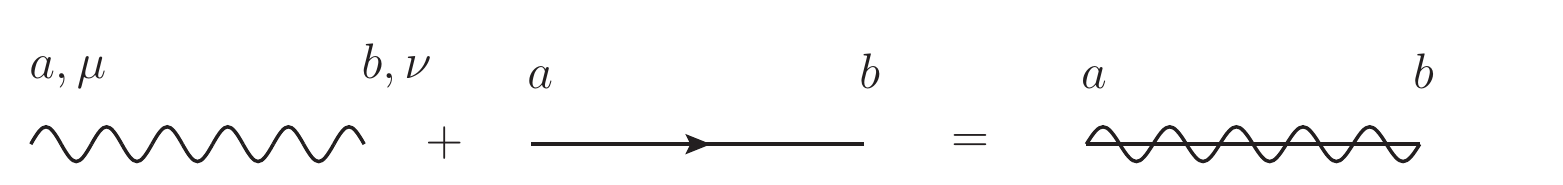}
	\end{center}
	\caption{Each building block is accompanied by its color coefficient of the type $C^\prime_{(n)}$.
	Here we used a generic dashed lines for hypermultiplets loops. In reality some part of 
	the loop should be dashed and some dotted, in accordance with the Feynman rules. 
	The wiggled/straight line stands for $V$ or $\Phi$ propagators, as explained in the second row 
	of the figure.}
	\label{tracesdraw}
\end{figure}

In the difference theory all diagrams up to order $g^6$ can be formed using the building 
blocks depicted in figure~\ref{tracesdraw}, and suitably contracting the adjoint lines, 
corresponding to $V$ or $\Phi$ propagators, inserted in the loops. 

As a matter of fact, we can also have quartic vertices with two gluon lines inserted in the 
same point along the hypermultiplet loop, each of which comes with a factor of $g^2$ and 
two generators. 
However, for the purpose of identifying the color factors, these contributions do not 
substantially differ from those produced by two separate insertions. Therefore, the possible 
color structures that occur up to the order $g^6$ can all be derived from the diagrams in 
figure~\ref{tracesdraw}.
Organizing the Feynman diagrams according to their color 
coefficients $C'_{(n)}$ in the way we have outlined facilitates the comparison with the 
matrix model. 

\begin{figure}[htb]
	\begin{center}
		\includegraphics[scale=0.47]{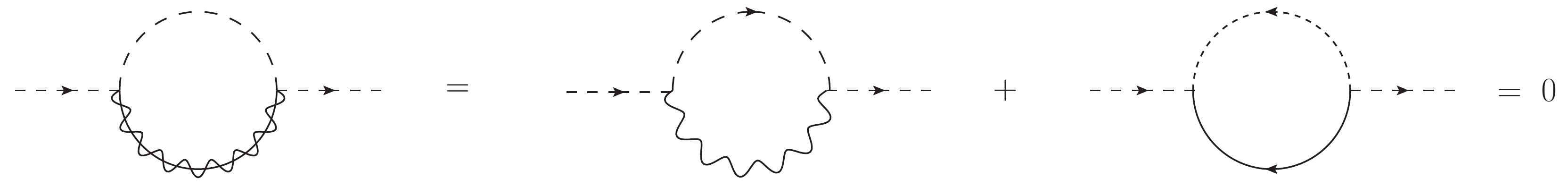}
	\end{center}
	\caption{The one-loop correction to hypermultiplet propagator vanishes.}
	\label{fig:hyper1loop}
\end{figure}

In constructing higher order diagrams we exploit a further simplification: in $\cN=2$ 
theories the one-loop correction to any hypermultiplet propagator vanishes. This is 
illustrated in figure~\ref{fig:hyper1loop}. Such one-loop corrections cannot therefore 
appear as sub-diagrams of higher loop diagrams. 

\subsubsection*{Two loops} 
\begin{figure}[htb]
	\begin{center}
		\includegraphics[scale=0.6]{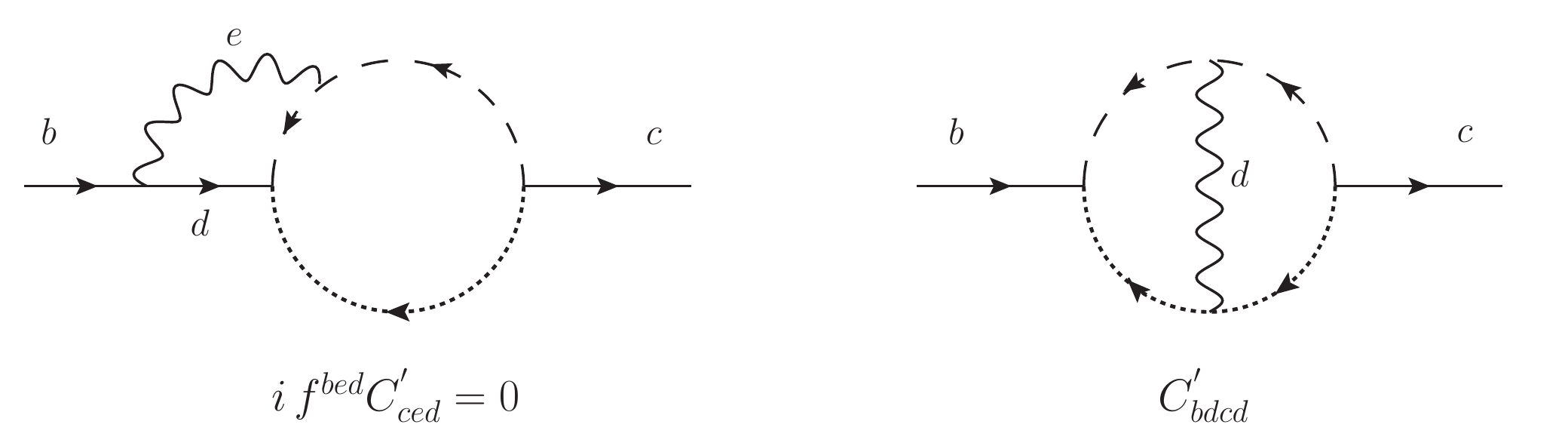}
	\end{center}
	\caption{Two-loop diagrams and their color factors}
	\label{fig:2loopp}
\end{figure}

At order $g^4$ there are two classes of diagrams that may contribute, whose color 
coefficients are proportional to $C^\prime_{(3)}$ or to $C^\prime_{(4)}$. 
The diagrams proportional to $g^3\,C^\prime_{(3)}$ always contain also an 
adjoint vertex proportional to $g$ with which they are contracted. 
This is the case represented on the left in  figure~\ref{fig:2loopp}. However, due the 
symmetry properties of the tensor $C^\prime_{(3)}$ 
(see  (\ref{C3confsym})), they vanish and one is left only with the diagrams with four 
adjoint insertions in the hypermultiplet loop.

Let us now consider these diagrams. As remarked before, a building block with four adjoint 
lines inserted on the hypermultiplet loop is proportional to $g^4\,C^\prime_{(4)}$, so at
this order we cannot add any other vertices to it. Moreover, there is a unique contraction 
allowed, since each hypermultiplet field has a vanishing one-loop propagator. Thus, the 
only diagram at this order is the one represented on the right in figure~\ref{fig:2loopp}. 
This has already been computed in \cite{Andree:2010na} (see also \cite{Billo:2017glv}). 
Performing the Grassmann algebra and the momentum integral, we obtain 
a finite result proportional to $\zeta(3)$, which explicitly reads
\begin{equation}
	\label{2loopprop}
		\parbox[c]{.33\textwidth}{\includegraphics[width 
		= .33\textwidth]{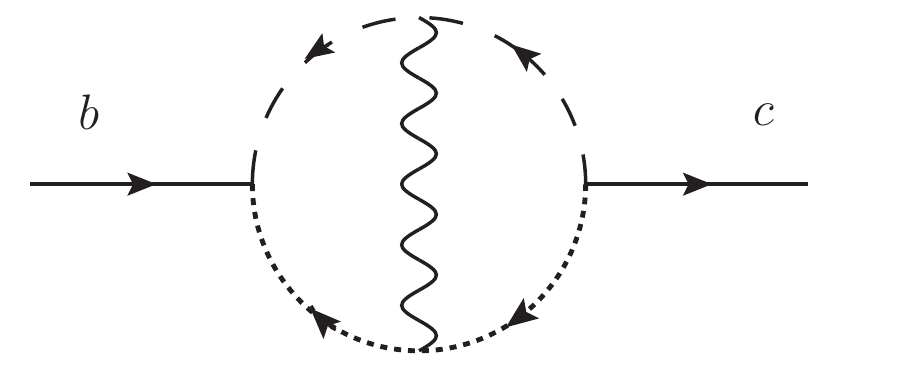}}
		\hspace{-0.4cm}= \frac{1}{q^2} \left(\frac{g^2}{8\pi^2}\right)^2\zeta(3)
		\times  6\, C^{\prime}_{bdcd}~.
\end{equation}
Using the properties of the $C^\prime$-tensors - see in particular (\ref{C4sw1}) and 
(\ref{C4sw12}) - we have
\begin{equation}
6 \,C^{\prime}_{bdcd} = 6 \,C^{\prime}_{(bcdd)}= \cC^{\prime}_{4}\,\delta^{bc}~.
\end{equation} 
Since this is the only correction to the propagator at this order, from (\ref{propPi}) we find 
\begin{align}
	\label{Pig2}
		\Pi = \left(\frac{g^2}{8\pi^2}\right)^2\zeta(3) \, \cC^{\prime}_{4} + \cO(g^6)~,
\end{align}
in perfect agreement with  the matrix model result reported in (\ref{c4c6delta}) and 
(\ref{Piis}). This is an extension to a generic $\cN=2$ SYM theory of the check 
originally performed in \cite{Andree:2010na} for conformal SQCD.

\subsubsection*{Three loops}
At order $g^6$ many diagrams survive also in the difference theory. Moreover, some
of them can be divergent in $d=4$. However, since we are dealing with conformal
field theories, all divergences cancel when one sums all contributing diagrams. 
Therefore, we can concentrate on extracting the finite part, which the matrix
model result (\ref{corrbcres}) suggests to be proportional to $\zeta(5)$.
Thus we only look for diagrams which provide $\zeta(5)$ contributions, and we check 
that their sum reproduces exactly the matrix model result.

To scan all the possible $\zeta(5)$-contributions we use the same approach we 
applied above. We start from the building blocks in figure~\ref{tracesdraw} and 
contract their adjoint lines in all the possible ways, introducing new vertices 
when necessary. It is quite simple to realize that many of the diagrams 
that are created in this way have a vanishing color factor.
For example, the diagrams proportional to $C'_{(3)}$ vanish for the same reason
we discussed before. As far as the diagrams with $C'_{(4)}$ are concerned, 
we can discard those containing as a sub-diagram the two-loop contribution 
on the right of figure \ref{fig:2loopp} since this latter is proportional to $\zeta(3)$, 
and no $\zeta(5)$-contribution can arise from this kind of diagrams. 
All other possible diagrams that one can construct using as building block a sub-diagram
with $C'_{(4)}$ vanish by manipulations of their color factors. 

We are left with diagrams whose color factor is proportional either to $C'_{(5)}$ 
or to $C'_{(6)}$. In the first case, the building block is proportional to $g^5$ and thus we
have insert a further cubic vertex to obtain the desired power of $g$; in the second case,
instead, the building block is already of order $g^6$, and so we can only contract its 
adjoint lines among themselves. 
We have made a systematic search of all diagrams that can be obtained in this way. 
Many of them vanish either because of their color factor or because of 
the $\theta$-algebra, while in other cases the momentum integral does not 
produce any $\zeta(5)$-contribution. 
In the following we list all of the diagrams that \emph{do} yield a 
$\zeta(5)$-term. There are seven such diagrams, named $\cW_{bc}^{(I)}(q)$
with $I=1,\ldots 7$, which are explicitly computed in appendix~\ref{app:diagrams}.  
Here we simply report the result in a schematic way, writing each of them in the
form
\begin{align}
	\label{resWI}
		\cW_{bc}^{(I)}(q) = 
		-\frac{1}{q^2} \left(\frac{g^2}{8\pi^2}\right)^3 \zeta(5)
		\times x^{(I)}\,\cT_{bc}^{(I)}
\end{align}  
where $\cT_{bc}^{(I)}$ is the color factor, which is in fact proportional to $\delta_{bc}$, 
and $x^{(I)}$ is a numerical coefficient determined by the explicit evaluation of the
 integrals over the loop momenta.
In detail, we have
\begingroup
\allowdisplaybreaks
\begin{align}
	\label{resz5-1}
		\cW_{bc}^{(1)} (q) = 
		\parbox[c]{.28\textwidth}{\includegraphics[width 
		= .28\textwidth]{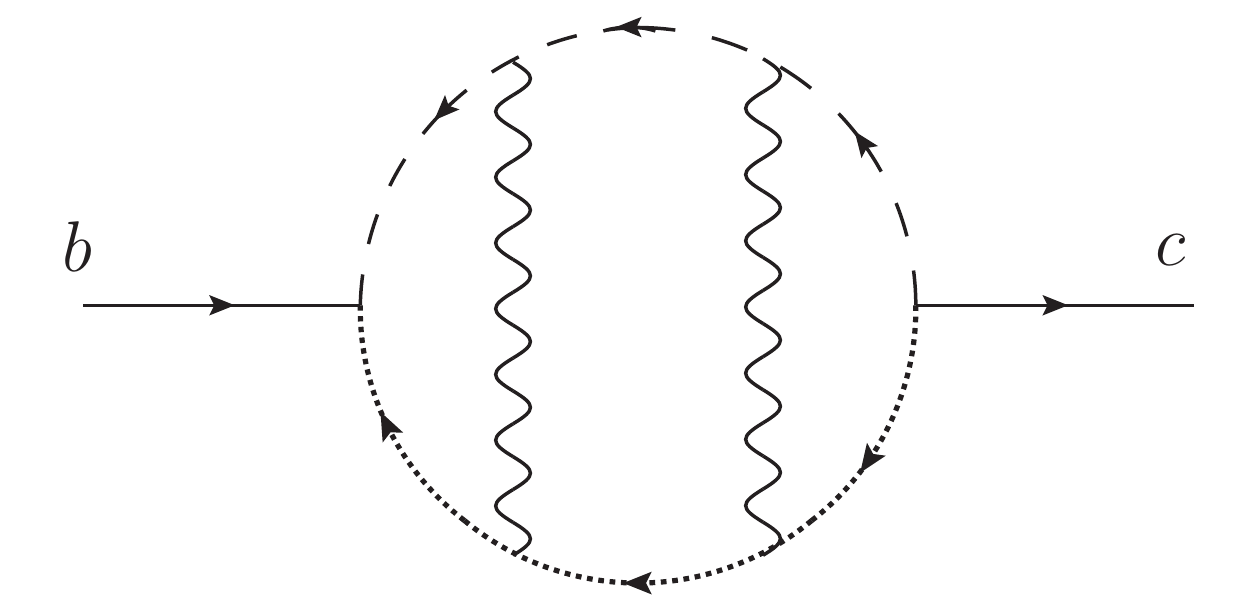}}
		~~~ \rightarrow ~~~  
		x^{(I)}\,\cT_{bc}^{(1)} & = 
		20\, C^\prime_{bdeced}~,\\
		\cW_{bc}^{(2)} (q) = 		
		\parbox[c]{.28\textwidth}{\includegraphics[width 
		= .28\textwidth]{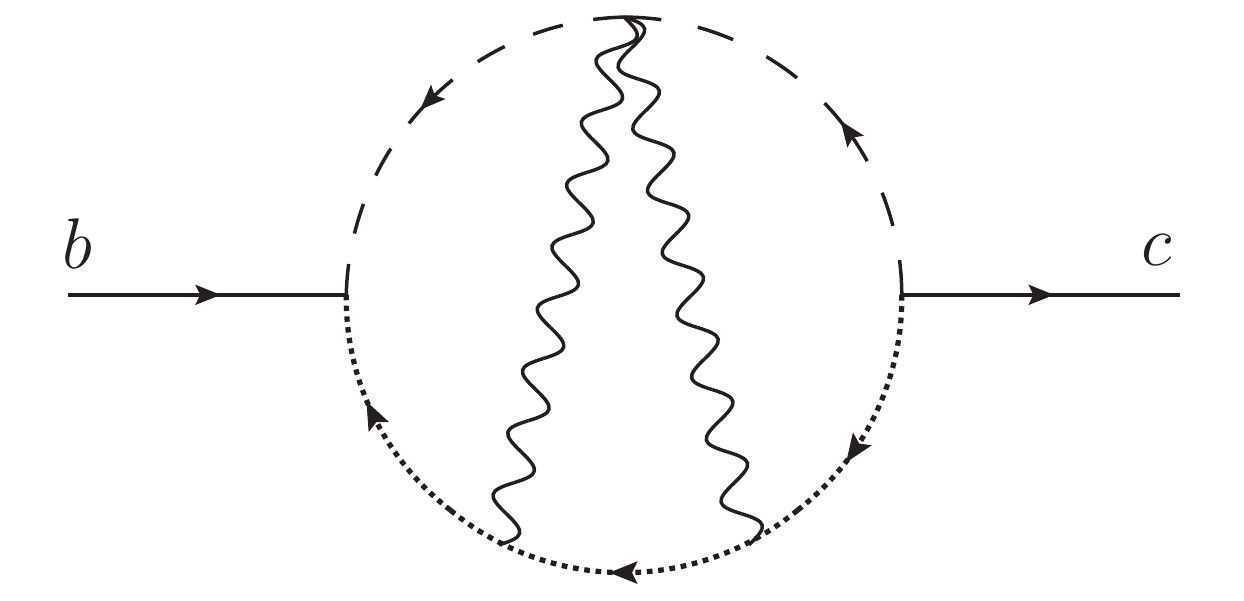}}
		~~~ \rightarrow ~~~  
		x^{(2)}\,\cT_{bc}^{(2)}
		& =  -20\,C^\prime_{bdeced}-20\,C^\prime_{bdecde}~,\\
		\cW_{bc}^{(3)} (q) = 		
		\parbox[c]{.28\textwidth}{\includegraphics[width 
		= .28\textwidth]{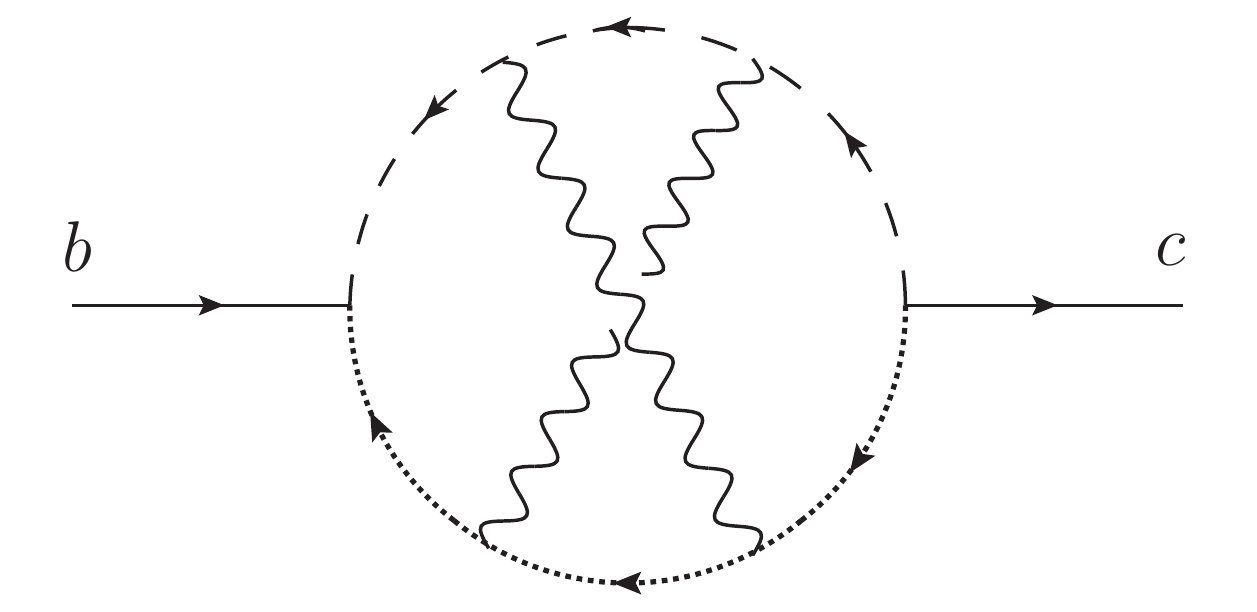}}
		~~~ \rightarrow ~~~  
		x^{(3)}\,\cT_{bc}^{(3)}
		& =  10\, C^\prime_{bdecde}~,\\
		\cW_{bc}^{(4)} (q) = 		
		\parbox[c]{.28\textwidth}{\includegraphics[width 
		= .28\textwidth]{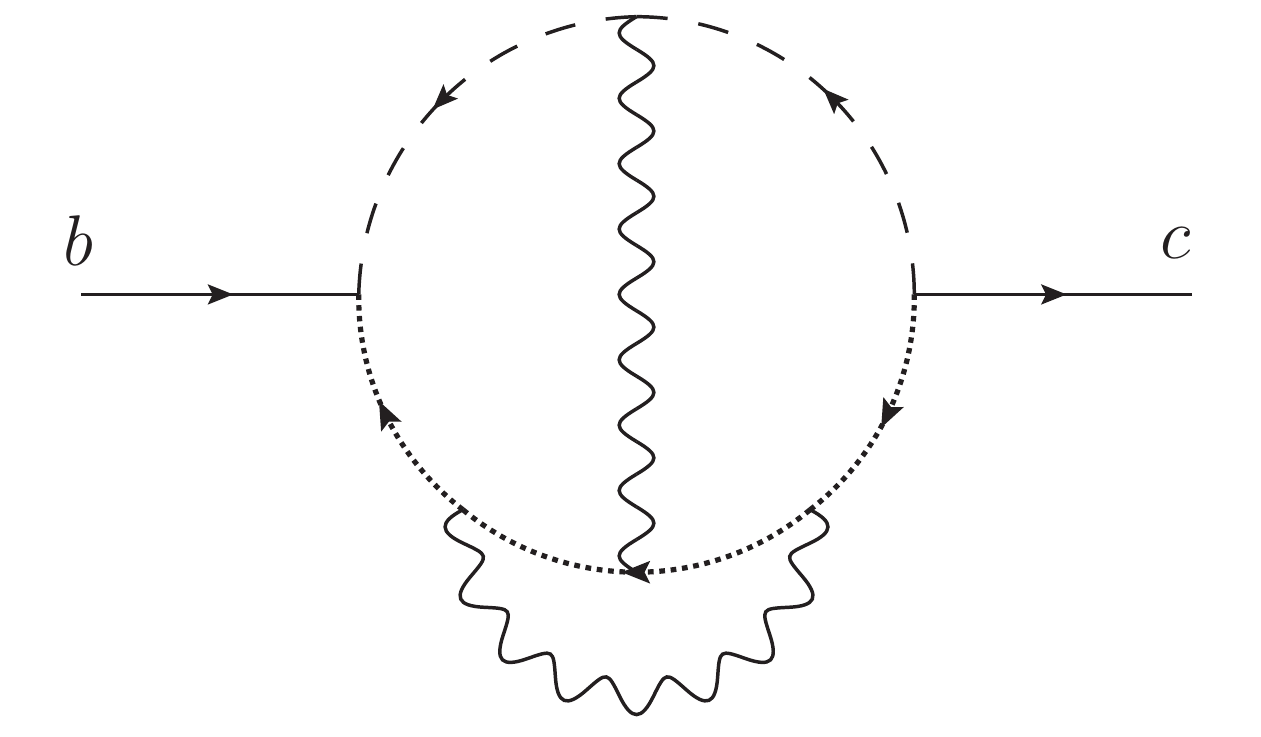}}
		~~~ \rightarrow ~~~  
		x^{(4)}\,\cT_{bc}^{(4)}
		& =  20\,C^\prime_{bdcede}+20\,C^\prime_{bedecd}~,\\
		\cW_{bc}^{(5)} (q) = 		
		\parbox[c]{.28\textwidth}{\includegraphics[width 
		= .28\textwidth]{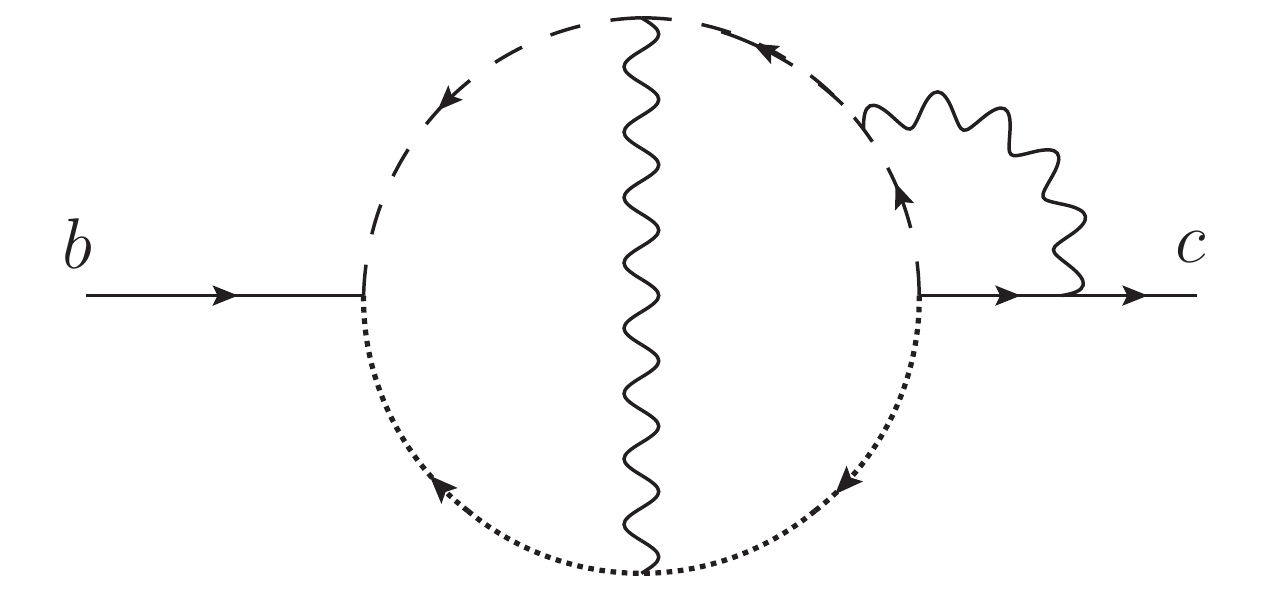}}
		~~~ \rightarrow ~~~  
		x^{(5)}\,\cT_{bc}^{(5)}
		& =  -40\,\ii f_{cef} C^\prime_{bdefd} -40\, \ii f_{bef} C^\prime_{cdefd}~,\\
		\cW_{bc}^{(6)} (q) = 		
		\parbox[c]{.28\textwidth}{\includegraphics[width 
		= .28\textwidth]{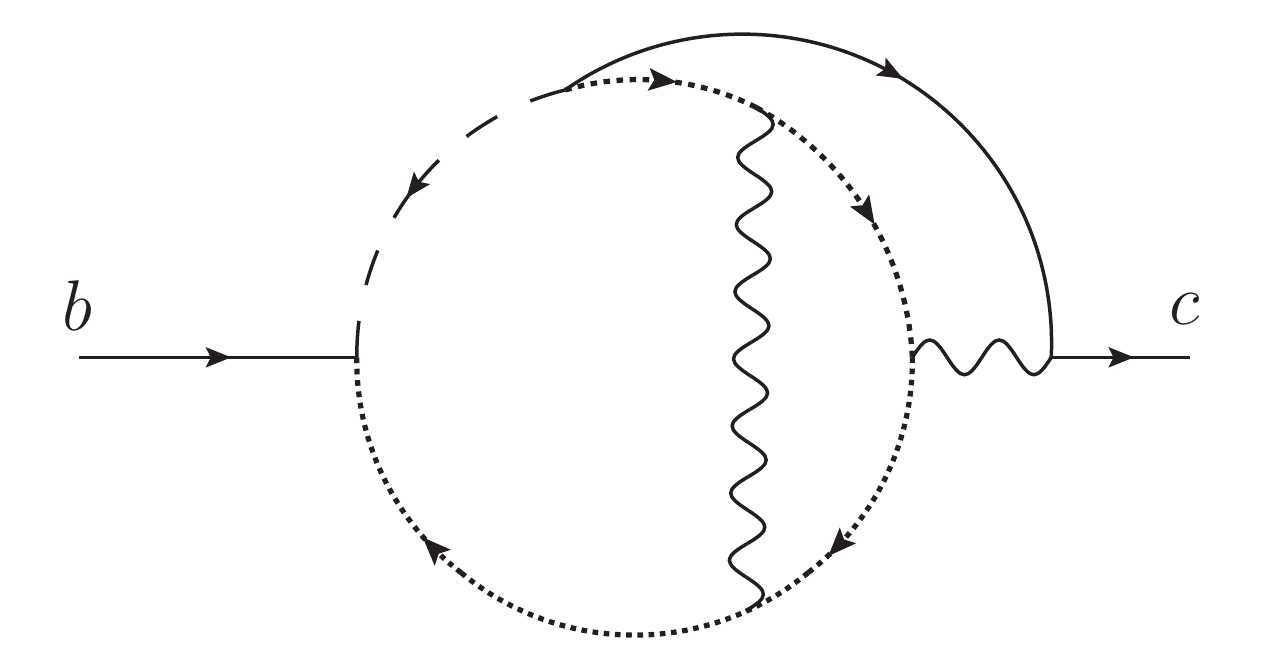}}
		~~~ \rightarrow ~~~  
		x^{(6)}\,\cT_{bc}^{(6)}
		& = -20\,\ii f_{ced} C^\prime_{bfdfe} +20\, \ii f_{ced} C^\prime_{befdf}
		\nonumber\\[-5mm]
		&~~~-20\, \ii f_{bed} C^\prime_{cfdfe} +20\, \ii f_{bed} C^\prime_{cefdf}~,\\
		\cW_{bc}^{(7)} (q) = 		
		\parbox[c]{.28\textwidth}{\includegraphics[width 
		= .28\textwidth]{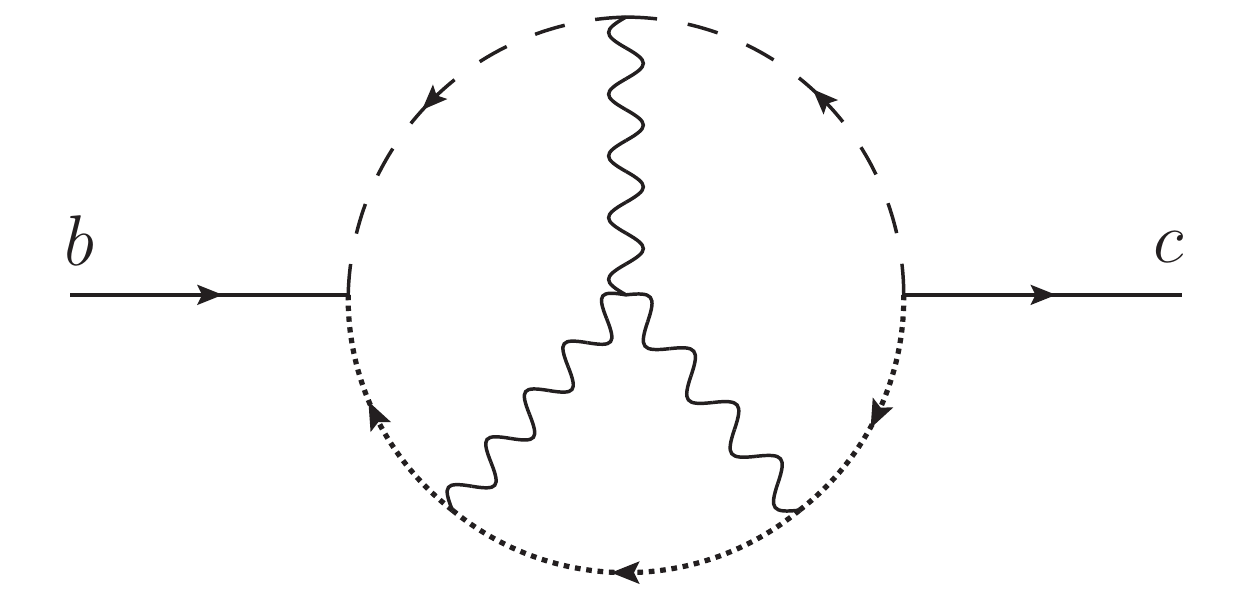}}
		~~~ \rightarrow ~~~  
		x^{(7)}\,\cT_{bc}^{(7)}
		& = 10\, \ii f_{def} C^\prime_{bfecd} + 10\,\ii f_{def} C^\prime_{cfebd}~.
\end{align}
\endgroup
Since each color factor is proportional to $\delta^{bc}$, we can identify terms that are equal up to an exchange of $b$ and $c$. In this way we get
\begin{equation}
	\label{sumTI}
		\sum_{I=1}^7 x^{(I)}\,\cT_{bc}^{(I)} = 
		- 80\, \ii f_{ced} C^\prime_{bfdfe} + 80\, \ii f_{ced} C^\prime_{bfdef} 
		- 10\, C^\prime_{bdecde} 
		+ 40 \,C^\prime_{bdcede} 
		+ 20\, \ii f_{def} C^\prime_{bfecd}~.
\end{equation}
Using the relation (\ref{switchC}), it is easy to see that the first two terms actually 
cancel, and that the remaining ones can be written as follows:
\begin{equation}
	\label{sumTIbis}
		\sum_{I=1}^7 x^{(I)}\,\cT_{bc}^{(I)} = 
		30\, C^\prime_{bdcede} -10 \ii f_{ced} C^\prime_{bfdfe} 
		+ 20\, \ii f_{def} C^\prime_{bfecd}~.
\end{equation}
This expression is apparently different from the color tensor in the $g^6$-term of the matrix model result (\ref{corrbcres}). In fact, the latter contains the totally symmetric 
combination $30 C^\prime_{(bdcede)}$ and does not contain any $C^\prime$ with five indices. However, using again (\ref{switchC}) and the properties of the $C^\prime$ tensors described in appendix~\ref{app:group}, it is possible to show that the last two terms in (\ref{sumTIbis}) 
precisely symmetrize the first term. 
The total three-loop contribution is therefore
\begin{align}
	\label{sumWI}
		\sum_{I=1}^7 \cW_{bc}^{(I)}(q) &
		=	-\frac{1}{q^2} \left(\frac{g^2}{8\pi^2}\right)^3 \zeta(5)
		\times 30\, C^\prime_{(bdcede)}\notag\\
		& = -\frac{1}{q^2} \left(\frac{g^2}{8\pi^2}\right)^3 \zeta(5)
		\times \cC^\prime_6\,\delta_{bc}~,
\end{align}
where in the last step we used (\ref{c4c6delta}). 
Altogether, adding the two-loop term (\ref{Pig2}), the quantum corrections
of the scalar propagator proportional to $g^4\,\zeta(3)$ and $g^6\,\zeta(5)$ are
\begin{align}
	\label{Pig3}
		\Pi = \zeta(3) \left(\frac{g^2}{8\pi^2}\right)^2 \cC^{\prime}_{4}
		- \zeta(5) \left(\frac{g^2}{8\pi^2}\right)^3 \cC^{\prime}_{6} + \cO(g^8)~. 
\end{align}
This result fully agrees with the matrix model prediction given in (\ref{Piis}).

As already mentioned at the end of section~{\ref{subsec:propagator}}, we observe that
the color tensors $C^\prime_{b_1\ldots b_n}$ and the coefficients $\cC^\prime_n$ can be 
defined for any representation of SU($N$) (or U($N$)). Moreover, the steps that we performed above
to show the agreement with the matrix model predictions only rely on the symmetry/anti-symmetry
properties of these tensors and their group-theory properties, and not on their specific expressions
for the SU($N$) theories with matter in the fundamental, symmetric or anti-symmetric representations. For this reason we believe that the same match could be proved and realized 
also in more general superconformal theories with other gauge groups and matter representations.

\subsection{Supersymmetric Wilson loop}
\label{subsec:swl}
We now consider the perturbative computation of the vacuum expectation value of a 
1/2 BPS circular Wilson loop in the fundamental representation. This composite operator, 
placed on a circle $C$ of radius $R$, is defined as
\begin{equation}
	\label{WLdef}
		W(C)=\frac{1}{N}\tr \: \mathcal{P}
		\exp \left\{g \oint_C d\tau \Big[\ii \,A_{\mu}(x)\,\dot{x}^{\mu}(\tau)
		+\frac{R}{\sqrt{2}}\big(\varphi(x) + \bar\varphi(x)\big)\Big]\right\}
\end{equation}
where $\mathcal{P}$ denotes the path-ordering.
We parametrize the loop as:

	\begin{equation}\label{circle}
		x^\mu(\tau)=R\,\big(\cos\tau,\sin\tau,0,0\,\big)
\end{equation}
with $\tau\in[\,0,2\pi\,]$. 

We compute $\vev{W(C)}$ in perturbation theory using the diagrammatic difference \eqref{difference}. This perturbative computation has been already performed up order 
$g^6$ in \cite{Andree:2010na}, where the term proportional to $\zeta(3)$ coming from the 
matrix model was reproduced using Feynman diagrams for the conformal SQCD case, namely
for theory $\mathbf{A}$ of table \ref{tab:scft}. 
Here we briefly review this result, generalizing it to a generic superconformal theory, and
extend it to an order higher, reconstructing the full $\zeta(5)$-coefficient at order $g^8$. 

\begin{figure}[htb]
	\begin{center}
		\includegraphics[scale=0.35]{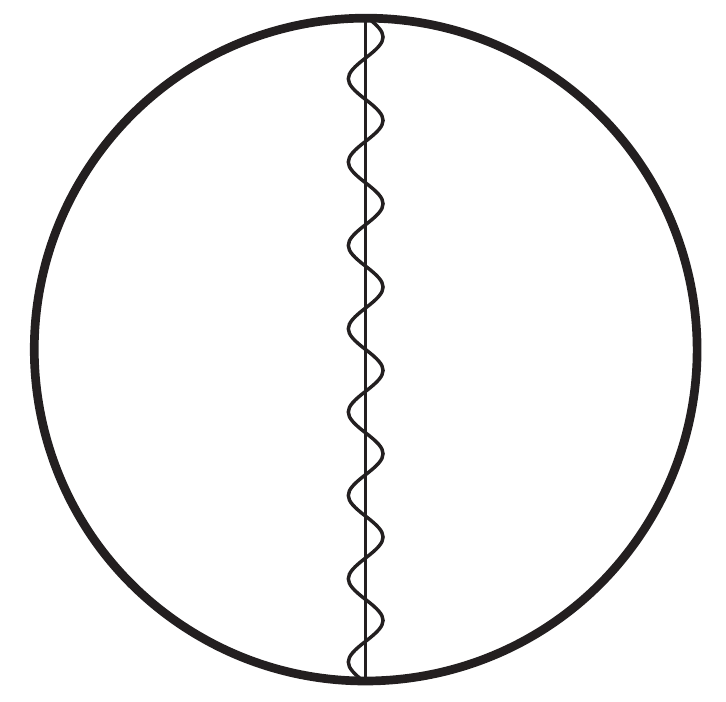}
	\end{center}
	\caption{The graphical representation of the $g^2$-correction to $\big\langle
	W(C)\big\rangle$.}
	\label{fig:WLtree}
\end{figure}

Let us recall first some remarkable properties of this observable that simplify 
the perturbative analysis. The tree-level propagators of the gauge field and of the adjoint 
scalar in configuration space are
\begin{equation}
	\label{propsAphi}
		\big\langle \bar\varphi^a (x_1)\, \varphi^b (x_2)\big\rangle_{\mathrm{tree}}
		= \frac{\delta^{ab}}{4\pi^2 x_{12}^2}~,~~~ \big\langle
		A_{\mu}^{a} (x_1)\, A_{\nu}^{b} (x_2)\big\rangle_{\mathrm{tree}}
		= \frac{\delta^{ab} \delta_{\mu\nu}}{4\pi^2 x_{12}^2}~.
\end{equation}
They are identical, a part from the different space-time indices. 
We will denote the sum of a scalar and a gluon propagator with the straight/wiggly line already
introduced in figure~\ref{tracesdraw}.
Expanding \eqref{WLdef} at order $g^2$, one gets an integral over $C$ of the sum of 
the tree-level propagators of the gluon and of the scalar fields between the points 
$x(\tau_1)$ and $x(\tau_2)$. This contribution is represented in figure \ref{fig:WLtree}.

Using (\ref{propsAphi}), one finds  
\begin{equation}
	\label{WLg2}
		\big\langle W(C)\big\rangle = 1+\frac{g^2 (N^2-1)}{4N} \oint \frac{d\tau_1 d\tau_2}{4\pi^2} \frac{R^2 -\dot{x}(\tau_1)\cdot \dot{x}(\tau_1)}{|x(\tau_1)-x(\tau_2)|^2} + \cO (g^4)~.
\end{equation}
Exploiting the parametrization \eqref{circle}, one can easily show that the integrand 
is $\tau$-independent; indeed
\begin{equation}
	\label{propwl}
		\frac{R^2 -\dot{x}(\tau_1)\cdot \dot{x}(\tau_1)}{4\pi^2 |x(\tau_1)-x(\tau_2)|^2}
		= \frac 12~.	
\end{equation}
Inserting this (\ref{WLg2}), one finally obtains
\begin{equation}
	\label{WLg2res}
		\big\langle W(C)\big\rangle = 1+\frac{g^2 (N^2-1)}{8N}+ \cO (g^4)~.
\end{equation}
At this order, this calculation is of course the same in $\cN=2$ and $\cN=4$, and 
thus there is no $g^2$- contribution to the vacuum expectation value of $W(C)$
in the difference theory. Also at order $g^4$ there are no
contributions in the difference, since the only possible sources for such contributions are
the one-loop corrections to the scalar and gluon propagators, which however vanish for
superconformal theories in the Fermi-Feynman gauge \cite{Grisaru:1979wc,Kovacs:1999rd}, 
see figure \ref{fig:1loop}. One begins to see a difference between the $\cN=4$ and the conformal
$\cN=2$ results at order $g^6$. Indeed, as we have seen in the previous section, in a generic
conformal $\cN=2$ theory the propagator of the adjoint scalar 
gets corrected by loop effect starting at order $g^4$.
Due to supersymmetry, also the gluon propagator in the Fermi-Feynman gauge
gets corrected in the same way and thus (\ref{propsAphi}) can be replaced by
\begin{equation}
	\label{propsAphicorr}
		\big\langle \bar\varphi^a (x_1) \varphi^b (x_2) \big\rangle
		= \frac{\delta^{ab}}{4\pi^2 x_{12}^2}\,
		 \big(1+ \Pi\big)~,~~~ 
		\big\langle A_{\mu}^{a} (x_1) A_{\nu}^{b} (x_2)\big\rangle
		= \frac{\delta^{ab} \delta_{\mu\nu}}{4\pi^2 x_{12}^2}\,\big(1+\Pi\big)~,
\end{equation}
where $\Pi$ is the quantity introduced in (\ref{propPi}).

\begin{figure}[htb]
	\begin{center}
		\includegraphics[scale=0.38]{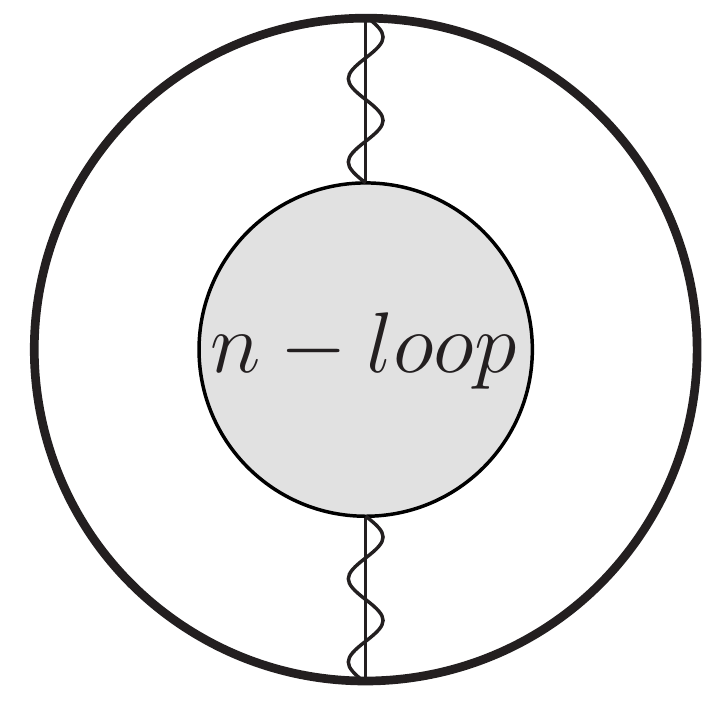}
	\end{center}
	\caption{The graphical representation of the contribution to $\big\langle
	W(C)\big\rangle$ arising from the $n$-loop correction of the gluon and scalar propagators.}
	\label{fig:WLnloops}
\end{figure}

Exploiting this fact, and repeating the same steps as before, we can easily compute the contribution
to the vacuum expectation value of $W(C)$ corresponding to the diagram in figure~\ref{fig:WLnloops}, which yields a term proportional to $g^{2n+2}\,\zeta(2n-1)$.

Using (\ref{Pig3}), for $n=2$ this calculation yields
\begin{align}
	\label{WC3}
		\frac{g^2 (N^2-1)}{8N} \left(\frac{g^2}{8\pi^2}\right)^2 \zeta(3)\, \cC^{\prime}_4~,
\end{align}
while for $n=3$ it gives
\begin{align}
	\label{WC5}
		-\frac{g^2 (N^2-1)}{8N} \left(\frac{g^2}{8\pi^2}\right)^3 \zeta(5)\,\cC^{\prime}_6~.
\end{align}
Comparing with (\ref{chi3expg}) and (\ref{chi5expg}), we find a perfect agreement with
the matrix model predictions for the lowest order terms in the $g$-expansion of
$\cX_3$ and $\cX_5$. The precise match with the matrix model results
suggests that in the vacuum expectation value of $W(C)$ the terms proportional 
to a given Riemann $\zeta$-value with the lowest power of $g$, namely the terms 
proportional to $g^{2n+2}\,\zeta(2n-1)$, are \emph{entirely} captured by the $n$-th loop
correction of a single gluon or scalar propagator inserted in the Wilson loop. Moreover,
the agreement with the matrix model also suggests that all diagrams contributing to $\big\langle
W(C)\big\rangle$ have an \emph{even} number of legs attached to the Wilson loop.
We shall now check that this is indeed true, at the first relevant orders. 

\subsubsection*{Absence of other contributions} 
\begin{figure}[htb]
	\begin{center}
		\includegraphics[scale=0.38]{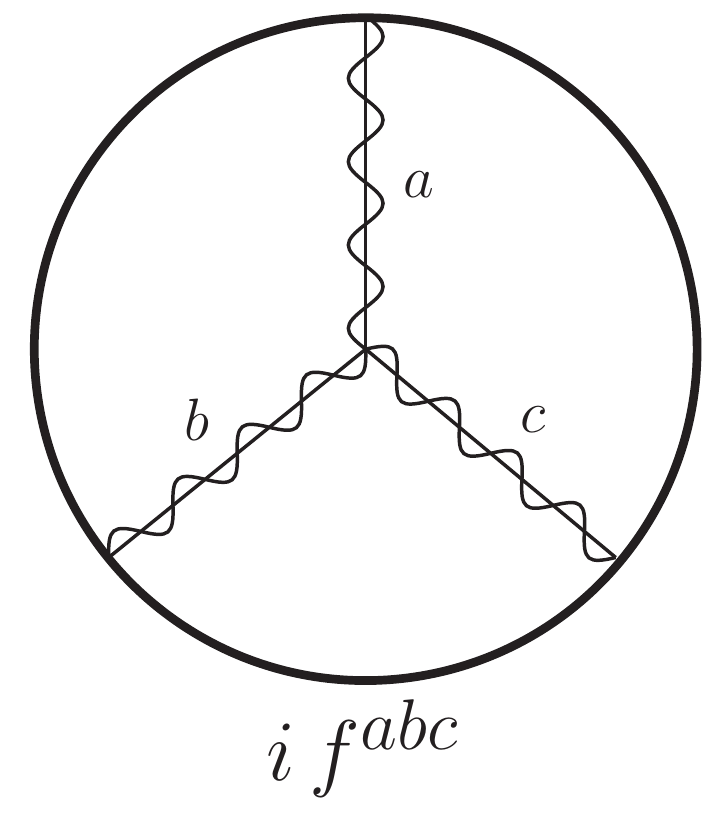}
	\end{center}
	\caption{The vertex correction to $\big\langle W(C)\big\rangle$ 
	in the $\cN=4$ theory at order $g^4$.}
	\label{fig:WLvertex0}
\end{figure}

Let us consider diagrams with three insertions on the Wilson loop contour. 
In the $\cN=4$ theory there is such a diagram already at order $g^4$ which is shown in figure~\ref{fig:WLvertex0}. Here the internal vertex can be with three gluons or with two scalars and
one gluon. In both cases it carries a color factor proportional to $f_{abc}$.
This contribution has been proven to vanish long ago \cite{Erickson:2000af,Bassetto:2008yf}. The 
cancellation is justified by symmetry properties of the (finite) integral over the insertion points 
along the circular loop\footnote{We thank L. Griguolo for a discussion on this point.}.

\begin{figure}[htb]
	\begin{center}
		\includegraphics[scale=0.38]{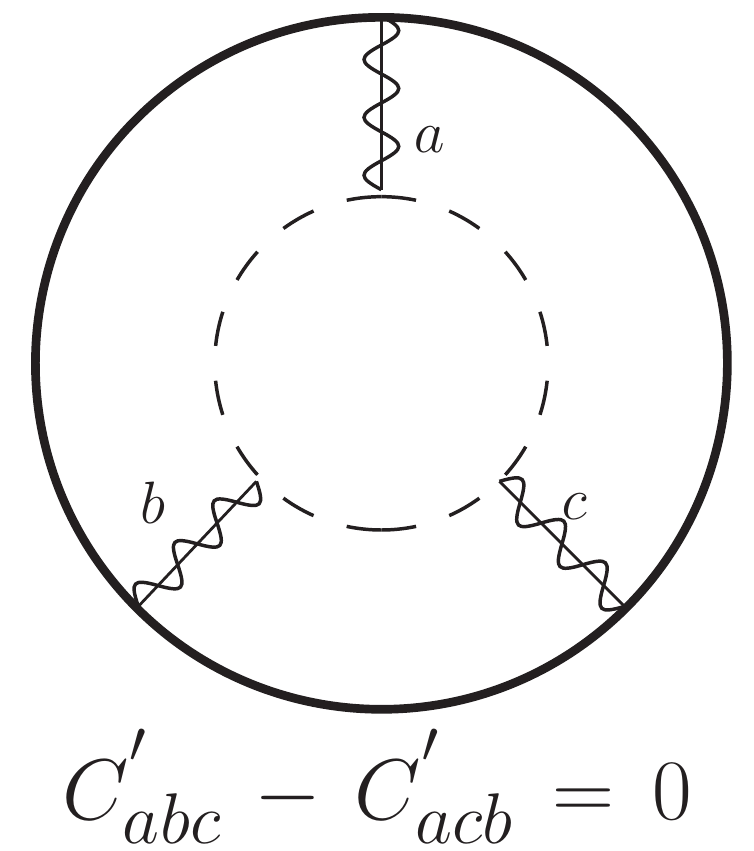}
	\end{center}
	\caption{The one-loop vertex corrections to $\big\langle W(C)\big\rangle$ at $g^6$ 
	order in the difference theory is vanishing.}
	\label{fig:WLvertex1}
\end{figure}

In the difference theory, instead, the first three-leg diagram appears at order $g^6$ 
and is depicted in figure~\ref{fig:WLvertex1}.
This contribution, however, has a vanishing color factor (see also \cite{Gomez:2018usu}). 
This is due to the different roles of the $Q$ or $H$ superfields, transforming in the 
representation $\cR$, and of the $\widetilde Q$ or $\widetilde H$ ones, transforming 
in the representation $\bar\cR$.
This implies that the color factor is
\begin{align}
\Tr_{\cR}^\prime T^aT^bT^c +\Tr_{\bar\cR}^\prime T^aT^bT^c = C^\prime_{abc} - C^\prime_{acb}~,
\label{Cabc}
\end{align}
which is automatically zero due to the complete symmetry of $C^\prime_{(3)}$ as shown in (\ref{C3confsym}).

\begin{figure}[h]
\begin{center}
\includegraphics[scale=0.38]{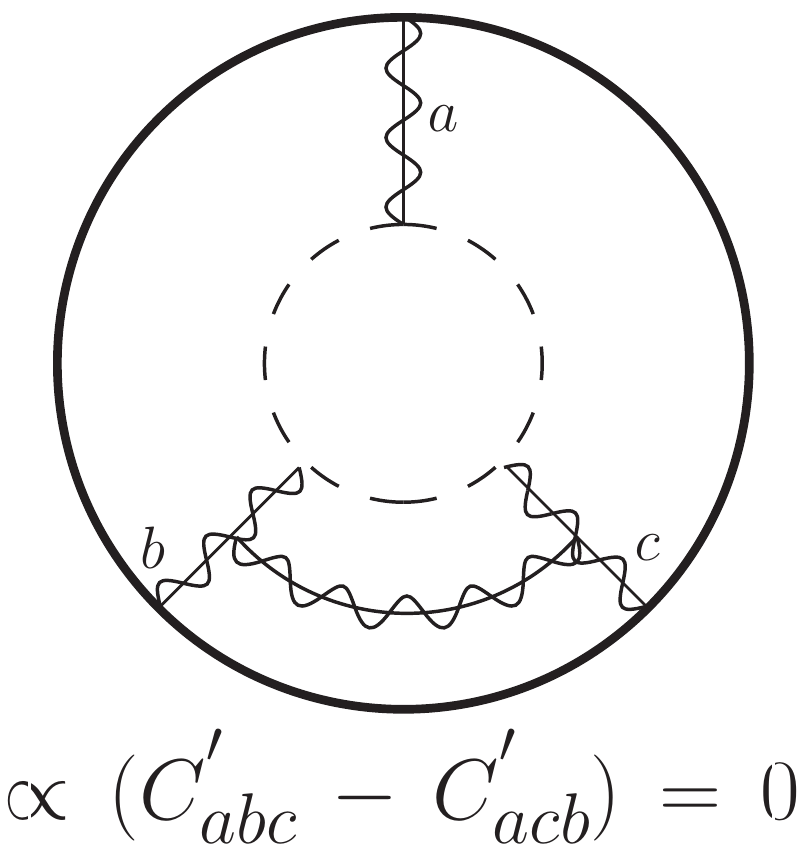}
\hspace{0.5cm}
\includegraphics[scale=0.38]{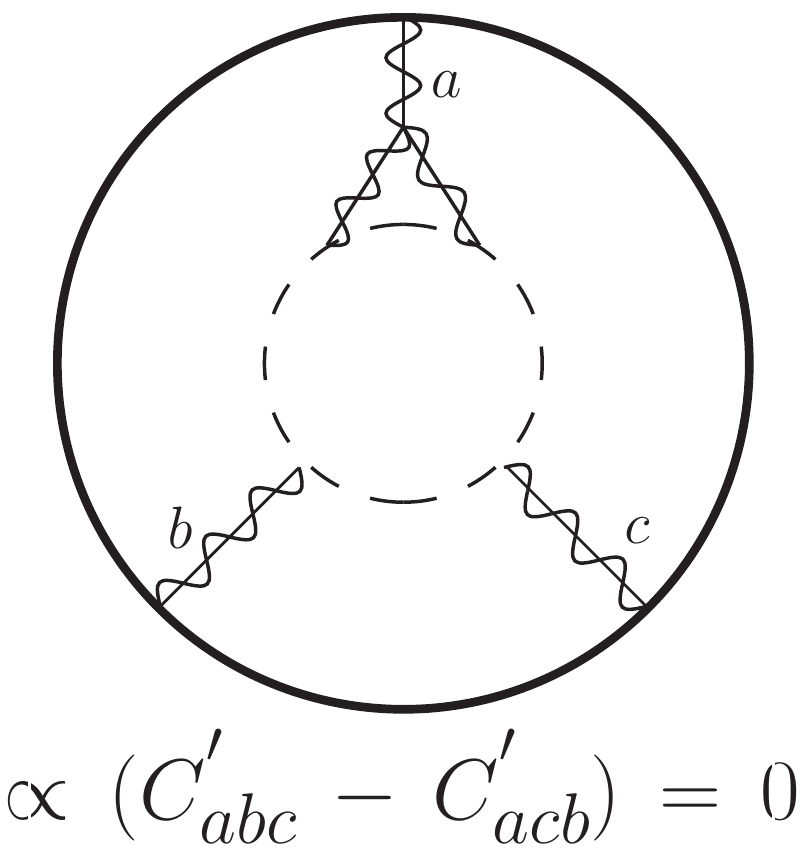}
\hspace{0.5cm}
\includegraphics[scale=0.38]{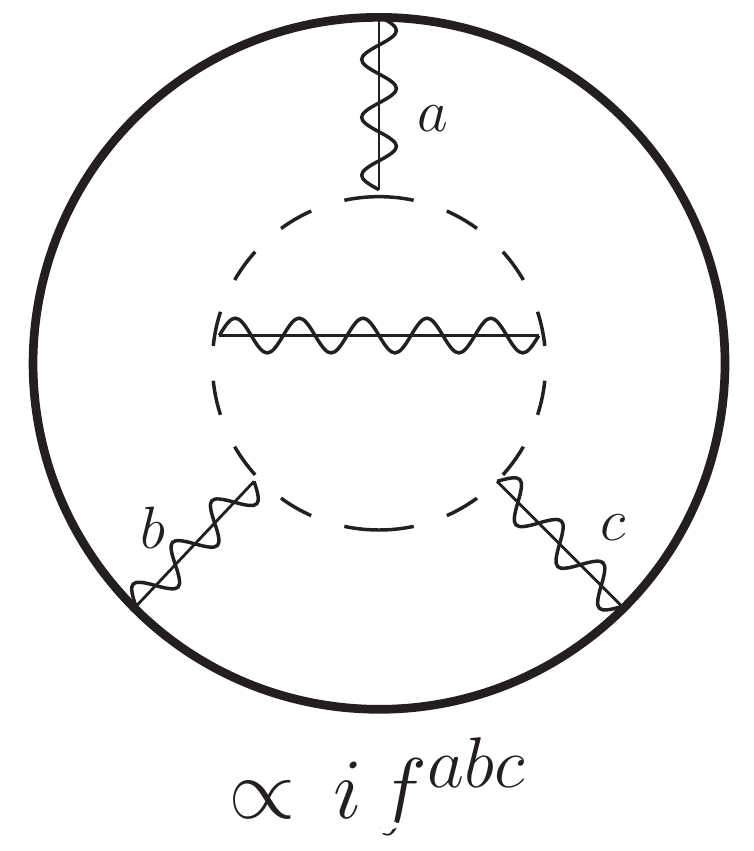}
\hspace{0.5cm}
\caption{Possible two-loop vertex corrections contributing to 
$\big\langle W(C)\big\rangle$ at order $g^8$ together with their color factors.}
\label{fig:WLvertex2}
\end{center}
\end{figure}

At order $g^8$ there are several possible three-leg diagrams. Again, if we classify them in terms 
of their color factor, we can distinguish three classes, 
represented in figure \ref{fig:WLvertex2}. The first two have again a color factor proportional to the combination
(\ref{Cabc}) which vanishes, while the last type has a color factor proportional to $f_{abc}$. 
We have not performed a detailed calculation of this class of diagrams, but it is natural 
to expect that they cancel by a mechanism analogous to the one at work 
in the $g^4$ diagrams of the $\cN=4$ theory represented in figure \ref{fig:WLvertex0}, since they have the same color structure and symmetry properties.
This concludes our analysis on the check of the agreement between the matrix model prediction and
the field theory results of $\big\langle W(C)\big\rangle$ at order $g^8$.

\section{Summary and conclusions}
\label{secn:concl}
We have considered the perturbative part of the matrix model, derived from localization, which 
describes a \emph{generic} conformal $\cN=2$ SYM theory with group SU($N$). 
We have described the color 
structure of the interactions in this matrix model in terms of the difference between 
the $\cN=2$ theory and the $\cN=4$ theory corresponding to a free Gaussian model. In this
set-up we have computed the matrix model counterpart of the propagator 
of the scalar field in the $\cN=2$ vector multiplet and of the vacuum expectation value 
of a 1/2 BPS circular Wilson loop, organizing the resulting expressions
according to their Riemann zeta-value structures. Having at our disposal generic expressions, we could focus on a class of conformal theories containing fundamental, symmetric and anti-symmetric matter multiplets and we singled out two classes of theories for which the Wilson loop 
in the large-$N$ limit approaches the $\cN=4$ value. 
Then, we have performed an explicit check of these matrix model results
against their field-theoretic perturbative evaluation by means of superdiagrams in the $\cN=1$ superfield formalism. We have done this up to order $g^6$ - three loops - for the propagator, 
which has allowed us to determine the four-loop terms of order $g^8$ proportional to 
$\zeta(5)$ in the Wilson loop vacuum expectation value. 
This is in itself a significant progress with respect to the checks previously available, 
namely those of order $g^6 \,\zeta(3)$ for the Wilson loop in the case of the conformal 
SQCD only. We think however that the relevance of this computation stays also in the fact 
that we have shown how the perturbative computations are made more efficient and 
tractable by organizing them in the way suggested by the matrix model, namely by focusing
on the color factors corresponding to traces of adjoint generators inserted on a loop of 
hypermultiplets. We think that such an organization is potentially useful also for 
different theories, for example non conformal ones or, maybe, even theories with less supersymmetry
for which localization techniques are not presently available. 
Beside the circular Wilson loop, it would be interesting also to study other observables in the various families of $\cN=2$ superconformal theories described in this paper and analyze their behavior in the large-$N$ limit to gain some insight on their holographic dual counterparts. 

\vskip 1.5cm
\noindent {\large {\bf Acknowledgments}}
\vskip 0.2cm
We thank A. Armato, L. Bianchi, M. Frau, P. Gregori, L. Griguolo, R.~R. John and J. Russo for 
many useful discussions and suggestions.

\noindent
The work of M.B.~and A.L. is partially supported by the MIUR PRIN Contract 
2015 MP2CX4 ``Non-perturbative Aspects Of Gauge Theories And Strings''.
The work of A.L. is partially supported also by ``Fondi Ricerca Locale dell'Universit\`a 
del Piemonte Orientale".
We thank the Galileo Galilei Institute for Theoretical Physics and INFN for hospitality and 
partial support during the workshop "String theory from a worldsheet perspective" where 
part of this work has been done.
\vskip 1cm
\begin{appendix}

\section{Useful group theory formul\ae~for SU$(N)$}
\label{app:group}
We denote by $T_a$, with $a=1,\ldots, N^2-1$, a set of Hermitean generators satisfying 
the $\mathfrak{su}(N)$ Lie algebra
\begin{align}
	\label{sunalgebrar}
		\big[\,T_a\,,\,T_b\,\big] = \ii f_{abc}\, T_c~.
\end{align} 
We indicate by $t_a$ the representative of $T_a$ in the fundamental representation; they are Hermitean, traceless $N\times N$ matrices that we normalize by setting 
\begin{equation}
	\label{norm}
		\tr\,t_a t_b = \frac{1}{2}\,\delta_{ab}~.
\end{equation}
In the conjugate fundamental representation the generators are
\begin{align}
	\label{genconj}
		\bar t_a = - t_a^{\,T}~.
\end{align}
The generators $t^a$ are such that the following fusion/fission identities hold
\begin{align}
	\label{fussion}
		\tr\left(t_a M_1 t_a M_2\right) & = \frac{1}{2}\,\tr\, M_1\, \tr\,M_2
		-\frac{1}{2N}\,\tr\left(M_1 M_2\right)~,\\
		\tr\left(t_a M_1\right) ~\tr\left(t_a M_2\right) & = \frac{1}{2}\,\tr\left(M_1 M_2\right) -\frac{1}{2N}\,\tr\,M_1 \,\tr\,M_2~,
\end{align}
for arbitrary $(N\times N)$ matrices $M_1$ and $M_2$.

In the enveloping matrix algebra, we have
\begin{align}
	\label{tatb}
		t_a\,t_b =\frac{1}{2}\,\left[\frac{1}{N}\,\delta_{ab}\,\mathbf{1}
		+\left(d_{abc}+\ii\,f_{abc}\right)\,t^c\right]~,
\end{align}
where $d^{abc}$ is the totally symmetric $d$-symbol of su$(N)$.
Using (\ref{norm}) and (\ref{tatb}), we obtain
\begin{align}
	\label{key}
		\tr\left(\big\{\,t_a\,,\,t_b\big\}\,t_c\right) = \frac{1}{2}\,d_{abc}~,\quad
		\tr\left(\big[\,t_a\,,\,t_b\big]\,t_c\right)= \frac{\ii}{2}\,f_{abc}~,
\end{align}
from which it follows that $d_{aac}=0$. 
We can write the $d$- and $f$-symbols as $(N^2-1)\times (N^2-1)$ matrices
\begin{equation}
\ii f^{abc} = (F^a)^{bc} , \hspace{0.5cm} d^{abc} = (D^a)^{bc}
\end{equation}
and derive the following useful identities:
\begin{equation}\label{adjtraces}
\begin{split}
&\Tr F^a=\Tr D^a=\Tr F^a D^b=0~,\\
&\Tr F^a F^b=N \delta^{ab}~,
\hspace{0.6cm}\Tr D^a D^b=\frac{N^2-4}{N} \delta^{ab}~,\\
&\Tr F^a F^b F^c=\frac{\ii N}{2} f^{abc}~,
\hspace{0.6cm}\Tr D^a F^b F^c=\frac{N}{2} d^{abc}~,\\
&\Tr F^a F^b F^c D^d=\frac{\ii N}{4}(d^{ade}f^{bce}-f^{ade}d^{bce})
\end{split}
\end{equation}
where $\Tr$ denotes the trace in the adjoint representation.

\subsection*{Traces of generators}
In any representation $\cR$ we have
\begin{equation}
	\label{indexR}
		\Tr_\cR T_a T_b = i_\cR\, \delta_{ab}~,
\end{equation}
where $i_\cR$ is the index of $\cR$, and is fixed once the generators have been normalized in
the fundamental representation (see (\ref{norm})). 
The quadratic Casimir operator in the representation $\cR$ is defined by 
\begin{equation}
	\label{casimirR}
		T_a \,T_a = c_\cR\, \mathbf{1}~.
\end{equation}  
By tracing this equation and comparing to (\ref{indexR}), we have
\begin{equation}
	\label{cRtoiR}
		c_\cR = \frac{N^2-1}{d_\cR} \, i_\cR~,
\end{equation}   
with $d_\cR$ being the dimension of the representation $\cR$. 

The traces of products of generators  define a set of cyclic tensors 
\begin{align}
	\label{defCapp}
		C_{a_1\ldots a_n} = \Tr_{\cR} T_{a_1}\ldots T_{a_n}
\end{align}
whose contractions are higher order invariants characterizing the representation $\cR$. 
Let us note that we can switch the order of any two consecutive indices using the Lie algebra relation (\ref{sunalgebrar}); indeed:
\begin{align}
	\label{switchC}
		C_{\ldots a b \ldots} = C_{\ldots b a \ldots} + \ii\, f_{abc}\, C_{\ldots c \ldots}~.
\end{align}

In our computations we encounter the particular combination of traces introduced in 
(\ref{defC}), namely
\begin{align}
	\label{defCagain}
		C^\prime_{a_1\ldots a_n} \,=\, \trp T_{a_1}\ldots T_{a_n} \,=\, 
		\Tr_{\cR} T_{a_1}\ldots T_{a_n}- 
		\Tr_{\mathrm{adj}} \,T_{a_1}\ldots T_{a_n}~.
\end{align}
These are of course also cyclic, and the relation (\ref{switchC}) applies to them as well.

If $\cR$ is the representation in which the matter hypermultiplets of a superconformal theory transform, one can prove that
\begin{equation}
C^\prime_{ab}= 0~,
\end{equation}
since $C^\prime_{ab}$ is proportional to the one-loop $\beta$-function coefficient. Therefore,
using this property and the relation (\ref{switchC}) one can easily show that for conformal theories
\begin{align}
\label{C3confsym}
C^\prime_{abc} = C^\prime_{acb} + \ii f_{abe} C^\prime_{ec} = C^\prime_{acb}
\end{align}
which, together with cyclicity, implies that the tensor $C^\prime_{abc}$ is totally symmetric. Thus, 
it is proportional to $d_{abc}$. Finally, one can prove that
\begin{equation}
C^\prime_{abcc} = C^\prime_{(abcc)}~.
\end{equation}
Indeed, if we exchange the two free indices we have
\begin{align}
	\label{C4sw1}
		C^\prime_{abcc} = C^\prime_{bacc} + \ii f_{abe} C^\prime_{ecc} = C^\prime_{bacc}~,
\end{align}
where the last step follows from the fact that $C^\prime_{ecc}=0$ since $d_{ecc} =0$. 
If instead we switch the position of a free and a contracted index, we have
\begin{align}
	\label{C4sw12}
		C^\prime_{abcc} = C^\prime_{acbc} + \ii f_{bce} C^\prime_{aec} = C^\prime_{acbc}~,
\end{align}
where have used the fact 
that $C^\prime_{aec}$, being symmetric, vanishes when contracted with $f_{bce}$. 

\subsection*{Some particular representations}
The generators in the direct product representation $\cR =  \Yfund \otimes \Yfund$ are given by
\begin{equation}
	\label{Tafunfun}
		T_a = t_a \otimes \mathbf{1} \oplus \mathbf{1} \otimes t_a~.
\end{equation} 
This representation is reducible into its symmetric and anti-symmetric parts:
\begin{equation}
	\label{funfundec}
		\Yfund \otimes \Yfund = \Ysymm \oplus \Yasymm~.	
\end{equation}
In the symmetric representation one has 
\begin{equation}
	\label{trsymm}
		\Tr_{\Ysymm}\big( X\otimes Y \big)=
		\frac{1}{2}\Big(\tr\,X\,\,\tr\,Y 
		+\tr\left(X\, Y\right)\Big)~,
\end{equation}
while in the anti-symmetric representation one has
\begin{equation}
	\label{trasymm}
		\Tr_{\Yasymm} \big(X\otimes Y\big) =
		\frac{1}{2}\Big(\tr\,X\,\,\tr\,Y 
		-\Tr\left(X\, Y\right)\Big)~.
\end{equation}

The adjoint representation is contained in the direct product of a fundamental and an anti-fundamental:
\begin{equation}
	\label{funbfun}
		\Yfund \otimes \overline{\Yfund} = \mathrm{singlet} \oplus \adj~.
\end{equation}
The generators in the adjoint can thus represented simply\,%
\footnote{They should be thought of as acting on the $N^2-1$-dimensional subspace orthogonal to the invariant vector $\sum_i e_i\otimes \bar e_i$, where $e_i$ and $\bar e_i$, for $i=1,\ldots N$, are basis vectors in the carrier spaces of the fundamental and anti-fundamental representations. This however makes no difference for the computation of the traces we are interested in.} by 
\begin{equation}
	\label{genadj}
		T_a = t_a \otimes \mathbf{1} + \mathbf{1} \otimes \overline{t}_a~.
\end{equation}
Using these relations it is easy to obtain the well-known results collected in table~\ref{tab:saad}. 
\begin{table}[ht]
	\begin{center}
		{\small
			\begin{tabular}{c|c|c}
				\hline
				\hline
				$\cR$ \phantom{\bigg|}& $d_\cR$ & $i_\cR$ \\
				\hline
				$\phantom{\bigg|}\Yfund\phantom{\bigg|}$ & $~~N~~$ & $~~\frac{1}{2}~~$\\
				$\phantom{\bigg|}\Ysymm\phantom{\bigg|}$ & $~~\frac{N(N+1)}{2}~~$ & $~~
				\frac{N+2}{2}~~$\\
				$\phantom{\bigg|}\Yasymm\phantom{\bigg|}$ & $~~\frac{N(N-1)}{2}~~$ & $~~
				\frac{N-2}{2}~~$\\
				$\phantom{\bigg|}\adj\phantom{\bigg|}$ & $~~N^2-1~~$ & $~~N~~$ \\
				\hline
				\hline
			\end{tabular}
		}
	\end{center}
	\caption{Dimensions and indices of the fundamental, symmetric, anti-symmetric and adjoint representations of SU($N$).}
	\label{tab:saad}
\end{table}

If we consider a representation $\cR$ made of $N_F$ fundamental, $N_S$ symmetric and $N_A$ anti-symmetric representations, namely
\begin{equation}
\cR = N_F\, \Yfund \oplus N_S\, \Ysymm \oplus N_A\, \Yasymm
\end{equation}
as in (\ref{RNFNASNA}), we immediately see that 
\begin{align}
	\label{RFSAb0}
		\trp T^a T^b =\big(N_F + N_S(N+2) + N_A(N-2) - 2N\big)\,\tr t^a t^b 
		= -{\beta_0}\,\tr t^a t^b 
	\end{align}
where 
$\beta_0$ is the one-loop $\beta$-function coefficient of the $\mathcal{N}=2$ SYM theory
(see (\ref{b0is})).

With a bit more work, but in a straightforward manner, one can compute traces of more 
generators. In particular, one can evaluate
\begin{align}
	\label{colfact1}
		\trp a^n & = 
		N_F\, \tr a^n 
		+ N_S\, \Tr_{\Ysymm}\big(a\otimes \bone + \bone \otimes a\big)^n
		+ N_A\, \Tr_{\Yasymm}\big(a\otimes \bone + \bone \otimes a\big)^n\nonumber\\[1mm]
		& ~~~- \Tr_\adj \big(a\otimes \bone + \bone \otimes (-a^{\,T})\big)^n~,
\end{align}
with the result
\begin{align}
	\label{colfact2}
		\trp a^n & = \Big[(N_F + 2^{n-1} \big(N_S - N_A\big) 
		+ N \big(N_S + N_A -(1 + (-1)^n)\big) \Big]\, \tr a^n
	\nonumber\\
	& ~~~+ \sum_{p=1}^{n-1} \binom{n}{p} \left(\frac{N_S + N_A}{2} - (-1)^{n-p}\right)\tr a^p\, \tr a^{n-p}~.		
\end{align}
In particular, when $n=2k$, this expression can be rewritten as in (\ref{S2n}) of the main text.
 
\subsection*{Traces in a generic representation} 
\label{secn:frob}
A representation $\cR$ is associated to a Young diagram $Y_R$; let $r$ be the number of boxes
in the tableau. Traces in the representation $\cR$ can be evaluated in terms of traces in the fundamental representation using the Frobenius theorem. 
For any group element $U$ in SU$(N)$, this theorem theorem states that 
\begin{equation}
	\label{frob}
		\Tr_{\cR} U = \sum_{M} \frac{1}{|M|}\, \chi^R(M)\,(\tr U)^{m_1}\, (\tr U^2)^{m_2}\,
		\ldots (\tr U^r)^{m_r}~.
\end{equation}  
We denote by $M$ a conjugacy class\,%
\footnote{$M$ is associated to a Young diagram with $r$ boxes, containing $m_j$ columns of length $j$.} of $S_r$ containing permutations made of $m_j$ cycles of length $j$, with $j=1,\ldots r$; the number of elements in the class is $r!/|M|$, with 
\begin{equation}
	\label{orderclassM}
		|M| = \prod_{_j=1}^{r} m_j!\, j^{m_j}~. 
\end{equation}  
With $\chi^R(M)$ we denote the character of the conjugacy class $M$ in the representation $R$ of the group $S_r$ associated to the tableau $Y_R$.
If we write $U = \rme^a$, with $a\in \mathfrak{su}(N)$, equation (\ref{frob}) reads
\begin{equation}
	\label{frob2}
		\Tr_{\cR} \rme^a = \sum_{M} \frac{1}{|M|} \,\chi^R(M)\,(\tr \rme^a)^{m_1}\, (\tr \rme^{2a})^{m_2}\,
		\ldots (\tr \rme^{ra})^{m_r}~,
\end{equation}  
and expanding it in powers of $a$, one can obtain the expression of all traces of the form
$\Tr_{R} a^k$ in terms of products of traces of powers of $a$ in the fundamental representation,
generalizing what we have seen before for the symmetric, anti-symmetric and adjoint representations.

\section{Spinors and Grassmann variables}
\label{app:grassmann}
\subsection*{Spinor notations}
We denote by $\psi$ a chiral spinor of components $\psi_\alpha$ with $\alpha=1,2$, 
and by $\bar \psi$ an anti-chiral one of 
components $\bar{\psi}^{\dot{\alpha}}$, with $\dot{\alpha=1,2}$.
The spinor indices are raised and lowered with the following rules:
\begin{align}
	\label{raislow}
		\psi^\alpha = \epsilon^{\alpha\beta}\,\psi_\beta~,~~~
		\psi_\alpha = \epsilon_{\alpha\beta}\,\psi^\beta~,~~~
		\bar\psi^{\dot{\alpha}} =\epsilon^{\dot{\alpha}\dot{\beta}}\,\bar\psi_{\dot{\beta}}~,~~~
		\bar\psi_{\dot{\alpha}} =\epsilon_{\dot{\alpha}\dot{\beta}}\,\bar\psi^{\dot{\beta}}~,
\end{align}
where
\begin{equation}
	\label{epsilons}
		\epsilon^{12} = \epsilon^{\dot{1}\dot{2}} =\epsilon_{21} = \epsilon_{\dot{2}\dot{1}} = 1~.
\end{equation}
We contract indices according to
\begin{align}
	\label{contractions}
		(\psi\chi) &\equiv \psi^\alpha\,\chi_\alpha =
		\epsilon^{\alpha\beta}\,\psi_\beta\,\chi_\alpha 
		= \psi^\alpha\,\chi^\beta\,\epsilon_{\alpha\beta}~,\\[1mm]
		(\bar\psi\bar\chi) &\equiv \bar\psi_{\dot{\alpha}}\,\bar\chi^{\dot{\alpha}} 
		=\epsilon_{\dot{\alpha}\dot{\beta}}\,\bar\psi^{\dot{\beta}}\,\bar\chi^{\dot{\alpha}}
		=\bar\psi_{\dot{\alpha}}\,\bar\chi_{\dot{\beta}}\,\epsilon^{\dot{\alpha}\dot{\beta}}~.
\end{align}
For the ``square'' of spinors, we use the notation
\begin{align}
	\label{t2tb2}
		\psi^2 \equiv (\psi\psi)~,~~~
		\bar{\psi}^2 \equiv (\bar{\psi}\bar{\psi})~.
\end{align}
{From} the previous relations, it is straightforward to obtain the Fierz identities
\begin{align}
	\label{2tist2}
		\psi^\alpha\psi^\beta = - \frac 12\, \epsilon^{\alpha\beta}\,\psi^2~,~~~
		\bar{\psi}^{\dot\alpha} \bar{\psi}^{\dot\beta} 
		= +\frac 12 \,\epsilon^{\dot\alpha\dot\beta}\,\bar{\psi}^{\,2}~.
\end{align}		
	
\subsection*{Clifford algebra}	
We realize the Euclidean Clifford algebra
\begin{equation}
	\label{cliff4}
		\sigma_\mu\bar\sigma_\nu + \sigma_\nu\bar\sigma_\mu =
		-2\,\delta_{\mu\nu}\,\mathbf{1}
\end{equation}
by means of the matrices $(\sigma^\mu)_{\alpha\dot\beta}$ and
$(\bar\sigma^{\mu})^{\dot\alpha\beta}$ that can be taken to be
\begin{equation}
	\label{sigmas}
		\sigma^\mu =
		(\vec\tau,-\ii\mathbf{1})~,\qquad
		\bar\sigma^\mu =
		-\sigma_\mu^\dagger = (-\vec\tau,-\ii\mathbf{1})~,
\end{equation}
where $\vec\tau$ are the ordinary Pauli matrices. They are such that
\begin{equation}
	\label{traspsigma}
		(\bar\sigma^{\mu})^{\dot\alpha\alpha}=\epsilon^{\alpha\beta}\,\epsilon^{\dot{\alpha}\dot{\beta}}(\sigma^\mu)_{\beta\dot\beta}~.
\end{equation}
With these matrices we can write the 4-vectors as bispinors:
\begin{equation}
	\label{bispinork}
		k_{\alpha\dot{\beta}} = k_\mu \, (\sigma^\mu)_{\alpha\dot{\beta}}~,~~~
		\bar k^{\alpha\dot\beta} = k^\mu \, (\bar\sigma_\mu)^{\dot\alpha\beta}~.		
\end{equation}
We will often use the notations $k$ and $\bar{k}$ to indicate the matrices $k_{\alpha\dot{\beta}}$ and  $\bar k^{\alpha\dot\beta}$
and form spinor bilinears of the type
\begin{equation}
	\label{tptb}
	\theta\, k\, \bar{\theta}= \theta^\alpha\, k_{\alpha\dot{\beta}} \,\bar{\theta}^{\dot{\beta}}~.
\end{equation}
The Clifford algebra, together with the property (\ref{traspsigma}),  allows to evaluate traces of $\sigma$ and $\bar\sigma$ matrices, which we can also write in terms of traces of matrices of the type (\ref{bispinork}). In our computations we will need the following traces:
\begin{align}
	\label{sigmatraces}
		\tr \big(k_1 \bar k_2\big)  = & - 2\, k_1\!\cdot\! k_2 ~,\nonumber\\[1mm]
		\tr \big(k_1 \bar k_2 k_3 \bar k_4\big) = &  
        +2 \,\Big[(k_1\!\cdot\!  k_2)\, (k_3\!\cdot\!  k_4) - (k_1\!\cdot\!  k_3)\,(k_2\!\cdot\!  k_4)
		+ (k_1\!\cdot\!  k_4)\, (k_2\!\cdot\!  k_3)\Big] + \ldots~,\nonumber\\[1mm]
		\tr \big(k_1 \bar k_2 k_3 \bar k_4 k_5 \bar k_6\big) = &  
		- 2\,k_1\!\cdot\!  k_2\, \Big[(k_3\!\cdot\!  k_4)\, (k_5\!\cdot\!  k_6) - (k_3\!\cdot\!  k_5)\, (k_4\!\cdot\!  k_6) + (k_3\!\cdot\!  k_6) \,(k_4\!\cdot\!  k_5)\Big]	\nonumber\\
		& +  2\, k_1\!\cdot\! k_3\, \Big[(k_2\!\cdot\!  k_4) \,(k_5\!\cdot\!  k_6) - (k_2\!\cdot\!  k_5)\, (k_4\!\cdot\!  k_6) + (k_2\!\cdot\!  k_6)\, (k_4\!\cdot\! k_5)\Big] \nonumber\\
		& -  2 \,k_1\!\cdot\! k_4\,\Big[(k_2\!\cdot\!  k_3) \,(k_5\!\cdot\!  k_6) - (k_3\!\cdot\! k_5)\, (k_3\!\cdot\!  k_6) + (k_2\!\cdot\!  k_6)\, (k_3\!\cdot\! k_5)\Big] \nonumber\\
		& +  2 \,k_1\!\cdot\! k_5\, \Big[(k_2\!\cdot\!  k_3)\, (k_4\!\cdot\!  k_6) - (k_3\!\cdot\!  k_4)\, (k_3\!\cdot\!  k_6) + (k_2\!\cdot\! k_6) \,(k_3\!\cdot\! k_4)\Big] \nonumber\\
		& -  2\, k_1\!\cdot\! k_6\,\Big[(k_2\!\cdot\! k_3)\, (k_4\!\cdot\!  k_5) - (k_3\!\cdot\! k_4)\, (k_3\!\cdot\!  k_5) + (k_2\!\cdot\! k_5) \,(k_3\!\cdot\!  k_4)\Big] \nonumber\\
		& + \ldots~,
\end{align}
where the ellipses in the second and last line stand for parity-odd terms containing contractions with a space-time $\varepsilon$-tensor that do not enter in our computations.

\subsection*{Grassmann integration formul\ae}
The basic integration formul\ae~ for Grassmann variables are
\begin{align}
	\label{intt2}
		\int d^2\theta\, \theta^2 = 1~,~~~
		\int d^2\bar{\theta}\, \bar{\theta}^2 = 1~.
\end{align} 
These imply that the $\theta^2$ and $\bar{\theta}^2$ act as fermionic $\delta$-functions; 
more in general, writing $\theta_{ij} = \theta_i - \theta_j$, we have
\begin{align}
	\label{deltat}
		\theta_{ij}^2 = \delta^2(\theta_{ij})~,~~~
		\bar\theta_{ij}^{\,2} = \delta^2(\bar\theta_{ij})~;
\end{align} 
we also use the notation
\begin{align}
	\label{delta4t}
		\theta_{ij}^2\, \bar\theta_{ij}^{\,2} = \delta^4(\theta_{ij})~.
\end{align}

\subsection*{Spinor derivatives} 
Writing $\partial_\alpha\equiv 
{\displaystyle{\frac{\partial}{\partial\theta^\alpha}}}$ and $\bar\partial_{\dot{\alpha}}\equiv{\displaystyle{\frac{\partial}{\partial\bar\theta^{\dot{\alpha}}}}}$, we have
\begin{equation}
\begin{aligned}
	\label{spdert2}
		&\partial_\alpha\,\theta^2 = 2\,\theta_\alpha~,~~~
		\partial\partial\,\theta^2 = - 4\\[1mm]
		&\bar\partial_{\dot{\alpha}}\,\bar\theta^{\,2} = -2\,\bar\theta_{\dot{\alpha}}~,~~~
		\bar\partial\bar\partial\,\bar\theta^{\,2}= - 4~.
\end{aligned}
\end{equation}
The covariant spinor derivatives are defined as 
\begin{align}
	\label{covspinder}
		D_\alpha=\partial_\alpha+\ii\,(\sigma^\mu)_{\alpha\,\dot{\alpha}}\,
		\bar\theta^{\dot{\alpha}}\,\partial_\mu
		\quad\mbox{and}\quad
		\bar D_{\dot{\alpha}}=-\bar\partial_{\dot{\alpha}}-
		\ii\,\theta^\alpha\,(\sigma^\mu)_{\alpha\,\dot{\alpha}}\,\partial_\mu~.
\end{align}
In momentum space, they become
\begin{align}
	\label{covspink}
		D_\alpha=\partial_\alpha-(k\,\bar\theta)_{\alpha}
		\quad\mbox{and}\quad
		\bar D_{\dot{\alpha}}=-\bar\partial_{\dot{\alpha}}+
		(\theta\,k)_{\dot{\alpha}}~,
\end{align}
where $k$ is the momentum flowing outward from the space-time point $x$, {\it{i.e.}} 
the Fourier transform is taken with the phase $\exp(+\ii \,k\!\cdot\! x)$.

\section{Grassmann integration in superdiagrams}
\label{app:grass-super}
We discuss a method to carry out the Grassmann integrations appearing in $\cN=1$ superdiagrams 
involving chiral/anti-chiral multiplet and vector multiplet lines.

\subsection*{Diagrams with only chiral/anti-chiral multiplet lines}
As we can see from the Feynman rules in figure~\ref{fig:Feynmatter}, the three-point 
vertex with incoming chiral lines carries a factor of $\theta^2$ and thus in the
integration over the fermionic variables associated to the vertex, one remains with
only an integral over $\bar{\theta}$. For the three-point vertex with outgoing anti-chiral 
lines, we remain instead with an integration over $\theta$ only.

\begin{figure}[ht]
	\begin{center}
	\includegraphics[width=0.92\textwidth]{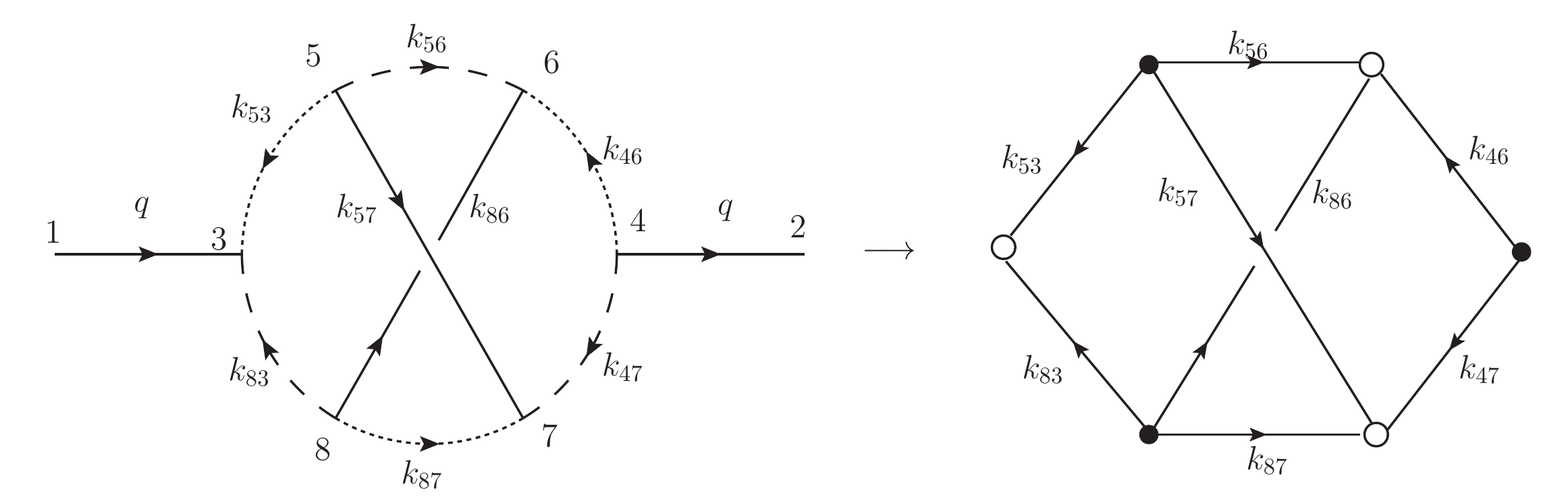}
	\end{center}
	\caption{On the left, a Feynman super-diagram involving only chiral/anti-chiral lines. 
	On the right, the corresponding $\theta$-graph encoding the Grassmann integrals. 
	The two ``external'' propagators with momentum $q$
	do not play a r\^ole in the bipartite graph because 
	the external states are the lowest components of the chiral and 
	anti-chiral superfields, and so the corresponding Grassmann variables are set to $0$.}
	\label{fig:1}
\end{figure}

We will use a graphical notation in which a black dot represents a $\theta$ variable and 
a white circle represents a $\bar{\theta}$ variable. {From} the point of view of the 
Grassmann integrations, superdiagrams with only hypermultiplet lines reduce to bipartite 
graphs, which we call ``$\theta$-graphs''. 
In these graphs a solid line connecting the $i$-th dot to the $j$-th circle corresponds to 
the factor 
\begin{equation}
	\label{exp2}
		\exp\Big(2\, \theta_i \,k_{ij} \,\bar{\theta}_j\Big)
		= 1 + 2\,\theta_i\, k_{ij}\, \bar{\theta}_j + \frac 12 \,\Big(2\,\theta_i\, k_{ij}\, 
		\bar{\theta}_j
		\Big)^2
\end{equation} 
coming from the chiral superfield propagator connecting two vertices at points $i$ and $j$ 
in a Feynman superdiagram. An example of a $\theta$-graph associated to a 
superdiagram is illustrated in figure~\ref{fig:1}, where the momenta respect momentum 
conservation at each node.

\begin{figure}[ht]
	\begin{center}
		\includegraphics[width=0.85\textwidth]{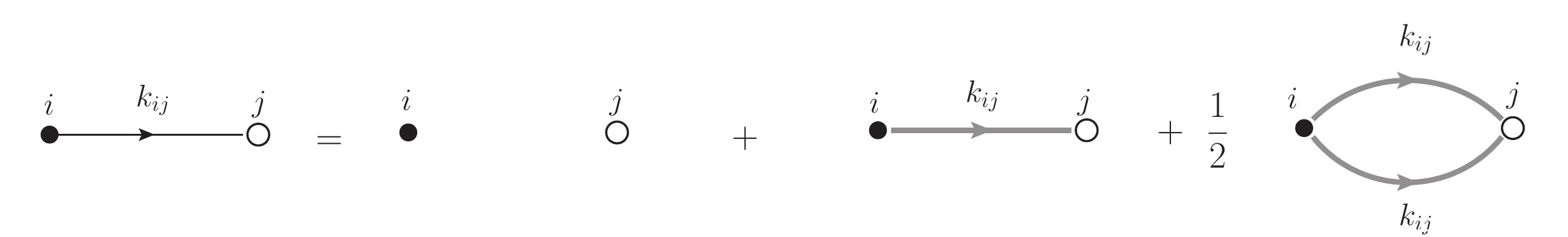}
	\end{center}
	\caption{Expansion of the exponential factor corresponding to a black line in 
	the $\theta$-graph. In the right hand side, each grey line corresponds to a 
	$\theta_i\, k\, \bar{\theta}_j$ term.}
	\label{fig:2}
\end{figure}

To compute the diagram we have to integrate over all $\theta_i$ and $\bar{\theta}_j$ 
variables. To do so, we expand the exponential factor corresponding to each line as in 
(\ref{exp2}); we graphically represent this expansion in figure~\ref{fig:2}.

\begin{figure}[ht]
	\begin{center}
		\includegraphics[width=\textwidth]{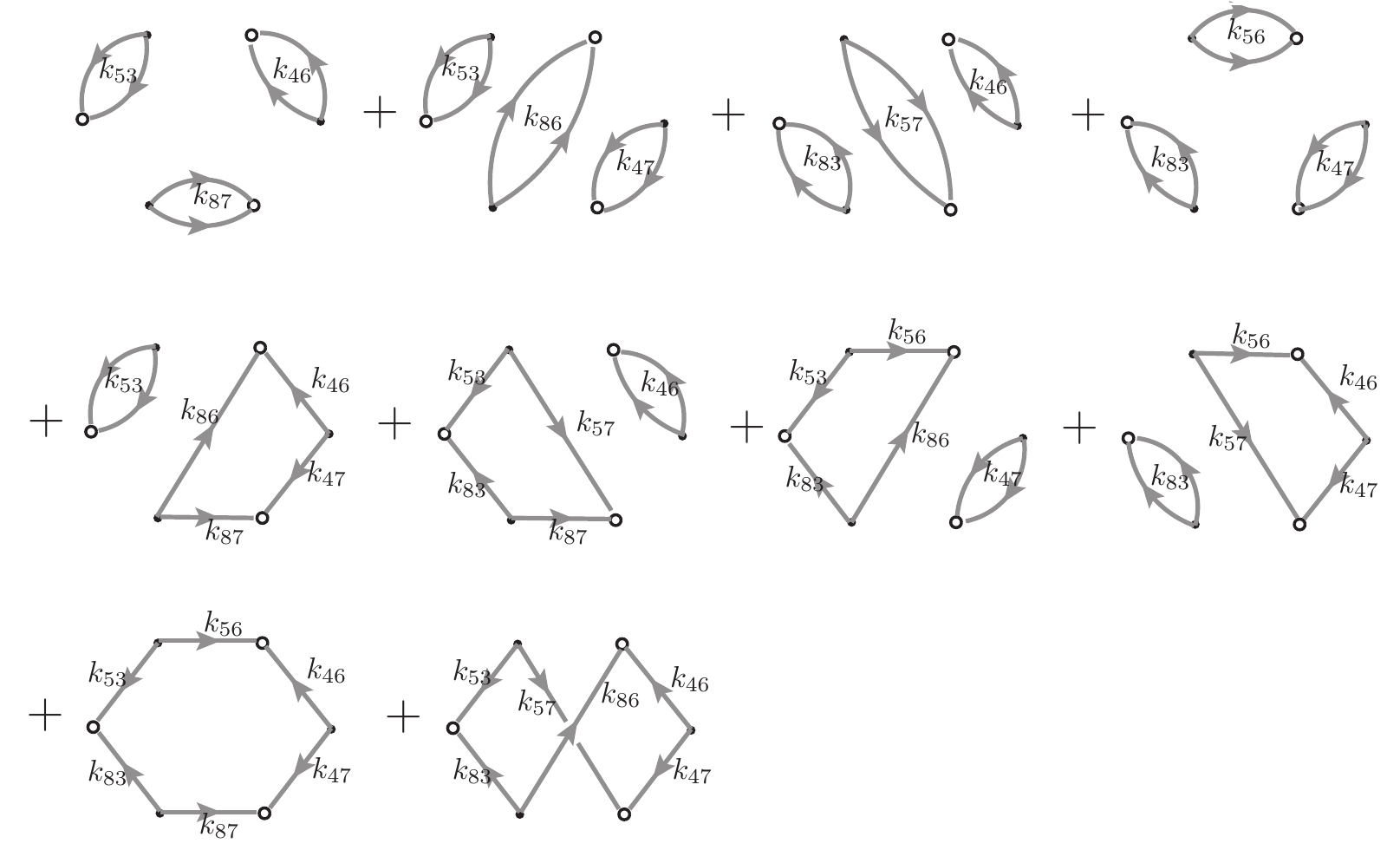}
	\end{center}
	\caption{The paths corresponding to non-vanishing contributions to the integral 
	encoded in the 
	diagram of figure \ref{fig:1}. Note that all cycles of length two are actually accompanied 
	by a factor of $1/2$ which, however, we did not write in the figure to avoid clutter.}
	\label{fig:3}
\end{figure}

Once this is done, it is easy to realize that one gets a non-zero contribution from the 
Grassmann integration if and only if in each black (or white) node one selects exactly two 
incoming (or outgoing) lines. As a consequence, one gets a contribution for each possible 
non-self-intersecting path passing through all the nodes that uses the edges present in the 
diagram. Such paths are collections of closed cycles. In the example of figure~\ref{fig:1} there are 
ten such paths, which are drawn in figure~\ref{fig:3}.

We can now integrate over all Grassmann variables belonging to a cycle. By using the Fierz 
identities (\ref{2tist2}) and the integration rules (\ref{intt2}), it is possible to show the following relation:
\begin{align}
\label{loopnruleg}
\parbox[c]{.3\textwidth}{\includegraphics[width = .3\textwidth]{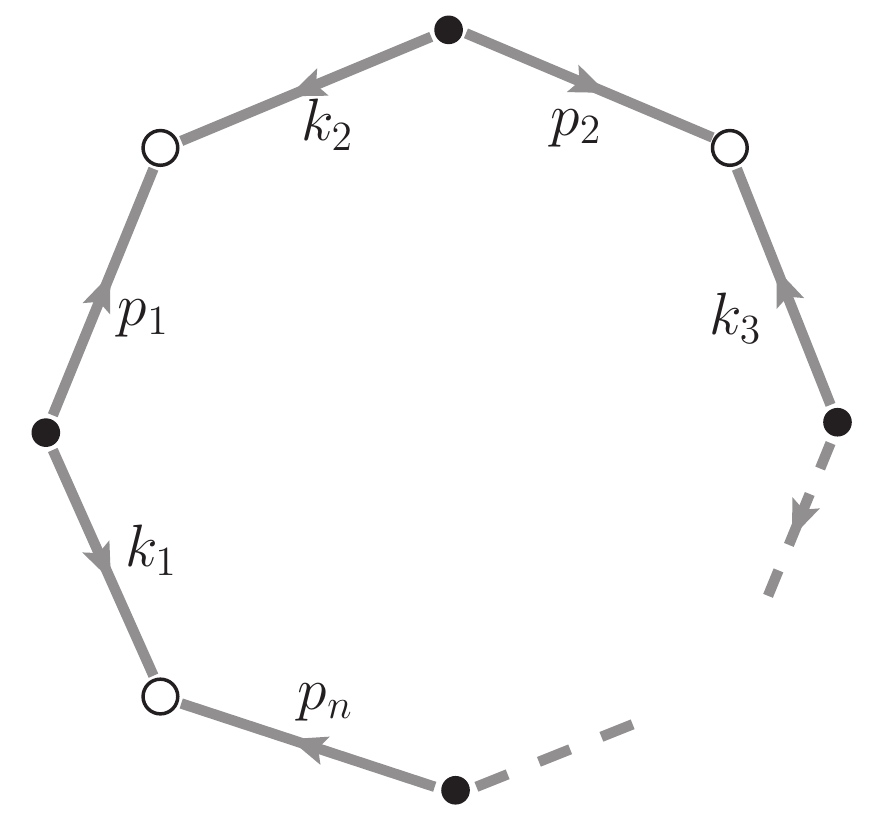}}
& = \int d^2\theta_1\, d^2\bar{\theta}_1 \ldots d^2\theta_n\,d^2\bar{\theta}_n\,
\big(2\,\theta_1\,k_1\,\bar{\theta}_1)\,\big(2\,\theta_1\,p_1\,\bar{\theta}_1\big)
\ldots
\nonumber\\[-15mm]
& 
= (-1)^{n+1} \, \tr\big(k_1\,\bar{p}_1\,k_2\,\bar{p}_2\,\ldots k_n\,\bar{p}_n\big)
\\
\nonumber
\end{align}
where the traces can be computed using (\ref{sigmatraces}) - or analogous formul\ae
~for $n>3$. This is the key Grassmann integration formula for the calculation of 
Feynman superdiagrams.

Applying this procedure to the $\theta$-graph of figure~\ref{fig:1}, we obtain
\begin{align}
\label{Fdefinition}
\parbox[c]{.35\textwidth}{\includegraphics[width = .35\textwidth]{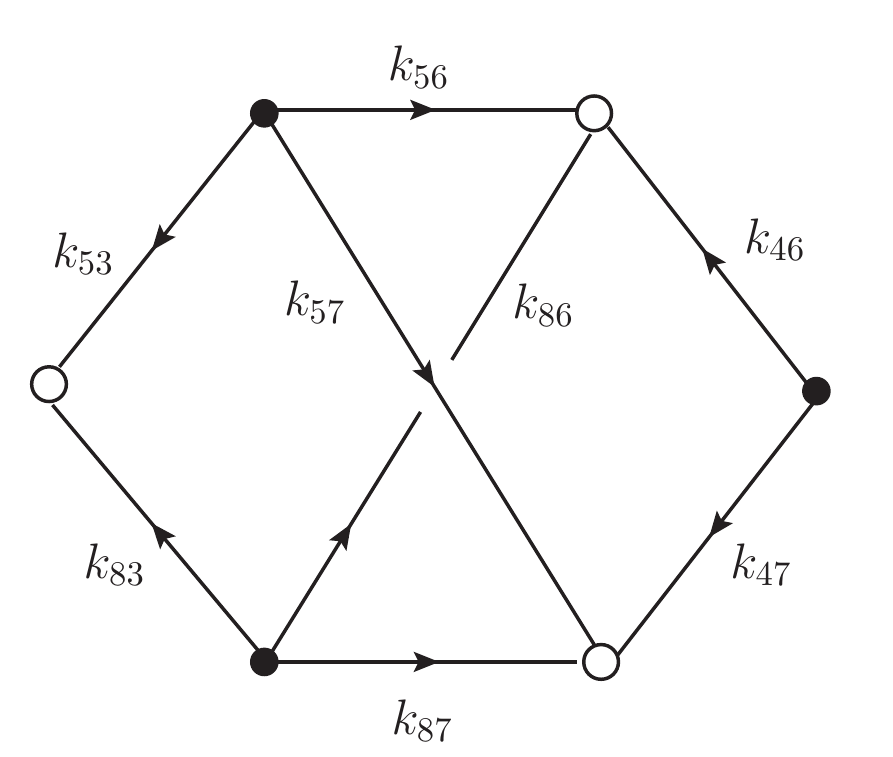}}
= F(k_{83},k_{87},k_{86},k_{53},k_{57},k_{56},k_{47},k_{46})~,
\end{align}
where we have introduced the function $F$ defined by
\begin{align}
F(p_1,p_2,p_3,p_4,p_5,p_6,p_7,p_8)&=
 -\,p_1^2 \,p_6^2 \,p_7^2 -\, p_2^2\,p_8^2\, p_4^2
 -\,p_3^2\, p_4^2\,p_7^2 - \,p_1^2 \,p_5^2\,p_8^2\nonumber\\[1mm]
& ~~~\,+ p_4^2 \,\tr\big(p_8\,\bar p_7\, p_2\, \bar p_3\big)
+ p_8^2\, \tr\big(p_2\,\bar p_1\, p_4\, \bar p_5\big) \nonumber\\[1mm]
&~~~\,+ p_7^2 \,\tr\big(p_1\,\bar p_4\, p_6\, \bar p_3\big) 
+ p_1^2 \,\tr\big(p_6\,\bar p_8\, p_7\, \bar p_5\big)\nonumber\\[1mm]
&~~~\,+ \tr\big(p_6\, \bar p_8\,p_7\, \bar p_2\, p_1\, \bar p_4\big)
+ \tr\big(p_8\, \bar p_7\, p_5 \,\bar p_4\, p_1 \bar p_3\big)~.
\label{Fisp}
\end{align}
With the momentum assignments as in (\ref{Fdefinition}), the ten terms in the right hand side
of (\ref{Fisp}) precisely reproduce the ten terms represented in figure~\ref{fig:3}. 
Computing the traces with the help of (\ref{sigmatraces}), one obtains in the end a polynomial 
of order six in the momenta entirely made of scalar products. 

We have explicitly worked out this example because this $\theta$-graph 
actually describes the prototypical example for the Grassmann factor associated to 
many of the Feynman superdiagrams that we will consider in detail in 
appendix~\ref{app:diagrams}, the only difference being in the different assignments 
of the momenta to the various lines.

\subsection*{Vector multiplet lines}

For Feynman superdiagrams containing vector multiplet lines, the most convenient strategy
to handle the Grassmann integration is first to eliminate the vector lines, so that one 
remains with graphs containing hypermultiplet lines only, which can then be computed as 
we have previously described. 

\begin{figure}[ht]
	\begin{center}
		\includegraphics[width=0.8\textwidth]{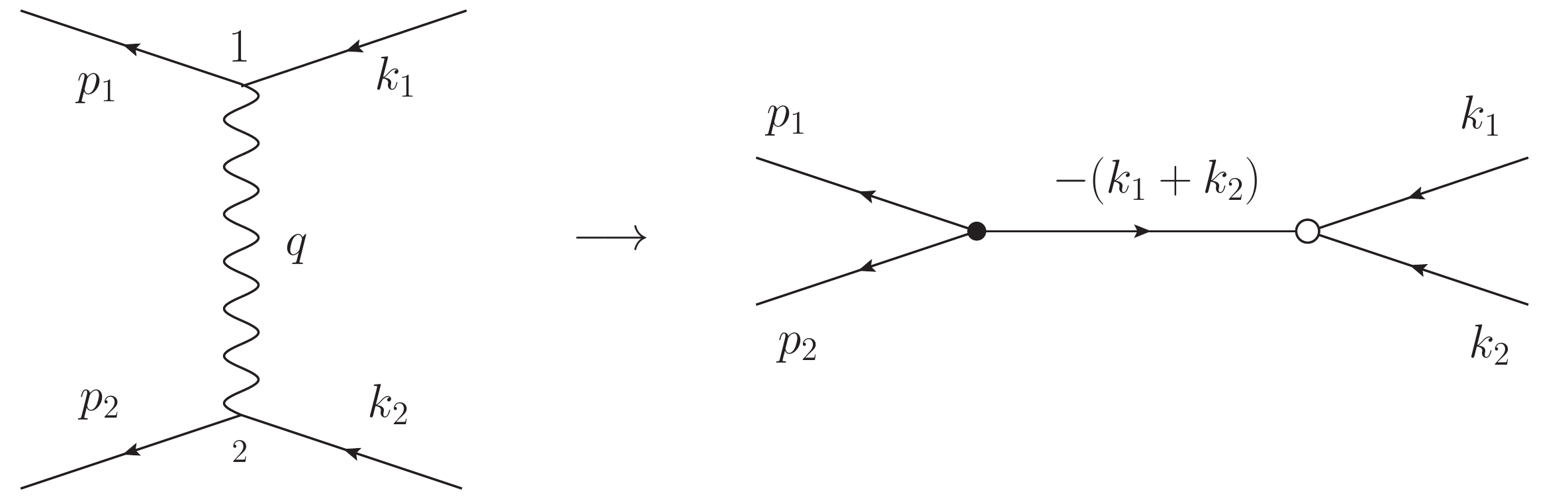}
	\end{center}
	\caption{How to associate a $\theta$-graph to a diagram with a vector line attached to matter current.}
	\label{fig:5}
\end{figure}

Let us first consider the graphs in which all vector lines are attached at both ends to a 
hypermultiplet line. In this case, for every vector line we have a sub-graph of the form 
described on the left of figure \ref{fig:5}, where the solid oriented lines indicate a generic 
chiral/anti-chiral multiplet propagator. 

As one can see from the Feynman rules listed in section~\ref{secn:fieldtheory}, at each cubic vertex,
labeled by 1 and 2, both $\theta_1$ and $\bar{\theta}_1$, and $\theta_2$
and $\bar{\theta}_2$ are present and have to be integrated.
However, the vector propagator contains a factor of $\theta_{12}^2\, \bar{\theta}_{12}^2$ 
which acts as a $\delta$-function identifying $\theta_2$ and $\bar{\theta}_2$ with 
$\theta_1$ and $\bar{\theta}_1$, respectively. Therefore, we remain with two 
Grassmann variables, say $\theta_1$ and $\bar{\theta}_1$, to be integrated. 
The hypermultiplet lines attached to these variables provide the factor
\begin{equation}
\label{expt1t1b}
\exp\Big[ -\theta_1 \big(k_1+p_1+k_2+p_2\big)\,\bar{\theta}_1\Big]
= \exp\Big[ - 2 \,\theta_1 \big(k_1 + k_2\big)\, \bar{\theta}_1\Big]
\end{equation}
where in the second step we have used momentum conservation. 
This is exactly the same type of exponential factor
that in a $\theta$-graph we associate to a solid line from the black dot 
representing $\theta_1$ to the white dot representing $\bar{\theta}_1$ (see (\ref{exp2})).  
Thus, we deduce the rule of figure \ref{fig:5} which allows us to write the portion of a 
$\theta$-graph corresponding to a vector line attached to matter lines.

\begin{figure}[ht]
	\begin{center}
	\vspace{0.3cm}
		\includegraphics[width=0.8\textwidth]{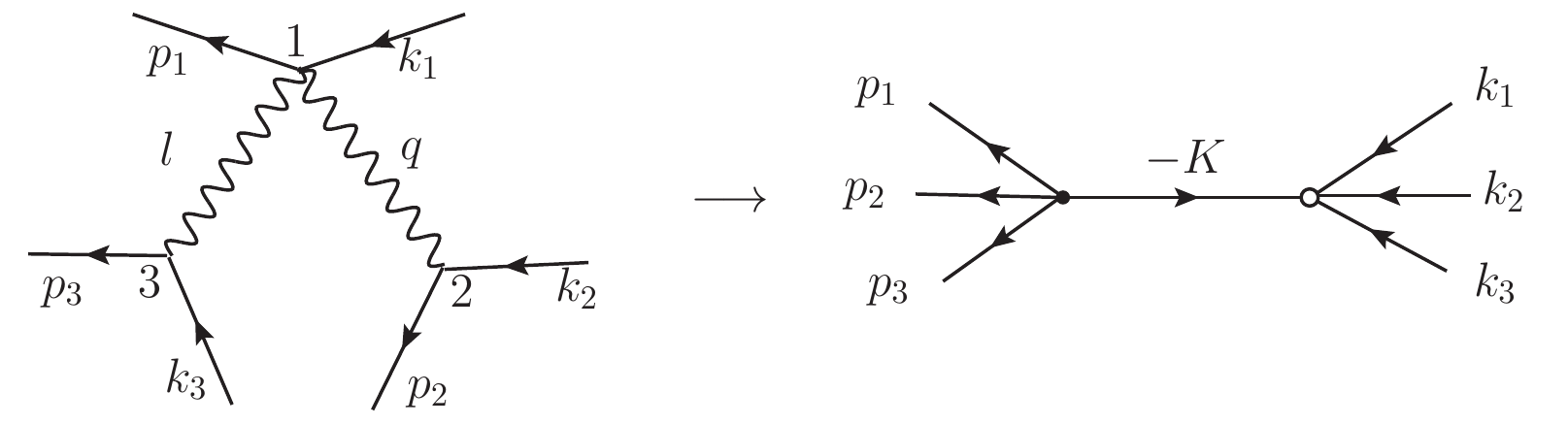}
	\end{center}
	\caption{The rule to replace a quartic vertex with two vector lines with the corresponding 
	$\theta$-graph. Here $K= k_1+k_2+k_3$.}
	\label{fig:6}
\end{figure}

Analogous rules can be worked out when there are vertices with the simultaneous emission of two vector lines from a scalar current line. The simplest case is the one represented in figure \ref{fig:6}.

Things proceed in a perfectly analogous way if there are more quartic vertices. In the end, the subdiagram gives rise to a $\theta$-subgraph with the same ``external'' lines. However now the outgoing lines are all attached to a single black dot - corresponding to an integration variable 
$\theta$ - and the incoming lines are all attached to a single white circle - corresponding to a variable $\bar{\theta}$. The dot and the circle are connected by a line, associated with 
the exponential factor $\exp\big(-2\,\theta\, K\,\bar{\theta}\,\big)$, where $K$ is the sum of the incoming momenta. 

When the diagram contains interaction vertices with three or more vectors, things are slightly 
more involved because of the presence of covariant spinor derivatives in such vertices.
We will not describe the procedure in general, because only one diagram with a three-vector vertex
is needed in our computations. Indeed, we find more convenient to deal directly with this case, 
in which it is again possible to rewrite the Grassmann integrals in terms of a $\theta$-graph of the type introduced above.

\section{Evaluation of the relevant superdiagrams}
\label{app:diagrams}
We report the computation of the Feynman superdiagrams that yield a contribution 
proportional to $\zeta(5)$ in the three-loop corrections to the propagator 
of the scalar field in the $\cN=2$ vector multiplet. 

Any diagram of this kind, with external adjoint indices $b$ and $c$, external 
momentum $q$ and $s$ internal lines, is written as
\begin{align}
	\label{gen-diag}
		\cW_{bc}(q) = \cN\times \cT_{bc}\times \!\int \!\prod_s \frac{d^d k_s}{(2\pi)^d}
		~\delta^{(d)}(\mathrm{cons}) ~\frac{\cZ(k)}{\prod_s k_s^2}~.
\end{align}
Here $\cN$ is the product of the symmetry factor of the diagram and all the factors (like 
the powers of the coupling constant $g$) appearing in the vertices - except for the color 
factors which give rise to the tensor $\cT_{bc}$. We have then the scalar integral over the 
internal momenta $k_s$ which we perform using dimensional regularization setting 
$d=4-2\varepsilon$. The momenta are subject to the appropriate momentum 
conservation relations enforced by the $\delta$-functions $\delta^{(d)}(\mathrm{cons})$. 
Beside the denominator coming from the massless propagators, the integrand 
contains also a numerator $\cZ(k)$ which is the result of the integration over all the 
Grassmann variables of the $\theta$-dependent expressions present in the superdiagram.

The massless scalar integrals at three loops with cubic or quartic vertices can be evaluated 
by various means; in particular, we use the FORM version of the program Mincer discussed 
in \cite{Larin:1991fz}, which classifies them according to different ''topologies'' 
described by diagrams in which a solid line indicates a massless scalar propagator, and 
momentum conservation is enforced at each vertex. 

\subsection*{Diagrams with six insertions on the hypermultiplet loop}
We start by considering the diagrams with six insertions of an adjoint generator on the hypermultiplet loop. The color factor of these diagrams is proportional to a doubly contracted $\cC^\prime$ tensor with six indices defined in (\ref{defC}).

The first diagram we consider is the following
\begin{align}
	\label{LA}
	\cW_{bc}^{(1)}(q) =
	\parbox[c]{.5\textwidth}{\includegraphics[width = .5\textwidth]{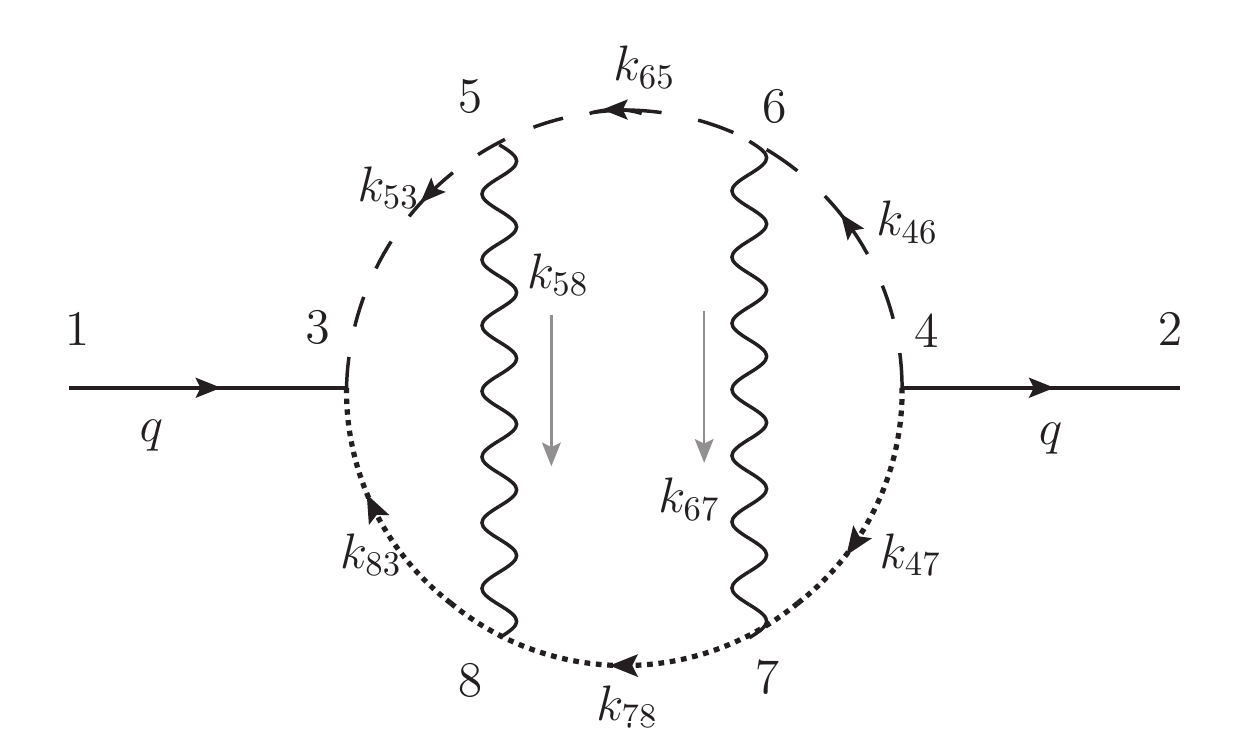}}
\end{align}
In this first diagram we set up the notation that we will use also in all subsequent ones. 
The external momentum is always denoted as $q$. Regarding the labeling of internal momenta,
we label the internal vertices (from $3$ to $8$ in this case) and we denote as $k_{ij}$ the 
momentum flowing in a propagator from the vertex $i$ to the vertex $j$, which is also 
the same convention introduced in (\ref{exp2}). Assuming it, from now on we will display in the 
figures only the labels of the vertices and not of the internal momenta.
The Feynman rules for propagators and vertices are given section~\ref{subsec:action-fr}. 
Using them, we get
\begin{align}
	\label{LAbis}
		\cW_{bc}^{(1)}(q)
		= 8 g^6 \times C^\prime_{bdeced} \times 
		\parbox[c]{.35\textwidth}{\includegraphics[width = .35\textwidth]{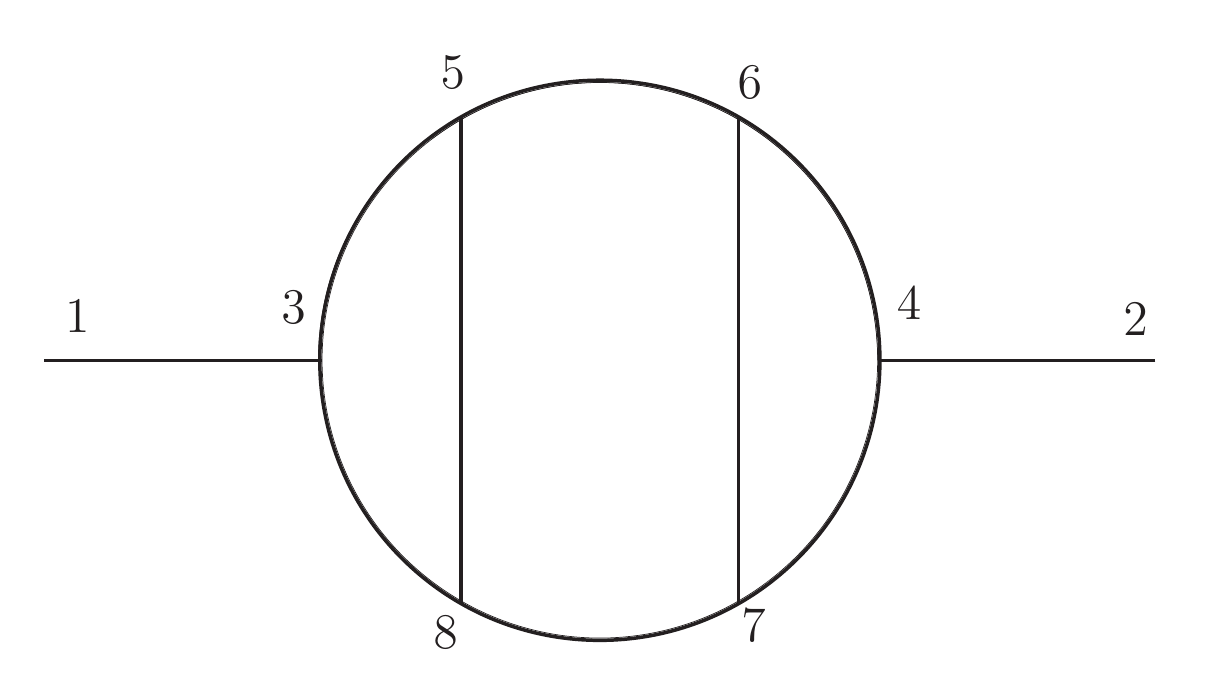}}
		\cZ^{(1)}(k).
\end{align}
The scalar diagram has the ladder topology denoted as LA in \cite{Larin:1991fz}.
The Grassmann factor $\cZ^{(1)}(k)$ is obtained integrating over $d^4\theta_i$ for $i=3,\ldots,8$ and is easily determined using the rule described in figure \ref{fig:5}. It is given by the following
$\theta$-diagram
\begin{align}
	\label{Z1is}
		\cZ^{(1)}(k) & = 
		\parbox[c]{.5\textwidth}{\includegraphics[width = .5\textwidth]{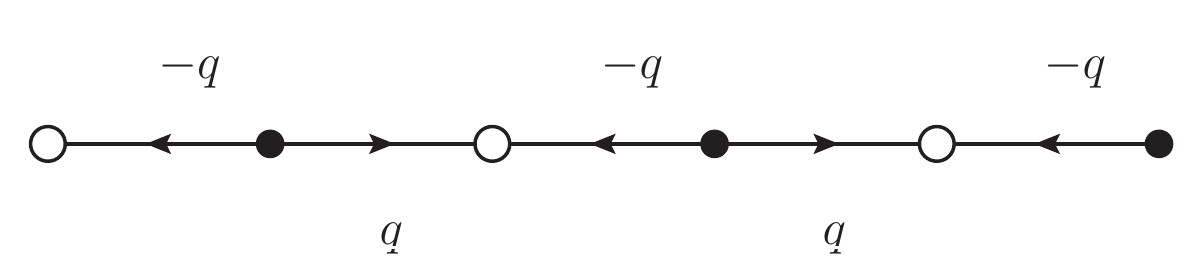}}
		= - q^6~.
\end{align}
The evaluation of this $\theta$-diagram by means of its cycle expansion, as explained after (\ref{exp2}) and illustrated in figure \ref{fig:1}, is immediate using (\ref{loopnruleg}).  
A factor of $q^4$ removes the two external propagators in the scalar diagram, so that it reduces to
\begin{align}
	\label{scal1}
		- q^2\,
		\parbox[c]{.12\textwidth}{\includegraphics[width = .12\textwidth]{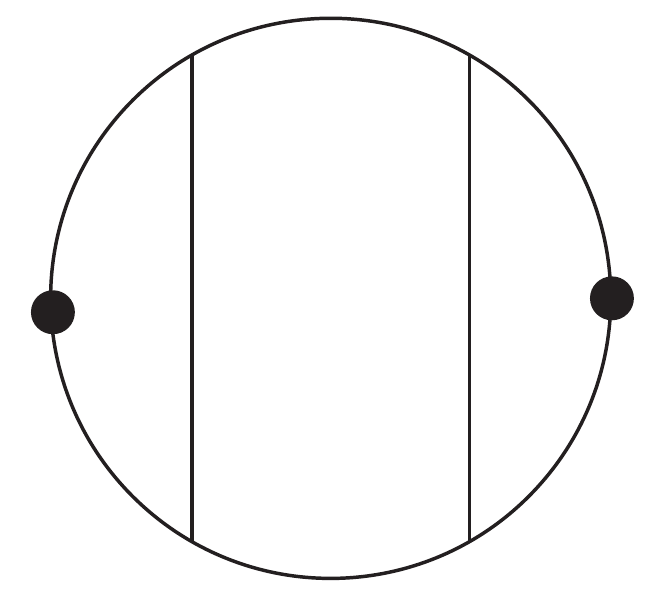}}
		= - \frac{20 \zeta(5)}{(4\pi)^6} \frac{1}{q^2}~.
\end{align}
Here we have employed the standard graphical notation for diagrams with canceled external propagators and we have given the value of this scalar integral, which is finite, directly in $d=4$. Altogether we get thus
\begin{align}
	\label{W1res}
		\cW_{bc}^{(1)}(q) = -\frac{1}{q^2}\left(\frac{g^2}{8\pi^2}\right)^3 \zeta(5) \times \left(20
		\,C^\prime_{bdeced}\right)~. 
\end{align}

The next diagram is 
\begin{align}
	\label{W2}
		\cW_{bc}^{(2)}(q) &=
		\parbox[c]{.40\textwidth}{\includegraphics[width = .40\textwidth]{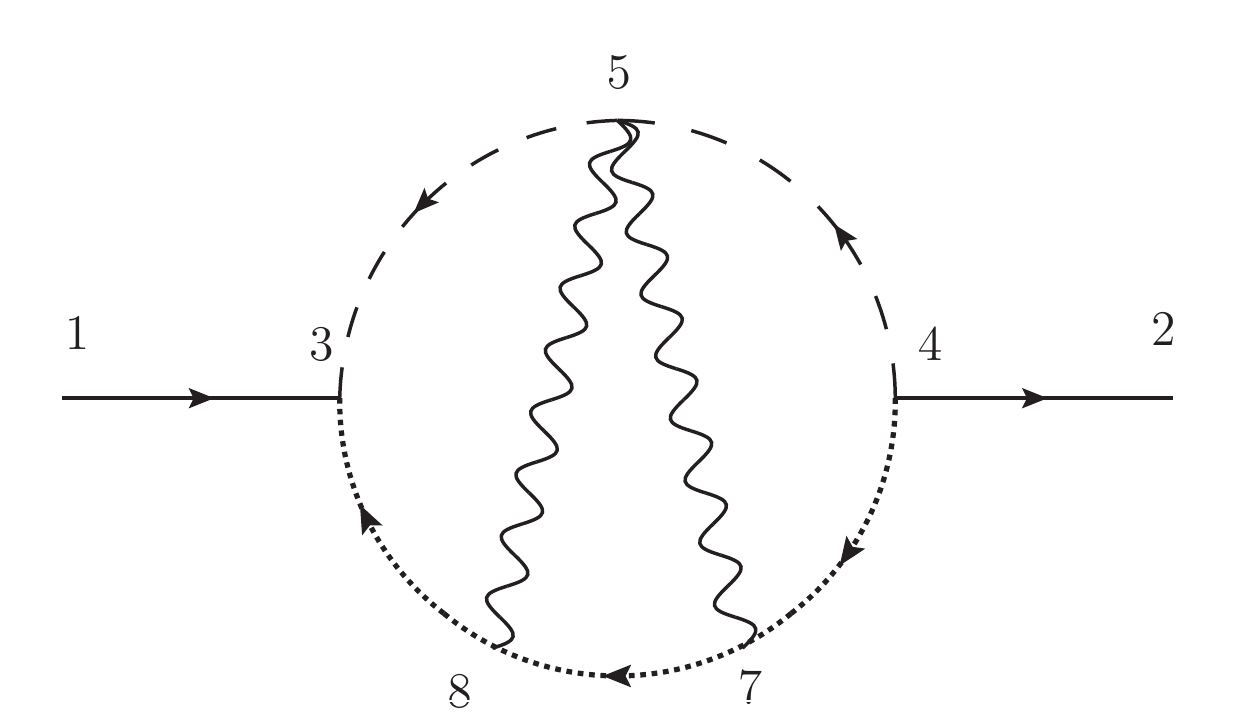}}
		\nonumber\\
		&=   4 g^6 \times 2 \,\cT^ {(2)}_{bc} \times 
		\parbox[c]{.33\textwidth}{\includegraphics[width = .33\textwidth]{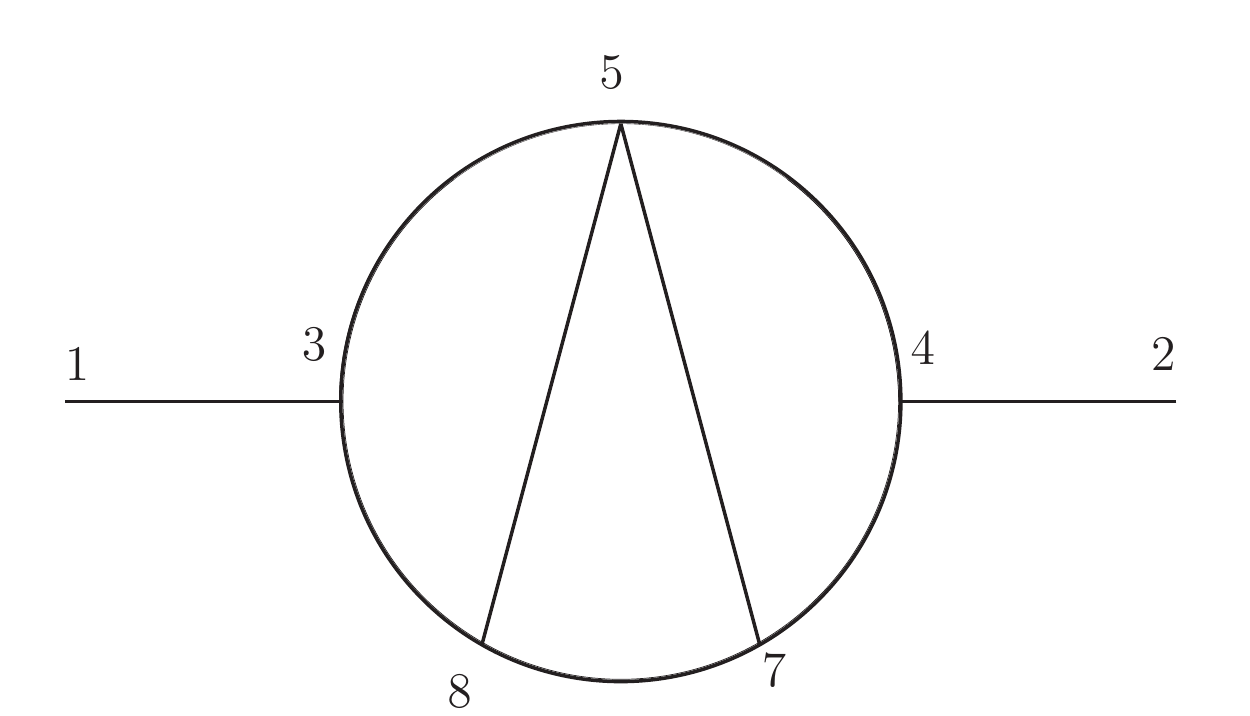}}
		\cZ^{(2)}(k)~.
\end{align}
Here the color tensor reads
\begin{equation}
	\label{T2bc}
		\cT_{bc}^{(2)} = C^\prime_{bdecde} + C^\prime_{bdeced}~, 
\end{equation}
the two terms stemming from the two ways to attach the gluon lines to the quartic vertex. This expression comes with a factor of $2$ in (\ref{W2}) to account for 
the diagram in which the dashed and dotted parts of the hypermultiplet loop are switched.    
The scalar diagram has the fan topology denoted as FA in \cite{Larin:1991fz}. 
The Grassmann factor can be determined using the rule described in figure \ref{fig:6} and 
it is given by
\begin{align}
	\label{Z2is}
		\cZ^{(2)}(k) =  
		\parbox[c]{.33\textwidth}{\includegraphics[width = .33\textwidth]{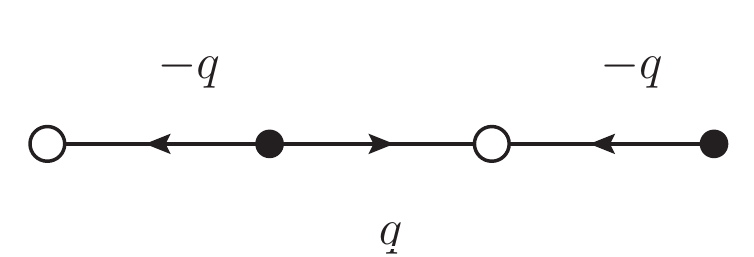}}
		= q^4~.
\end{align}
This factor removes the two external propagators in the scalar diagram, so that it reduces to
\begin{align}
	\label{scal2}
		\parbox[c]{.12\textwidth}{\includegraphics[width = .12\textwidth]{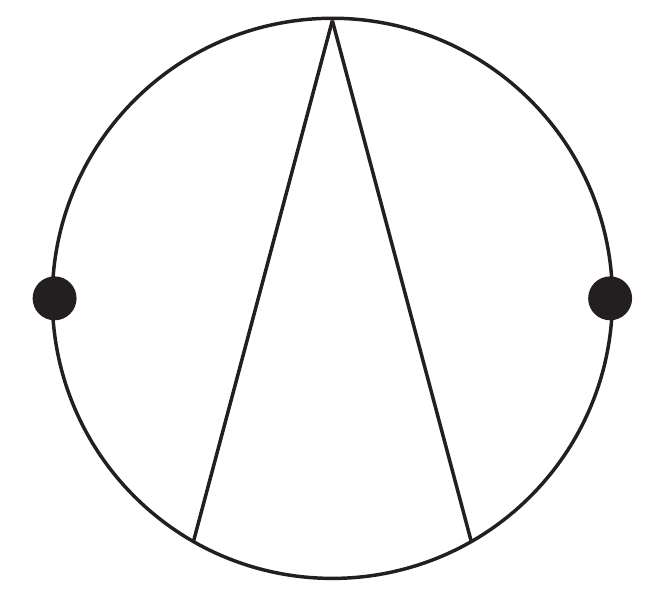}}
		=  \frac{20 \zeta(5)}{(4\pi)^6} \frac{1}{q^2}~.  
\end{align}
Altogether we find thus
\begin{align}
	\label{W2res}
		\cW_{bc}^{(2)}(q) = -\frac{1}{q^2}\left(\frac{g^2}{8\pi^2}\right)^3 \zeta(5) \times 
		\left(-20\,C^\prime_{bdecde} -20\,C^\prime_{bdeced}\right)~. 
\end{align}

The third diagram that contributes is 
\begin{align}
	\label{W3}
		\cW_{bc}^{(3)}(q) &=
		\parbox[c]{.4\textwidth}{\includegraphics[width = .4\textwidth]{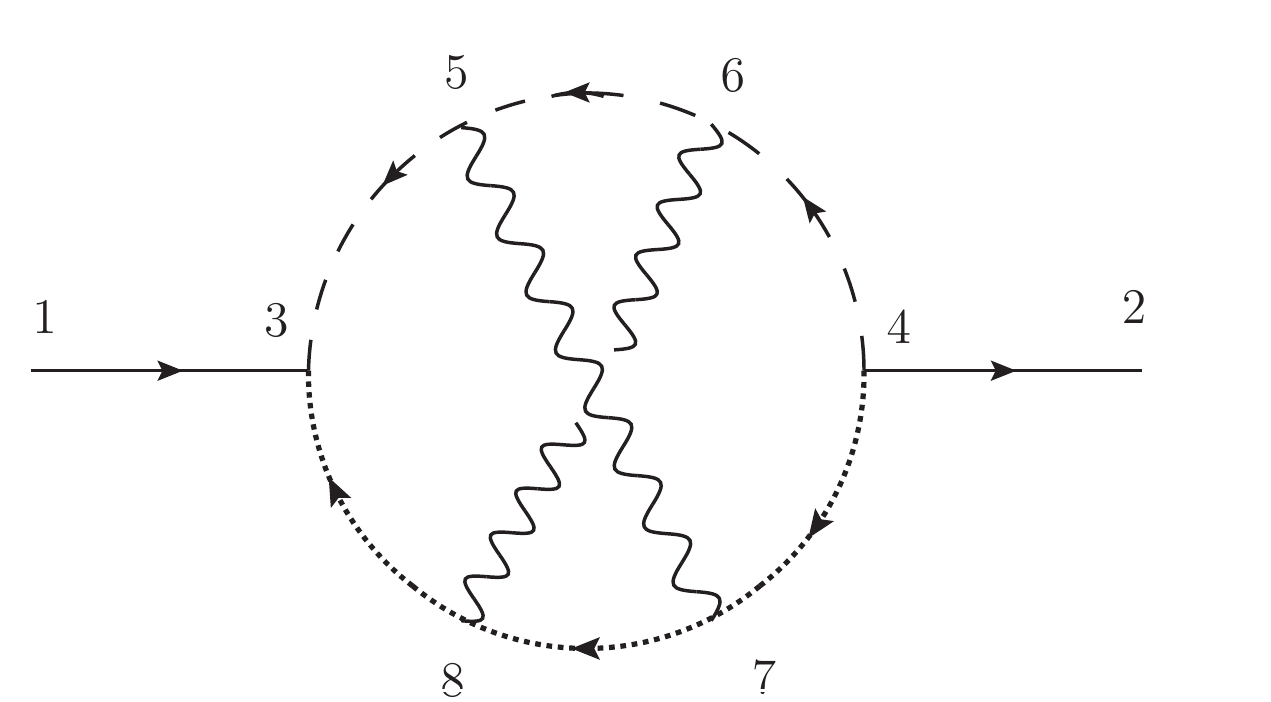}}   
		\nonumber\\
		&= 8 g^6 \times C^\prime_{bdecde} \times 
		\parbox[c]{.33\textwidth}{\includegraphics[width = .33\textwidth]{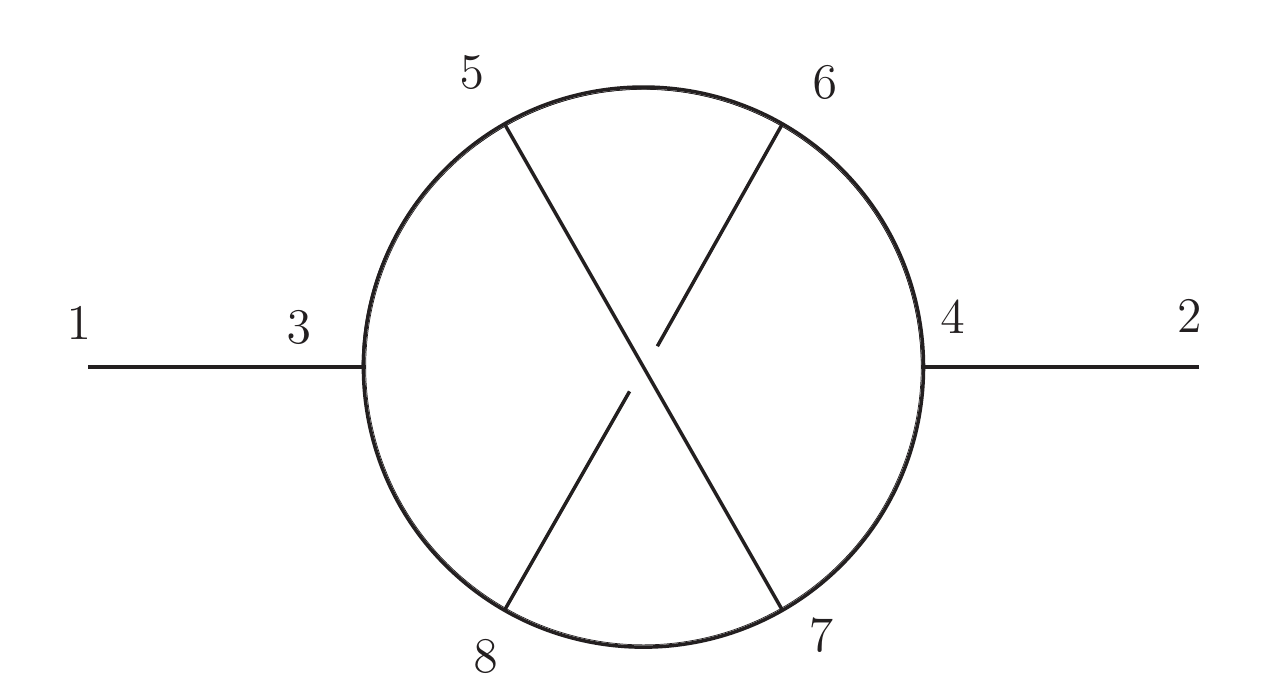}}
		\cZ^{(3)}(k)~.
\end{align}
The scalar diagram has the non-oriented topology denoted as NO in \cite{Larin:1991fz}. 
The Grassmann factor is found applying the rule of figure \ref{fig:5} and it is given by
a $\theta$-diagram of the type depicted in \ref{Fdefinition}, but with a different assignment of momenta. In particular, one has 
\begin{align}
	\label{Z3is}
		\cZ^{(3)}(k) = F\big(k_{83},-(k_{46} + k_{78}),k_{65},k_{53},k_{78},-(k_{47}+ k_{65}),k_{46},k_{47}\big)~.
\end{align}
Evaluating this and inserting it in the scalar momentum integral, we find that the results 
contains a $\zeta(5)$-contribution. Indeed we have
\begin{align}
\label{scal3}
		\parbox[c]{.33\textwidth}{\includegraphics[width = .33\textwidth]{NO.pdf}}
		\cZ^{(3)}(k)
		=  - \frac{10 \zeta(5)}{(4\pi)^6} \frac{1}{q^2} + \ldots
\end{align}
where the ellipses stand for terms that do not contain $\zeta(5)$.
Putting together the various factors, we find
\begin{align}
	\label{W3res}
		\cW_{bc}^{(3)}(q) = -\frac{1}{q^2}\left(\frac{g^2}{8\pi^2}\right)^3 \zeta(5) \times 
		\left(10 \,C^\prime_{bdecde}\right) + \ldots~. 
\end{align}

Next we consider
\begin{align}
	\label{W4}
		\cW_{bc}^{(4)}(q) &=
		\parbox[c]{.4\textwidth}{\includegraphics[width = .4\textwidth]{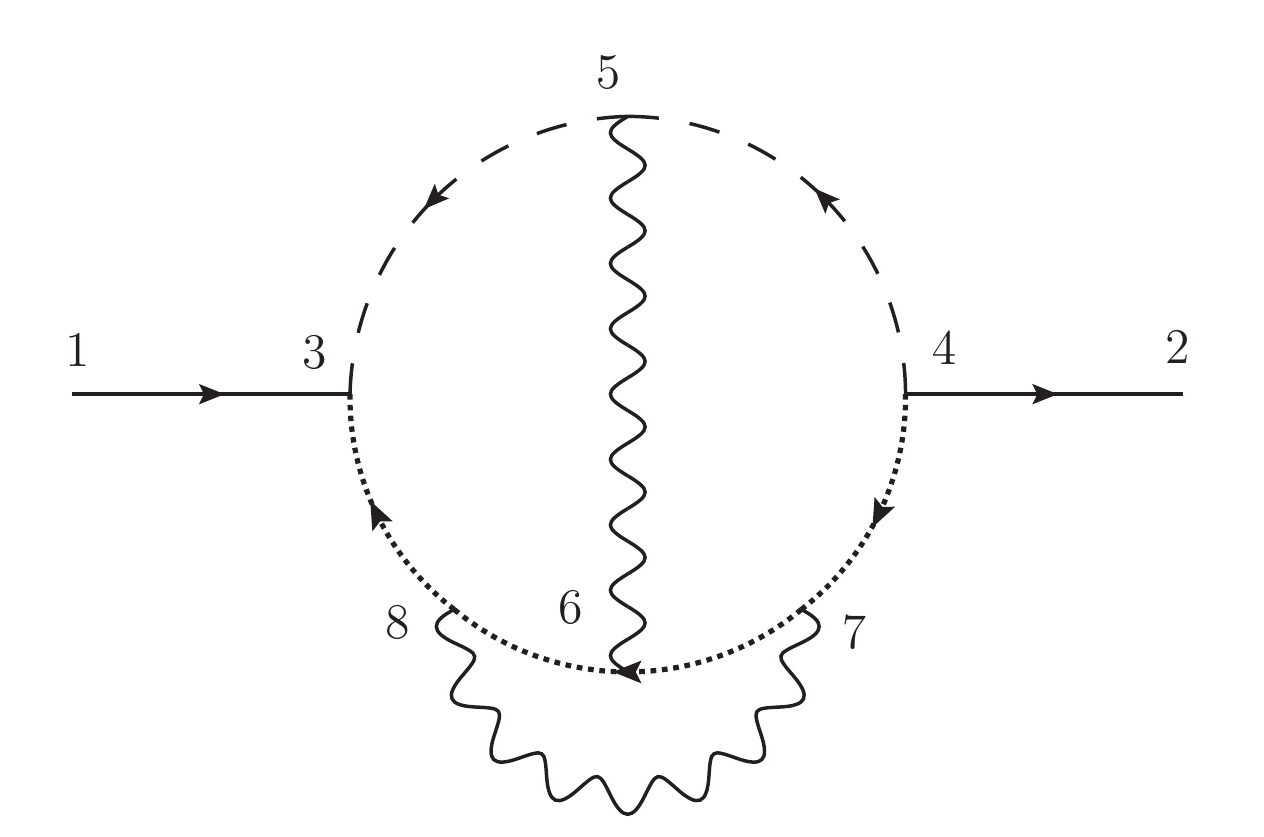}}
		\nonumber\\
		&= - 8 g^6 \times \cT_{bc}^{(4)}\times 
		\parbox[c]{.33\textwidth}{\includegraphics[width = .33\textwidth]{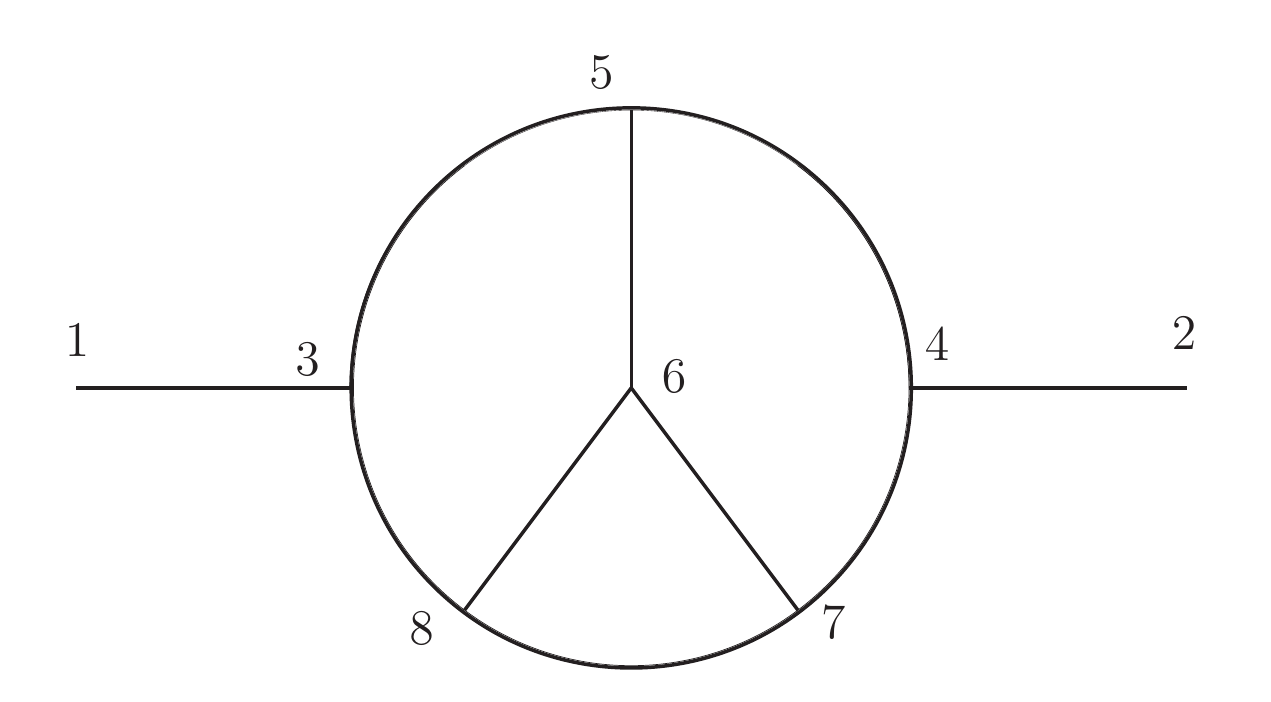}}
		\cZ^{(4)}(k)~.
\end{align}
where the color tensor reads
\begin{equation}
	\label{T4bc}
		\cT_{bc}^{(4)} = C^\prime_{bdedce} + C^\prime_{bdcede}~.
\end{equation}
Here the second term comes from the diagram where the dashed and dotted parts of the 
hypermultiplet loop are exchanged. The scalar diagram has the ``Benz'' topology denoted as BE 
in \cite{Larin:1991fz}. The Grassmann factor is found using the rule of figure \ref{fig:5} and it is given by
\begin{align}
	\label{Z4is}
		\cZ^{(4)}(k) = F\big(k_{83},-(k_{46} + k_{68}),k_{67},k_{53},k_{68},-(k_{45}+ k_{67}),
		k_{47},k_{45}\big)~.
\end{align}
The corresponding scalar momentum integration contains a $\zeta(5)$ contribution; indeed
\begin{align}
	\label{scal4}
		\parbox[c]{.33\textwidth}{\includegraphics[width = .33\textwidth]{BE.pdf}}
		\cZ^{(4)}(k)
		=   \frac{20 \zeta(5)}{(4\pi)^6} \frac{1}{q^2} + \ldots~.  
\end{align}
Altogether we have thus
\begin{align}
	\label{W4res}
		\cW_{bc}^{(4)}(q) = -\frac{1}{q^2}\left(\frac{g^2}{8\pi^2}\right)^3 \zeta(5) \times 
		 \left( 20\,C^\prime_{bdedce} + 20\,C^\prime_{bdcede}\right) + \ldots~. 
\end{align}

\subsection*{Diagrams with five insertions on the hypermultiplet loop}
We now consider the diagrams with five insertions of an adjoint generator on the hypermultiplet loop. The first diagram of this kind we consider is
\begin{align}
	\label{W5}
		\cW_{bc}^{(5)}(q) &=
		\parbox[c]{.4\textwidth}{\includegraphics[width = .4\textwidth]{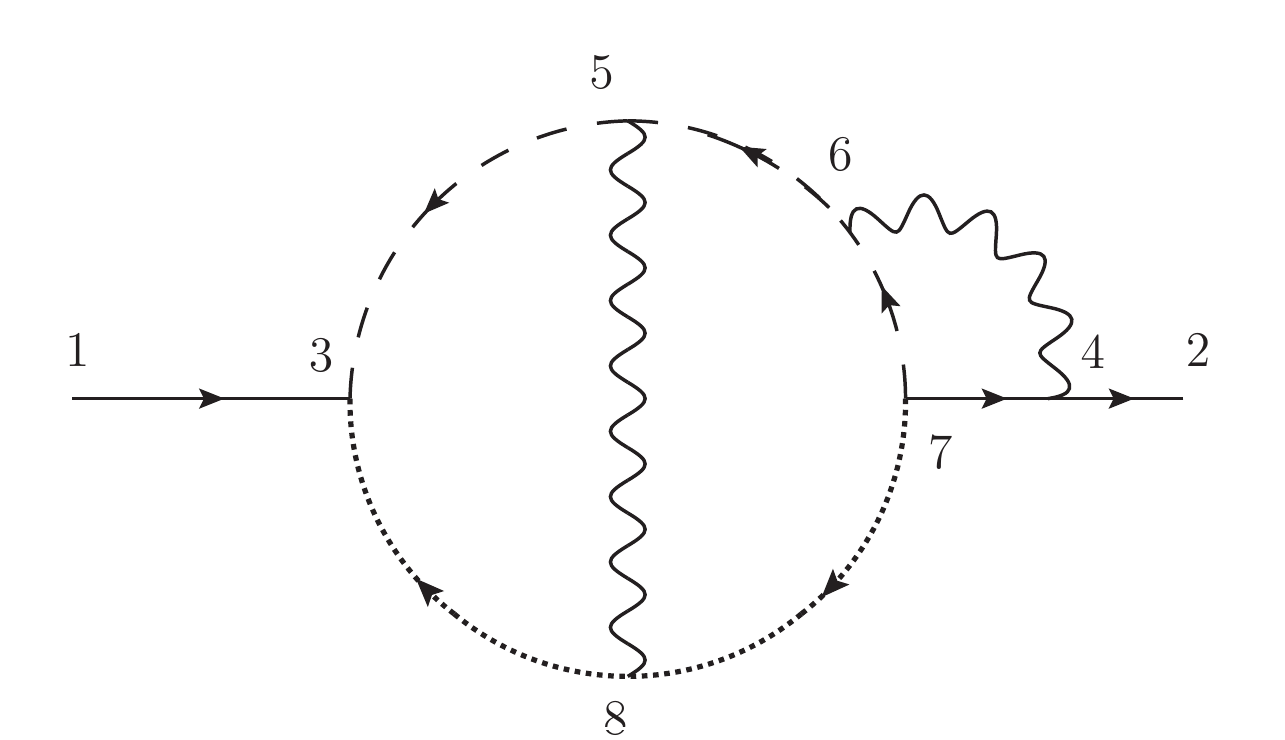}}
		\nonumber\\
		&= -8 g^6 \times \cT_{bc}^{(5)} \times 
		\parbox[c]{.33\textwidth}{\includegraphics[width = .33\textwidth]{LA.pdf}}
		\cZ^{(5)}(k)~.
\end{align}
The color factor is given by 
\begin{align}
	\label{T5bc}
		\cT_{bc}^{(5)} & = 
		\ii f_{cef} C^\prime_{bdefd} - \ii f_{cef} C^\prime_{bdfed}
		+ \ii f_{bef} C^\prime_{cdefd} - \ii f_{bef} C^\prime_{cdfed}
		\nonumber\\[1mm]
		& = 2 \,\ii f_{cef} C^\prime_{bdefd} + 2 \,\ii f_{bef} C^\prime_{cdefd}~,
\end{align}
where the four terms that appear in the first line correspond to the four possible ways to 
attach the ``external'' vector multiplet line. The Grassmann factor is again found using 
the rule of figure \ref{fig:5} and it is given by
\begin{align}
	\label{Z5is}
		\cZ^{(5)}(k) = F\big(0,-k_{78},k_{78},-q,0,q,k_{78},-(k_{78}+ q)\big)~.
\end{align}
Using this result inside the scalar momentum integral, which has the LA topology, one finds
\begin{align}
	\label{scal5}
		\parbox[c]{.33\textwidth}{\includegraphics[width = .33\textwidth]{LA.pdf}}
		\cZ^{(5)}(k)
		=   - \frac{20 \zeta(5)}{(4\pi)^6} \frac{1}{q^2} + \ldots~.  
\end{align}
The final result for this diagram is then
\begin{align}
	\label{W5res}
		\cW_{bc}^{(5)}(q) = -\frac{1}{q^2}\left(\frac{g^2}{8\pi^2}\right)^3 \zeta(5) \times 
		\left(-40\,\ii f_{cef} C^\prime_{bdefd} - 40\,\ii f_{bef} C^\prime_{cdefd} \right) + \ldots~. 
\end{align}

Another diagram in this class is
\begin{align}
\label{W6}
	\cW_{bc}^{(6)}(q) &=
		\parbox[c]{.4\textwidth}{\includegraphics[width = .4\textwidth]{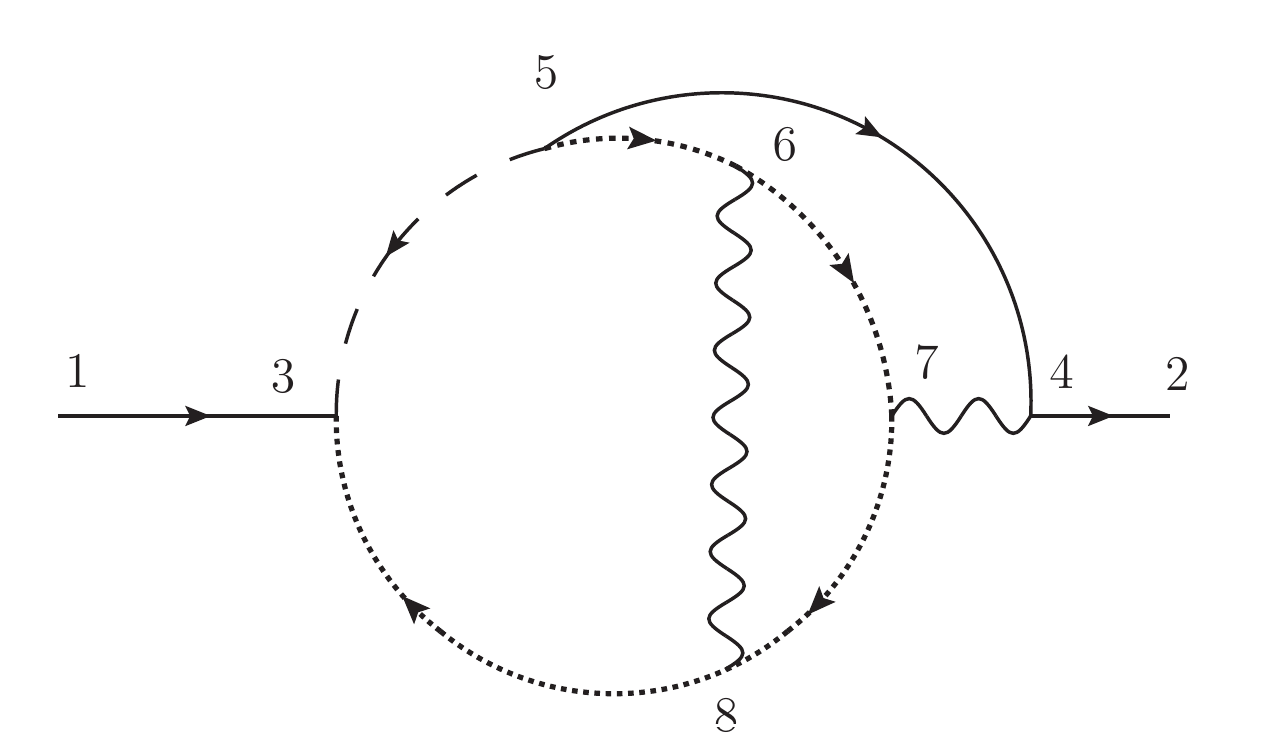}}
		\nonumber\\		
		&= -8 g^6 \times \cT_{bc}^{(6)} \times 
		\parbox[c]{.33\textwidth}{\includegraphics[width = .33\textwidth]{BE.pdf}}
		\cZ^{(6)}(k)~,
\end{align}
where the color factor is
\begin{align}
	\label{T6bc}
		\cT_{bc}^{(6)} & = 
		\ii f_{ced} C^\prime_{bfdfe} - \ii f_{ced} C^\prime_{befdf}
		+ \ii f_{bed} C^\prime_{cfdfe} - \ii f_{bed} C^\prime_{cefdf}~.
\end{align}
Here the four terms correspond to the four possible ways to attach the ``external'' adjoint 
chiral multiplet line. Using the by-now familiar procedure, the Grassmann factor is found to be
\begin{align}
	\label{Z6is}
		\cZ^{(6)}(k) = F\big
		(k_{73},-(k_{56} + k_{87}),k_{68},k_{53},k_{56},k_{54},k_{87},-(k_{87}+ q)\big)~.
\end{align}
The scalar integral, which has the BE topology, yields the result
\begin{align}
	\label{scal6}
		\parbox[c]{.33\textwidth}{\includegraphics[width = .33\textwidth]{BE.pdf}}
		\cZ^{(6)}(k)
		=   - \frac{20 \zeta(5)}{(4\pi)^6} \frac{1}{q^2} + \ldots~.  
\end{align}
The total result is thus
\begin{align}
	\label{W6res}
		\cW_{bc}^{(6)}(q) & = -\frac{1}{q^2}\left(\frac{g^2}{8\pi^2}\right)^3 \zeta(5)\nonumber\\
		& ~~~\times 
		\left(	-20\,\ii f_{ced} C^\prime_{bfdfe} + 20\,\ii f_{ced} C^\prime_{befdf}
		-20\, \ii f_{bed} C^\prime_{cfdfe} + 20\,\ii f_{bed} C^\prime_{cefdf}\right) + \ldots~. 
\end{align}

Among the diagrams with five insertions that give a $\zeta(5)$ contribution, there is one
whose Grassmann factor cannot be computed simply by using the rules illustrated in 
appendix~\ref{app:grass-super}. It is the following:
\begin{align}
	\label{W7}
		\cW_{bc}^{(7)}(q) &=
		\parbox[c]{.4\textwidth}{\includegraphics[width = .4\textwidth]{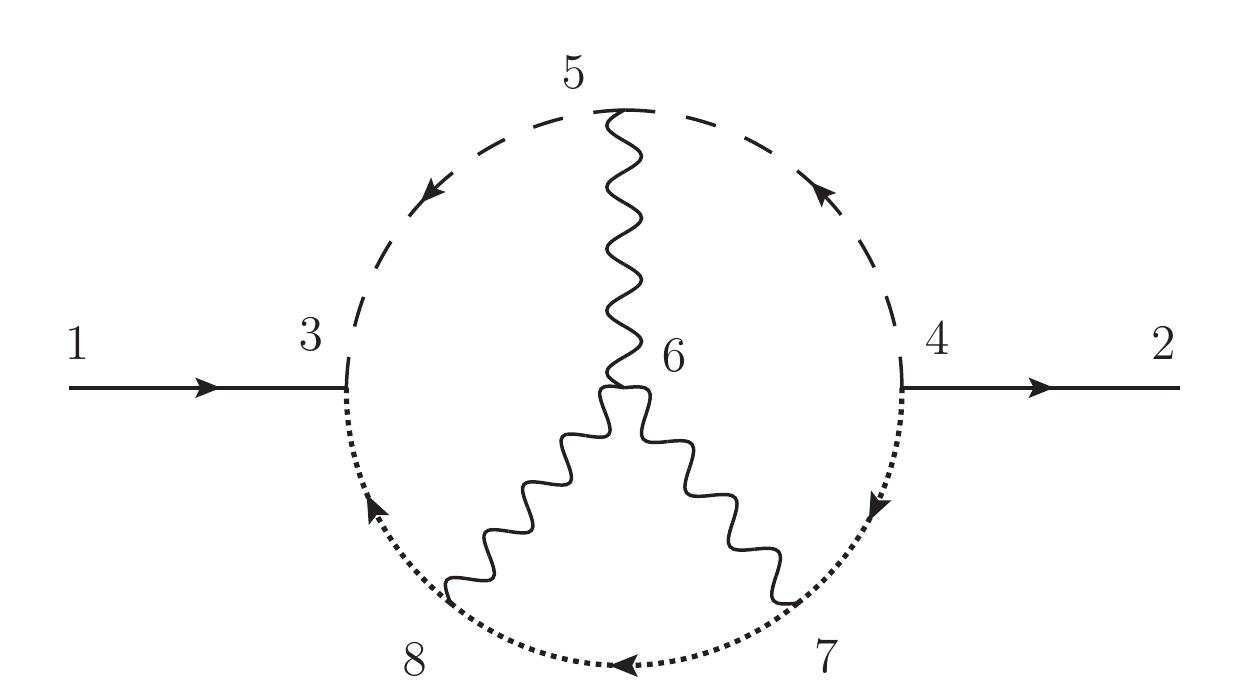}}
		\nonumber\\
		&= -\frac{1}{16}\, (8 g^6) \times \cT_{bc}^{(7)} \times 
		\parbox[c]{.33\textwidth}{\includegraphics[width = .33\textwidth]{BE.pdf}}
		\cZ^{(7)}(k)~.
\end{align}
The color factor reads
\begin{align}
	\label{T7bc}
		\cT_{bc}^{(7)} & =
		\ii f_{def} C^\prime_{bfecd} + \ii f_{def} C^\prime_{cfebd}~,
\end{align}
with the two terms corresponding to the fact that in the hypermultiplet loop the dashed or dotted parts can be exchanged. Since the cubic vector vertex contains covariant spinor derivatives and is not symmetric in the three vector lines that it contains, the diagram gets six distinct contributions 
arising from the six different ways it is contracted with the other vertices of the diagram. We write
these six terms as follows
\begin{align}
	\label{Z7def}
		\cZ^{(7)} = \cZ^{(7)}_{578} + \cZ^{(7)}_{758} + \cZ^{(7)}_{785} + \cZ^{(7)}_{875}
		+ \cZ^{(7)}_{857} + \cZ^{(7)}_{587}~.
\end{align}
The first term above is
\begin{align}
	\label{Z578}
		\cZ^{(7)}_{578}(k) & = 
		\Big[\big(\overline{D}_6\big)^2 D_6^\alpha\, \delta^4(\theta_{65}) \Big]
		\, \delta^4(\theta_{67})\, \Big[ D_{6,\alpha} 
		\, \delta^4(\theta_{68})\Big]\,\exp\big[\cA(k)\big]~.
\end{align}
Here we have denoted by $D_{6,\alpha}$ and $\overline{D}_{6,\dot{\alpha}}$ 
the covariant spinor derivatives defined in (\ref{covspink}) with respect to $\theta_6$ and
$\bar\theta_6$.  The last exponential factor $\exp\big[\cA(q,k)\big]$ 
contains all other contributions which amount to
\begin{align}
	\label{CA3g}
		\cA(k) & 
		= 2 \,\theta_4 \,k_{45}\, \bar{\theta}_5 + 2 \,\theta_5 \,k_{53} \,\bar{\theta}_3
		- \theta_5\,\big(k_{45} + k_{53}\big)\, \bar{\theta}_5
		+ 2 \,\theta_4\, k_{47}\, \bar{\theta}_7
		+ 2 \,\theta_7\, k_{78}\, \bar{\theta}_8 \nonumber\\
		&~~~ - \theta_7 \,\big(k_{47} + k_{78}\big)\, \bar{\theta}_7
		+ 2 \,\theta_8\, k_{83} \,\bar{\theta}_3
		- \theta_8 \,\big(k_{78} + k_{83}\big)\, \bar{\theta}_8~. 		 
\end{align}
Using the identity
\begin{align}
	\label{D6toD8}
		D_{6,\alpha} \,\delta^4(\theta_{68}) = 
		\big(\partial_{6,\alpha} - k_{68} \bar{\theta}_6\big)\, \delta^4(\theta_{68}) 
		= - \big(\partial_{8,\alpha} + k_{68} \bar{\theta}_6\big)\, 
		\delta^4(\theta_{68})
\end{align}
and then integrating by parts with respect to $\theta_8$, we can rewrite (\ref{Z578})
as follows 
\begin{align}
	\label{Z578bis}
		\cZ^{(7)}_{578}(k) & = 
		\delta^4(\theta_{67})\, \delta^4(\theta_{68})
		\Big[\big(\overline{D}_6\big)^2 D_6^\alpha\, \delta^4(\theta_{65}) \Big]
		\big(\partial_{8,\alpha} - (k_{68}\,\bar{\theta}_8)_\alpha\big)
		\exp\big[\cA(k)\big]~.
\end{align} 
By direct evaluation one can show that
\begin{align}
	\label{D3id}
		\big(\overline{D}_6\big)^2 D_6^\alpha\, \delta^4(\theta_{65}) = -4\,
		\rme^{-\theta_6\, k_{65}\, \bar{\theta}_{65}} \,
		\left[2 \,\theta_{65}^\alpha + (k_{65}\,\bar{\theta}_5)^\alpha 
		 \left(\theta_{65}\right)^2\right]~,
\end{align}
and
\begin{align}
	\label{DA}
			\big(\partial_{8,\alpha} - (k_{68}\,\bar{\theta}_8)_\alpha\big)\, 
		\exp\big[\cA(k)\big]= 2 (k_{83} \,\bar{\theta}_{38})_\alpha\,
			\exp\big[\cA(k)\big]
\end{align}
where in the last step we used momentum conservation.
Substituting (\ref{D3id}) and (\ref{DA}) into (\ref{Z578bis}), after a Fierz rearrangement we 
arrive at 
\begin{align}
	\label{Z578tris}
		\cZ^{(7)}_{578}(k) & = 
		-16\, \delta^4(\theta_{67})\, \delta^4(\theta_{68}) \,
		\left(\theta_{65}\,k_{83}\,\bar{\theta}_{38}\right)
		\left(1 + \theta_{65}\, k_{65}\, \bar{\theta}_5\right)
		\exp\big[\cA(k) - \theta_6\, k_{65}\,\bar{\theta}_{65}\big]
		\nonumber\\[1mm]
		& = -16 \,\delta^4(\theta_{67})\, \delta^4(\theta_{68}) \,
		\left(\theta_{65}\,k_{83}\,\bar{\theta}_{38}\right)
		\exp\big[\cA(k) - \theta_6\, k_{65}\,\bar{\theta}_{65}
		+ \theta_{65}\, k_{65}\, \bar{\theta}_5\big]~,
\end{align}	
where in the second step we could replace the factor $\big(1 + \theta_{65}\, k_{65} \,
\bar{\theta}_5\big)$ with $\exp\big[\theta_{65}\, k_{65} \,\bar{\theta}_5\big]$ because 
it is multiplied by $\theta_{65}$. 

We now perform the $\theta$-integrations using the $\delta$-functions present in 
(\ref{Z578tris}) and keep as remaining independent variables $\theta_4$, 
$\bar{\theta}_{63}$, $\theta_{65}$, $\theta_6$ and $\bar{\theta}_6$; with straightforward manipulations, involving also the use of momentum conservation, we rewrite
$\big[\cA(k) - \theta_6\, k_{65}\,\bar{\theta}_{65}
		+ \theta_{65}\, k_{65}\, \bar{\theta}_5\big]$ as
\begin{align}
	\label{exprepl}
-2\,\theta_4 \,q \,\bar{\theta_6} - 2\, \theta_4\, k_{45}\,\bar{\theta}_{65}
+2\,\theta_5 \,q \,\bar{\theta}_{63}+ 2 \,\theta_{65}\, k_{53}\,\bar{\theta}_{63}
+2\,\theta_6\, k_{45}\, \bar{\theta}_{65}
-2\,\theta_{65}\, k_{53}\, \bar{\theta}_{65}~.
\end{align}	
We also have
\begin{align}
	\label{floorrepl}
		2\,\theta_{65}\,k_{83}\,\bar{\theta}_{38}
		= -2 \,\theta_{65}\,k_{83}\,\bar{\theta}_{63}\, 
		\equiv\, \exp\big[-2 \lambda \,\theta_{65}\,k_{83}\,\bar{\theta}_{63}\big]
		\Big|_{\lambda}~
\end{align}		
where the notation $X\big|_\lambda$ means the term of $X$ that is linear in $\lambda$.
Altogether we have managed to express $\cZ^{(7)}_{578}(k)$ as an exponential:
\begin{align}
	\label{Z578quater}
		\cZ^{(7)}_{578}(k)  = 
		& -8\, \exp\big[\!-2 \,\theta_4 \,q\, \bar{\theta}_6
		- 2 \,\theta_4\, k_{45}\,\bar{\theta}_{65} 
		+ 2 \,\theta_5\, q \,\bar{\theta}_{63} \nonumber\\
		&\qquad + 2\, \theta_{65}\, (k_{53}-\lambda k_{83})\, \bar{\theta}_{63}
		+ 2 \,\theta_6\, k_{45}\, \bar{\theta}_{65} 
		- 2 \,\theta_{65}\, k_{53}\, \bar{\theta}_{65}\big]
		\Big|_{\lambda}~.
\end{align}
This exponential can be interpreted as a $\theta$-graph\,%
\footnote{Since we use as Grassmann variables the differences $\bar{\theta}_{63}$
and $\theta_{65}$ of original variables, in the resulting $\theta$-graph momentum conservation is not realized at each node. However, this is does not cause any problem.}:
\begin{align}
	\label{Z578fifth}
		\cZ^{(7)}_{578}(k)  =-8
		\parbox[c]{.32\textwidth}{\includegraphics[width = .32\textwidth]{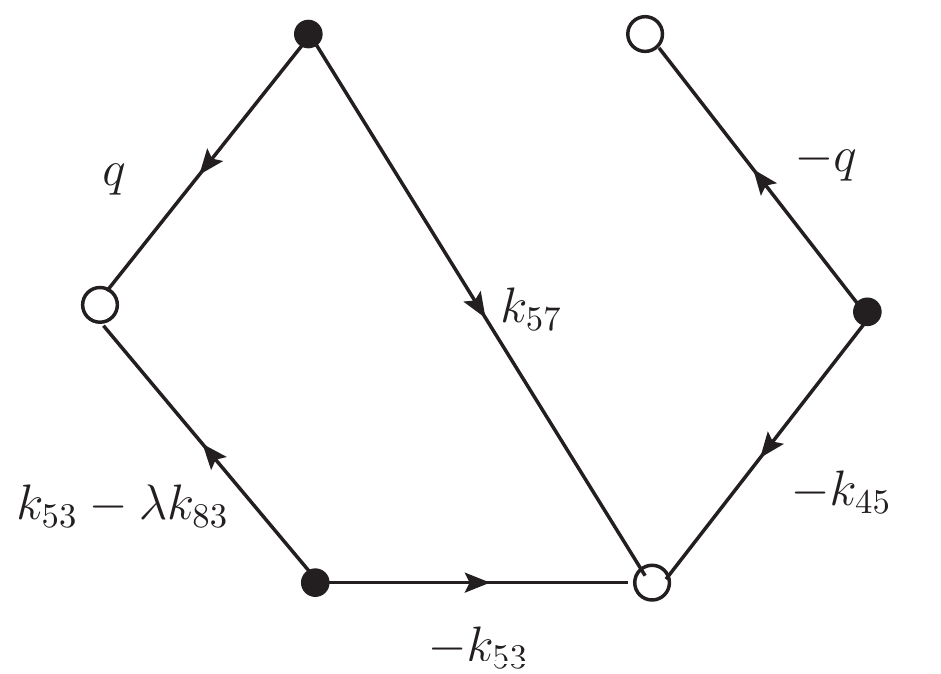}}\Bigg|_{\lambda}
		=  - 8 \,F\big(k_{53}-\lambda k_{83},-k_{53},0,q,k_{45},0,-k_{45},-q\big)\Big|_\lambda~.
\end{align}		 

We can apply this same procedure to evaluate the other five terms in (\ref{Z7def}) and obtain
\begin{align}
	\label{Z758}
		\cZ^{(7)}_{758}(k) & = 
		\delta^4(\theta_{65})\,
		\Big[\big(\overline{D}_6\big)^2 D_6^\alpha\, \delta^4(\theta_{67}) \Big]\,  
		\Big[ D_{6,\alpha}\, \delta^4(\theta_{68})\Big]\,
		\exp\big[\cA(k)\big]\nonumber\\
		&= - 8 \, F\big(-\lambda k_{83},-k_{78},0,q,k_{47},0,-k_{47},-q\big)\Big|_\lambda~,
		\\[1mm]
		\cZ^{(7)}_{785}(k) & = 
		\Big[\big(\overline{D}_6\big)^2 D_6^\alpha \,\delta^4(\theta_{67}) \big]\, 
		\delta^4(\theta_{68})\,		
		\big[ D_{6,\alpha} \,\delta^4(\theta_{65})\big]\,
		\exp\big[\cA(k)\big]\nonumber\\
		& = - 8 \, F\big(-\lambda k_{53},-k_{78},0,q,k_{47},0,-k_{47},-q\big)\Big|_\lambda~,
		\label{Z785}\\[1mm]
		\cZ^{(7)}_{875}(k) & = 
		\delta^4(\theta_{67})\,		
		\Big[\big(\overline{D}_6\big)^2 D_6^\alpha\, \delta^4(\theta_{68}) \big]\, 
		\Big[ D_{6,\alpha} \,\delta^4(\theta_{65})\Big]\,
		\exp\big[\cA(k)\big]\nonumber\\
		& = - 8 \, F\big(k_{83}-\lambda k_{53},-k_{83},0,q,0,0,0,-q\big)\Big|_\lambda = 0~,
		\label{Z875}\\[1mm]
		\cZ^{(7)}_{857}(k) & = 
		\delta^4(\theta_{65})\,		
		\Big[\big(\overline{D}_6\big)^2 D_6^\alpha\, \delta^4(\theta_{68}) \Big]\, 
		\Big[ D_{6,\alpha} \,\delta^4(\theta_{67})\Big]\,
		\exp\big[\cA(k)\big]\nonumber\\
		& = - 8 \, F\big(-k_{83},-k_{83}-\lambda k_{78},k_{83},-q,0,q,0,-q\big)\Big|_\lambda~,
		\label{Z857}\\[1mm]
		\cZ^{(7)}_{587}(k) & = 
		\Big[\big(\overline{D}_6\big)^2 D_6^\alpha\, \delta^4(\theta_{65}) \Big]\, 
		\Big[ D_{6,\alpha} \,\delta^4(\theta_{67})\Big]\,
		\delta^4(\theta_{68})\,		
		\exp\big[\cA(k)\big]\nonumber\\
		& = 0~.	\label{Z587}
\end{align}
The vanishing of the last contribution is due to the fact that in the step analogous to the one in
(\ref{DA}) we compute
\begin{align}
	\label{DAzero}
		\Big(\partial_{7,\alpha} - (k_{67}\,\bar{\theta}_7)_\alpha\Big) \exp\big[\cA(k)\big] 
		= 2 \big(k_{78}\, \bar{\theta}_{87}\big)_\alpha\,		\exp\big[\cA(k)\big] = 0~;
\end{align}
indeed in presence of $\delta^4(\theta_{68})\, \delta^4(\theta_{67})$, the difference 
$\bar{\theta}_{87}$ is null. The vanishing of this factor makes zero the entire expression. 

Now that we have computed all six terms of (\ref{Z7def}), we can insert the resulting expression 
for $\cZ^{(7)}(k)$ in the momentum integration, which has the BE topology, obtaining 
\begin{align}
	\label{scal7}
		\parbox[c]{.33\textwidth}{\includegraphics[width = .33\textwidth]{BE.pdf}}
		\cZ^{(7)}(k)
		=    \frac{160 \zeta(5)}{(4\pi)^6} \frac{1}{q^2} + \ldots~.  
\end{align}
Putting everything together, we finally get
\begin{align}
	\label{W7res}
		\cW_{bc}^{(7)}(q) & = -\frac{1}{q^2}\left(\frac{g^2}{8\pi^2}\right)^3 \zeta(5)
		\times 
		\left( 10\,\ii\, f_{def} C^\prime_{bfecd} + 10\,\ii\, f_{def}
		C^\prime_{cfebd} \right) + \ldots~. 
\end{align}

We have made a thorough analysis of all diagrams that can contribute to the propagator at order
$g^8$ and the ones we have listed above are the only ones that yield a term proportional to 
$\zeta(5)$ in the difference theory for a generic superconformal matter content. 
Other diagrams, indeed, either vanish due their color structure or give contributions that do not
contain $\zeta(5)$.

\end{appendix}

%\providecommand{\href}[2]{#2}\begingroup\raggedright
%\bibliography{references}
%\bibliographystyle{JHEP}
%\endgroup

\providecommand{\href}[2]{#2}\begingroup\raggedright\endgroup

\end{document}